\documentclass[11pt,aps,a4paper,eqsecnum,amsmath,amssymb,longbibliography,notitlepage,nofootinbib]{revtex4-1}

\usepackage{graphicx}
\graphicspath{ {./images/} }
\usepackage{dsfont}
\usepackage{hyperref}
\usepackage{xcolor}
\usepackage{tikz}
\usetikzlibrary{decorations.markings}

\usepackage{mathtools}
\usepackage{float}
\usepackage{subfigure}
\usepackage{enumitem}

\newcommand{\kommu}[2]{\left[#1,#2 \right] }
\newcommand{\dOBar}[0]{\delta_{\overline{0}}}
\newcommand{\dO}[0]{\delta_0}
\newcommand{\DA}[0]{\Delta_{0\overline{0}}}
\newcommand{\Dsame}[0]{\Delta}
\newcommand{\FlowParameter}[0]{\vartheta}
\newcommand{\DQ}[0]{\Delta_{12}}

\newcommand{\tr}[0]{\text{tr}}

\newcommand{\bNull}{0}
\newcommand{\bbNull}{\overline{0}}
\newcommand{\cross}[2]{
	\draw[-,thick](#1-0.1,#2-0.1)--(#1+0.1,#2+0.1);
	\draw[-,thick] (#1-0.1,#2+0.1)--(#1+0.1,#2-0.1);
}

\newcommand{\Incross}[2]{
	\draw[-,thick](#1-0.1,#2-0.1)--(#1+0.1,#2+0.1);
	\draw[-,thick] (#1-0.1,#2+0.1)--(#1+0.1,#2-0.1);
	\draw[-,thick] (#1-0.12,#2-0.12)--(#1+0.12,#2-0.12);
}

\newcommand{\ecelll}[3]{
	\draw[<-,thick] (#1,#2+1) -- (#1,#2-1);
	\draw[<-,thick] (#1-1,#2) -- (#1+1,#2);
	\node[anchor=south west] at (#1+0.1,#2+0.1) {\footnotesize#3};
}
\newcommand{\ecellr}[3]{
	\draw[<-,thick] (#1,#2+1) -- (#1,#2-1);
	\draw[->,thick] (#1-1,#2) -- (#1+1,#2);
	\node[anchor=south west] at (#1+0.1,#2+0.1) {\footnotesize#3};
}

\newcommand{\permucellr}[2]{
	\draw[->,thick] (#1-1,#2) .. controls (#1,#2) .. (#1,#2+1);
	\draw[->,thick] (#1,#2-1) .. controls (#1,#2)  .. (#1+1,#2);
}
\newcommand{\permucelll}[2]{
	\draw[<-,thick] (#1-1,#2) .. controls (#1,#2) .. (#1,#2-1);
	\draw[->,thick] (#1+1,#2) .. controls (#1,#2)  .. (#1,#2+1);
}

\newcommand{\ri}{\mathrm{i}\color{black}}

\newcommand{\tn}{|}

\makeatletter
\def\@bibdataout@aps{%
	\immediate\write\@bibdataout{%
		@CONTROL{%
			apsrev41Control%
			\longbibliography@sw{%
				,author="08",editor="1",pages="1",title="0",year="1"%
			}{%
				,author="08",editor="1",pages="1",title="",year="1"%
			}%
		}%
	}%
	\if@filesw \immediate \write \@auxout {\string \citation {apsrev41Control}}\fi 
}
\makeatother

\begin{document}

	\title{Integrable boundary conditions for staggered vertex models}

	\author{Holger Frahm}
	\author{Sascha Gehrmann}
	\affiliation{%
		Institut f\"ur Theoretische Physik, Leibniz Universit\"at Hannover,
		Appelstra\ss{}e 2, 30167 Hannover, Germany}

	\date{\today}
	
	\begin{abstract}
		Yang-Baxter integrable vertex models with a generic $\mathbb{Z}_2$-staggering can be expressed in terms of composite $\mathbb{R}$-matrices given in terms of the elementary $R$-matrices.  Similarly, integrable open boundary conditions can be constructed through generalized reflection algebras based on these objects and their representations in terms of composite boundary matrices $\mathbb{K}^\pm$. We show that only two types of staggering yield a local Hamiltonian with integrable open boundary conditions in this approach. The staggering in the underlying model allows for a second hierarchy of commuting integrals of motion (in addition to the one including the Hamiltonian obtained from the usual transfer matrix), starting with the so-called quasi momentum operator. In this paper, we show that this quasi momentum operator can be obtained together with the Hamiltonian for both periodic and open models in a unified way from enlarged Yang-Baxter or reflection algebras in the composite picture.
		For the special case of the staggered six-vertex model, this allows constructing an integrable spectral flow between the two local cases.  
	\end{abstract}
	\maketitle

	\section{Introduction}
	Integrable lattice models based on representations of the Yang-Baxter algebra have proven to be extremely useful in the understanding of non-perturbative phenomena in one-dimensional many-body systems.  Given a particular $R$-matrix, it is possible to consider variations in the spectral parameter leading to local inhomogeneities preserving the integrability of such models.  This has first been used by Baxter in the context of the six-vertex model \cite{Baxt71}.  In particular, models with periodically repeating inhomogeneities (or staggered models) have proven to be applicable to a wide range of problems.  Apart from the construction of integrable spin chains  with larger unit cells \cite{PoZv93,FrRo96} inhomogeneous vertex models have been used, e.g., to formulate the Potts model as a $\mathbb{Z}_2$-staggered six-vertex model \cite{Baxter73}, for the lattice regularization of field theories such as the principal chiral model \cite{FaRe86,DeVe89b}, in the quantum transfer matrix approach to the thermodynamics of integrable models \cite{Kluemper92}, and to study integrable perturbations of conformal field theories \cite{ReSa94}.
	Alternatively, the staggering can be realized by choosing alternating local representations of the underlying symmetry algebra.  Such a staggering appears quite naturally in the superspin formulation of network models describing the disorder induced plateau transition in integer quantum Hall systems \cite{ChCo88,Zirn97} which can be made integrable by fine-tuning of the coupling constants \cite{Gade99,EsFS05}.  Extensions of such staggered models to open boundary conditions with their integrability encoded in representations of the corresponding reflection algebra \cite{Ch84, Sklyanin:1988yz} allow for the construction of spin chains with soliton non-preserving boundary conditions \cite{An00}. Similar algebraic structures emerge in the study of certain AdS/CFT-type integrability theories \cite{PrRT15,Bai_etal19}.

	Interestingly, finite-size studies of certain staggered models based both on variations of the spectral parameter or the local representations have revealed that their continuum limit -- in spite of the compact formulation as a spin chain -- is described by conformal field theories with a non-compact target space \cite{EsFS05,IkJS08,FrMa11,FrMa12}.
	Among these the most studied example is the periodically staggered six-vertex model whose low energy effective theory has been identified to be the $SL(2,\mathbb{R})/U(1)$ black hole CFT \cite{IkJS08,IkJS12,CaIk13,FrSe14,BKSL21,BKKL21b}. More recently, the influence of open $U_q(sl(2))$-invariant boundary conditions in this model has been studied, both for the self-dual staggering related to the Potts model \cite{RoPaJaSa20,RoJS21} and also away from the self-dual line \cite{FrSg2022}. These studies have shown that boundary conditions have a profound effect: depending on their choice the symmetry of the ground state  may be spontaneously broken or the continuous component of the conformal spectrum disappears completely.
	
	A conserved quantity existing in these models which has been particularly useful for the identification of the conformal field theory is the so-called quasi momentum operator. Its role in staggered models without a non-compact continuum limit, however, has not been studied yet.
	The definition of the quasi momentum relies on the possibility to introduce a staggering in the vertical direction of the vertex model which is compatible with the horizontal one \cite{IkJS12,FrSe14,FrHo17,FrSg2022}.
	That such an operator cannot be defined in the homogeneous case has impeded progress in the analysis of the spectrum other models where indications for a continuous spectrum of conformal weights have been observed, namely the $a_{N-1}^{(2)}$ models and a family of orthosymplectic superspin chains
	\cite{VeJS14,VeJS16a,MaNR98,FrMa15,FrMa18,FrHM19,FrMa22}. 
	
	In this work, we will use a different perspective on the staggered models to address the question of whether the quasi momentum can be constructed in an alternative approach which may be applicable for homogeneous models, too.
	After a brief review of the construction of integrable models with periodic and open boundary conditions based on an 'elementary' $R$-matrix solving the Yang-Baxter equation and corresponding boundary matrices, we construct 'composite' $\mathbb{R}$-matrices using the co-multiplication property of the Yang-Baxter algebra.  These $\mathbb{R}$-matrices satisfy a generalized Yang-Baxter equation (\ref{Generalized_YBE}) and depend on the staggering parameters through additional arguments. For periodic boundary conditions this allows to define a homogeneous transfer matrix generating both the local integrals of motions such as the Hamiltonian and the quasi momentum operator. 
	
	In Section~\ref{Reversed_Factorization} we generalize this procedure to the open case, where we express composite boundary matrices in terms of the elementary ones.  Depending on the properties of the elementary $R$-matrices (and unlike in the periodic case) we identify two different choices of the staggering parameters leading to a transfer matrix constructed from the composite $\mathbb{R}$- and boundary matrices which generates a Hamiltonian with local interactions in the bulk (similar as in Refs.~\cite{NeRe21a,Li:2022clv}). For one of these choices a second homogeneous transfer matrix with boundary matrices satisfying a different reflection equation generates the quasi momentum.  The commutativity of these objects is guaranteed by a set of intertwining relations between the two sets of boundary matrices.
	
	Finally we apply our findings to the self-dual staggered six-vertex model.  Based on this construction the spectral flow between the models with compact and non-compact continuum limits can be studied in a family of integrable models.  Although one has to give up locality at the intermediate steps one finds that the two endpoints of this scheme are separated by two first-order transitions where massive degeneracies lead to a reordering of levels. Based on the numerical solution of the Bethe equations we provide some insights into the role of the quasi momentum in the model with compact continuum limit. 
	\section{Basic Ingredients}
	\subsection{Yang-Baxter integrable models}
	\label{sec:BasicModel}
	Let $\mathbb{V}=\mathcal{V}_0\otimes \mathcal{V}_{\bar{0}}\otimes\bigotimes^{L'}_{j=1} \mathcal{V}_j$ be the tensor product of $2+L'$ copies of a  vector space $\mathcal{V}$. Given an operator $A$ acting on the space $\mathcal{V}^{\otimes n}$ we define $A_{j_1,\dots,j_n}$ to be the operator on $\mathbb{V}$ which acts as $A$ on  $\bigotimes^{n}_{m=1}\mathcal{V}_{j_m}\cong \mathcal{V}^{\otimes n}$ and as the identity on all the other factors (assuming implicitly that all $j_k$ are different). We will use this notation throughout the study. 
	Denote by $R(u)$  a linear operator depending meromorphically on $u\in \mathbb{C}$ and acting on the twofold tensor product $\mathcal{V}\otimes \mathcal{V}$ which satisfies the Yang-Baxter equation (YBE)
	\begin{align}
		R_{i,j}(u-v)R_{i,k}(u)R_{j,k}(v)=R_{j,k}(v) R_{i,k}(u)R_{i,j}(u-v)\,.\label{YBE}
	\end{align}
	In the following we will call $R$ the $R$-matrix. We assume it to satisfy the initial condition
	\begin{subequations}
		\label{R_props}
		\begin{align}
			R_{i,j}(0)&=P_{i,j}\,,\label{Reg}
		\end{align}
		with the permutation operator $P_{i,j}$ on $\mathcal{V}_i\otimes \mathcal{V}_j$. Note that initial condition guarantees that $R_{i,j}(u)$ is differentiable near $u=0$. In addition to this regularity condition we require several properties of the $R$-matrix throughout this paper, namely unitarity, $PT$-symmetry, crossing symmetry and crossing unitarity\footnote{The stated properties are not independent from each other. The initial condition paired with the Yang-Baxter equation gives unitarity. In turn, the combination of (\ref{R_Uni}),(\ref{R_PT_Sym}) and (\ref{R_Crossing_Sym}) imply (\ref{R_Crossing_Uni}).}
		\begin{align}
			R_{i,j}(u)R_{j,i}(-u)&=\xi(u)\mathbf{1}\,,\label{R_Uni}\\
			R^{t_it_j}_{i,j}(u)&=R_{j,i}(u)\,,\label{R_PT_Sym}\\
			R_{i,j}(u)&=V_iR^{t_j}_{i,j}(-u-\eta)V^{-1}_i\label{R_Crossing_Sym}\\
			R^{t_i}_{i,j}(u)M_iR^{t_j}_{i,j}(-u-2\eta)M^{-1}_i&=\xi(u+\eta)\,\mathbf{1}\,,  \label{R_Crossing_Uni}
		\end{align}
		with the crossing parameter $\eta \in \mathbb{C}$ , a scalar function $\xi(u)$ and some invertible matrix $V \in \mathrm{End}(\mathcal{V})$ and $M=V^tV=M^t$ being a symmetry transformation of the $R$-matrix
		\begin{align}
			M^{-1}_{i}R_{i,j}(u)M_i&=M_jR_{i,j}(u)M^{-1}_j\,.\label{Important_Iden_Small_R}
		\end{align}
	\end{subequations}
	The first study of such type of $R$-matrix in the context of open spin chains was carried out in \cite{Nepomechie_Mezincescu_1991}.
	
	In a later section we will assume that the $R$-matrix is quasi periodic i.e.\ 
	\begin{align}
		\label{R_Qper}
		R_{i,j}(u+p)=f_pG_{i}R_{i,j}(u)G^{-1}_i, 
	\end{align}
	where $p \in \mathbb{C}$ is non-zero constant. Using the properties (\ref{R_props}) of the $R$-matrix, we find that $f^2_p= 1$ and $G$ is an invertible matrix, in fact $G^{-1}\propto V^{-1}GV$. Further, one concludes that $G$ must be a symmetric or anti-symmetric matrix. Using its (anti-)symmetry one deduces by using PT-symmetry in the transposed version of (\ref{R_Qper}) with $i$ and $j$ interchanged that $R$ is $G$-invariant, i.e. 
	\begin{align}
		G_iG_j R_{i,j}(u)G^{-1}_iG^{-1}_j=R_{i,j}(u).\label{R_G_in}
	\end{align}
	Trigonometric and elliptic $R$-matrices which obey the quasi periodicity condition (\ref{R_Qper})  have been constructed for example in \cite{Be81,Ba87}. We will state explicitly whenever we use the quasi periodicity assumption in addition to Eqs.~(\ref{R_props}). 
	
	Based on the Yang-Baxter equation (\ref{YBE}) an algebraic structure equipped with a coproduct can be introduced.  This allows for the construction of a monodromy matrix
	
	\begin{align}
		\label{Monodr}
		T_0(u,\{u_\ell\})=&R_{0,L'}(u+u_{L'})R_{0,L'-1}(u+u_{L'-1}) \cdots R_{0,1}(u+u_{1})\,,
	\end{align}
	which obeys the following equivalent ($RTT$-)relations  
	\begin{subequations}
		\begin{align}
			R_{i,j}(u-v)T_i(u,\{u_\ell\})T_j(v,\{u_\ell\})&=T_j(v,\{u_\ell\})T_i(u,\{u_\ell\})R_{i,j}(u-v)\,,\label{RTT}\\
			T_i(u,\{u_\ell\})R_{i,j}(u+v)T^{-1}_j(-v,\{u_\ell\})&=T^{-1}_j(-v,\{u_\ell\})R_{i,j}(u+v)T_i(u,\{u_\ell\})\,,\label{TRT}\\
			R_{i,j}(u-v)T^{-1}_i(u,\{u_\ell\})T^{-1}_j(v,\{u_\ell\})&=T^{-1}_j(v,\{u_\ell\})T^{-1}_i(u,\{u_\ell\})R_{i,j}(u-v)\,.\label{Inv_RTT}  
		\end{align}
	\end{subequations}
	Each $R$-matrix in (\ref{Monodr}) acts on the auxiliary space $\mathcal{V}_0$ and one of the factors in the quantum space $\mathcal{H}=\otimes_{j=1}^{L'} \mathcal{V}_j$.  The parameters $\{u_\ell\}$ are called inhomogeneities.
	Taking the trace of (\ref{Monodr}) one obtains the transfer matrix
	\begin{align}
		\label{tm1_pbc}
		\tau^{\mathrm{pbc}}(u,\{u_\ell\})&=\tr_0\bigg(T_0(u,\{u_\ell\})\bigg)\,,
	\end{align}
	which, as a consequence of (\ref{RTT}), commutes for different values of the spectral parameter $u$. Therefore, it generates integrals of motion for a model defined on the Hilbert space $\mathcal{H}$ corresponding to an $L'$ site lattice subject to periodic boundary conditions.
	
	For models with integrable open boundary condition one needs, in addition, representations $K^\pm(u)\in \text{End}(\mathcal{V})$ of the reflection algebras 
	\cite{Ch84}
	\begin{subequations}
		\label{Reflection_algebra}
		\begin{align}
			R_{i,j}(u-v) K^{-}_{i} (u) R_{j,i}(u+v)K^{-}_{j} (v)&=K^{-}_{j} (v)R_{i,j}(u+v)K^{-}_{i} (u)R_{j,i}(u-v)\,,\label{Reflection_algebra_Small_K_Minus}
		\end{align}
		and 
		\begin{equation}
			\label{Reflection_algebra_Small_K_Plus}
			\begin{aligned}
				R_{i,j}(-u+v)&\left(K^{+}_i(u)\right)^{t_i}M^{-1}_{i}R_{j,i}(-(u+v)-2\eta)M_{i}\left(K^{+}_j(v)\right)^{t_j}\\
				&=\left(K^{+}_j(v)\right)^{t_j}M_{i}R_{i,j}(-(u+v)-2\eta)M^{-1}_{i}  \left(K^{+}_i(u)\right)^{t_i} R_{j,i}(-u+v)\,.
			\end{aligned}
		\end{equation}
	\end{subequations}
	For the six- and eight-vertex models (or spin-$1/2$ chains) the most general $c$-number solutions to these equation have been constructed in Refs.~\cite{Sklyanin:1988yz,Nepomechie_Mezincescu_1991,deVe_1994}.  $K$-matrices for more general cases of quantum affine algebras have been constructed \cite{Delius2002,Delius_2003,ApVl20}.
	In the following we will assume the unitarity property 
	\begin{align}
		\label{K_Uni}
		K^-(u)K^-(-u)\propto \mathbf{1}\,,
	\end{align}
	which can be shown to hold for a large number of $K$-matrices, see e.g.\ \cite{ApVl22} and references therein. 
	Given the form of the  reflection equation (\ref{Reflection_algebra_Small_K_Minus}) with a quasi periodic $R$-matrix (\ref{R_Qper}) we further assume that\footnote{Note that both conditions (\ref{K_Qper}) and (\ref{K_G}) are  satisfied by the general $K$-matrices for the anisotropic spin-$1/2$ chains \cite{deVe_1994}. }
	\begin{align}
		\label{K_Qper}
		K^{-}(u+p)\propto GK^{-}(u)G\,.
	\end{align}
	Finally, it is natural, especially with regard to \emph{local} Hamiltonian's discussed in section \ref{Boundary_terms},  to assume that $K^-(u)$ is meromorphic in $u$, too, and obeys
	\begin{align}
		K^{-}(0)\propto \mathbf{1}\,, \quad    K^{-}(\frac{p}{2})\propto G\,. \label{K_G}
	\end{align}
	The corresponding properties for $K^+$ follow from the isomorphisms of the algebras given in \cite{Sklyanin:1988yz}. 
	Given representations of (\ref{Reflection_algebra}) one obtains the transfer matrix 
	\begin{align}
		\tau(u,\{u_\ell\})=\tr_{0}\big(K^{+}_0(u)T_0(u,\{u_\ell\})K^{-}_0(u)T^{-1}_0(-u,\{u_\ell\})\big)\,, \label{Standard Form}
	\end{align}
	which can be shown \cite{Sklyanin:1988yz,Nepomechie_Mezincescu_1991} to commute for different values of the spectral parameter i.e. 
	\begin{align}
		\kommu{\tau(u,\{u_\ell\})}{\tau(v,\{u_\ell\})}=0\,.
	\end{align}

	Hence, it generates commuting integrals of motion of a lattice model on $\mathcal{H}$ with boundary conditions defined by $K^\pm(u)$.
	
	Note that the unitarity relation (\ref{R_Uni}) allows to rewrite the inverse monodromy matrix $T_0^{-1}$ in (\ref{Standard Form}) as
	\begin{align}
		T_0^{-1}(-u,\{u_\ell\}) = R_{1,0}(u-u_1)R_{2,0}(u-u_2) \cdots R_{L',0}(u-u_{L'})\,\left(\prod^{L'}_{j=1}\xi(-u+u_j)\right)^{-1}\,.
	\end{align}
	In the following of this study we will consider lattices of even length $L'=2L$ and restrict to a $\mathbb{Z}_2$-staggering of the inhomogeneities, i.e.\
	\begin{align}
		u_{2j}=\delta_1,\quad u_{2j-1}=\delta_2\,,  \qquad j=1,\dots,L\,,\label{Z_2_Staggering}
	\end{align}
	where $\delta_1,\delta_2\in \mathbb{C}$. The tuple of inhomogeneities $\{\delta_2,\delta_1,.....\delta_2,\delta_1\}$ will be abbreviated by $\{\delta_1,\delta_2\}$ below. Further, we define the function
	\begin{align}
		c_{\tau}(u,\{\delta_1,\delta_2\})=\left[\xi(-u+\delta_1)\xi(-u+\delta_2)\right]^{-L}\,.
	\end{align}.
	
	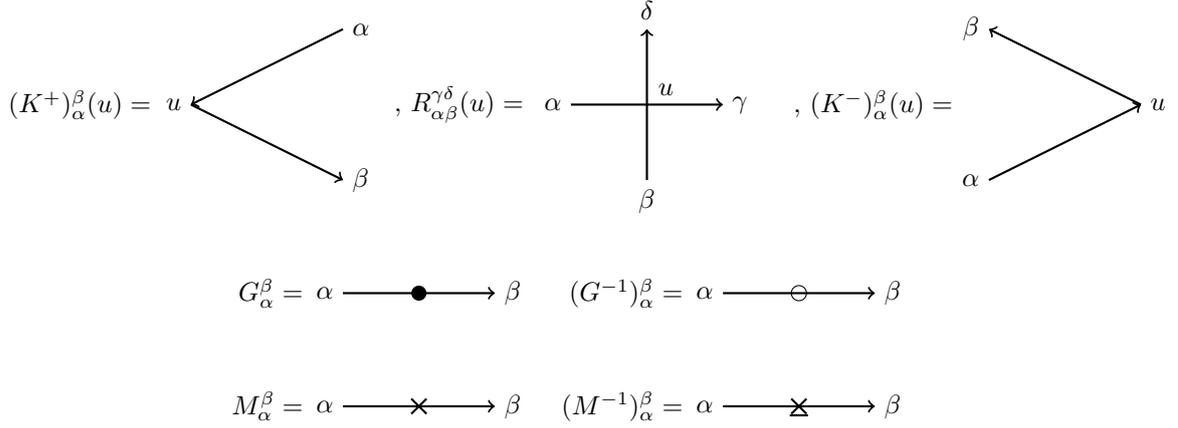
\begin{figure}[t]
		\centering
		\begin{tikzpicture}
			\node[right] at (-4,1) {$\alpha$}; 
			\node[right] at (-4,-1) {$\beta$}; 
			\draw[thick,<-] (1-5,-1)--(-1-5,0);
			\draw[thick,<-] (-1-5,0)--(1-5,1);
			\node[left] at (-1-5,0) {$u$};
			\node[left] at (-1.4-5,0) {$(K^+)^\beta_\alpha(u)=$};
			
			\node[left] at (4.5,1) {$\beta$}; 
			\node[left] at (4.5,-1) {$\alpha$}; 
			\draw[thick,->] (4.5,-1)--(6.5,0);
			\draw[thick,->] (6.5,0)--(4.5,1);
			\node[right] at (6.5,0) {$u$};
			\node[left] at (3.65+0.5,0) {,\,\,$(K^-)^\beta_\alpha(u)=$};

			\node[left] at (-1.5,0) {,\,\,$R^{\gamma\delta}_{\alpha\beta}(u)=$}; 
			\node[above] at (0.25,0) {$u$}; 
			\node[left] at (-1,0) {$\alpha$}; 
			\draw[->,thick] (-1,0)--(1,0);
			\node[right] at (1,0) {$\gamma$}; 
			\node[below] at (0,-1) {$\beta$}; 
			\draw[->,thick] (0,-1)--(0,1);
			\node[above] at (0,1) {$\delta$}; 
			
			\draw[->,thick] (-6+2,-2.5)--(-4+2,-2.5);
			\node[left] at (-6+2,-2.5) {$\alpha$};
			\node[right] at (-4+2,-2.5) {$\beta$};
			\fill (-5+2,-2.5) circle (0.1);
			\node[left] at (-1.4-5+2,-2.5) {$G^\beta_\alpha=$};
			
			\draw[->,thick] (-6+5+2,-2.5)--(-4+5+2,-2.5);
			\node[left] at (-6+5+2,-2.5) {$\alpha$};
			\node[right] at (-4+5+2,-2.5) {$\beta$};
			\draw(-5+5+2,-2.5) circle (0.1);
			\node[left] at (-1.4-5+5+2,-2.5) {$(G^{-1})^\beta_\alpha=$};
			\draw[->,thick] (-6+2,-2.5-1.5)--(-4+2,-2.5-1.5);
			\node[left] at (-6+2,-2.5-1.5) {$\alpha$};
			\node[right] at (-4+2,-2.5-1.5) {$\beta$};
			\cross{-5+2}{-2.5-1.5};
			\node[left] at (-1.4-5+2,-2.5-1.5) {$M^\beta_\alpha=$};
			
			\draw[->,thick] (-6+5+2,-2.5-1.5)--(-4+5+2,-2.5-1.5);
			\node[left] at (-6+5+2,-2.5-1.5) {$\alpha$};
			\node[right] at (-4+5+2,-2.5-1.5) {$\beta$};
			\Incross{-5+5+2}{-2.5-1.5};
			\node[left] at (-1.4-5+5+2,-2.5-1.5) {$(M^{-1})^\beta_\alpha=$};
			
		\end{tikzpicture}
		\caption{$R$, $G$, $G^{-1}$, $M$, $M^{-1}$ and $K$-matrices in graphical notation}
		\label{Graphical Notation}
	\end{figure}
	
	\subsection{The composite \texorpdfstring{$\mathbb{R}$}{R}-matrix}
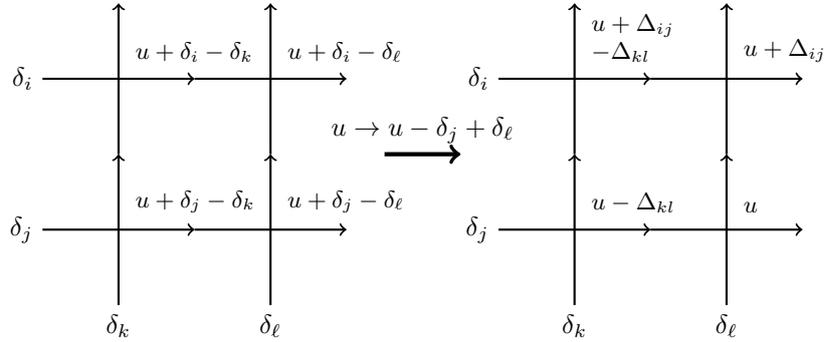
\begin{figure}
		\centering
		\begin{tikzpicture}
			\ecellr{2}{2}{$u+\delta_i-\delta_\ell$}
			\ecellr{2}{0}{$u+\delta_j-\delta_\ell$}
			\ecellr{0}{2}{$u+\delta_i-\delta_k$}
			\ecellr{0}{0}{$u+\delta_j-\delta_k$}
			
			\node[left] at (-1,2) {$\delta_i$};
			\node[left] at (-1,0) {$\delta_j$};
			\node[below] at (0,-1) {$\delta_k$};
			\node[below] at (2,-1) {$\delta_\ell$};
			
			\draw[ultra thick,->] (3.5,1)--(4.5,1) node[midway,sloped,above] {$u\to u-\delta_j+\delta_\ell$};
			
			\ecellr{8}{2}{$u+\Delta_{ij}$}
			\ecellr{8}{0}{$u$}
			\ecellr{6}{2}{$-\Delta_{kl}$}
			\ecellr{6}{0}{$u-\Delta_{kl}$}
			
			\node[right] at (6.1,2.7) {\footnotesize$u+\Delta_{ij}$};
			
			\node[left] at (5,2) {$\delta_i$};
			\node[left] at (5,0) {$\delta_j$};
			\node[below] at (6,-1) {$\delta_k$};
			\node[below] at (8,-1) {$\delta_\ell$};
		\end{tikzpicture}
		\caption{The composite $\mathbb{R}$-matrix (\ref{Def_Big_R}), where $\Delta_{ij}=\delta_i-\delta_j$}
		\label{Graphical_Rep_Big_R}
	\end{figure}

	Using the coproduct of the Yang-Baxter algebra a solution of the YBE (\ref{YBE}) can be extended to act as an endomorphism on $\mathcal{W}\otimes\mathcal{W}$ with $\mathcal{W}=\mathcal{V}\otimes\mathcal{V}$ (or, more generally, $n$-fold tensor products of the vector space $\mathcal{V}$).  Specifically, we define
	
	\begin{align}
		\mathbb{R}_{i,j\tn k,\ell}(u,\Delta_{ij},\Delta_{k\ell})=R_{i,\ell}(u+\Delta_{ij})R_{i,k}(u+\Delta_{ij}-\Delta_{k\ell})R_{j,\ell}(u)R_{j,k}(u-\Delta_{k\ell})\,,\label{Def_Big_R}
	\end{align}
	where $\Delta_{ij},\Delta_{k\ell}$ are arbitrary parameters. The explicit form is motivated by a general choice of inhomogeneities in both the horizontal and vertical direction as displayed in Fig.~\ref{Graphical_Rep_Big_R} using the graphical notation introduced in Fig.~\ref{Graphical Notation}. 
	
	The index notation implies that $\mathbb{R}_{i,j\tn k,\ell}$ acts on the tensor product of the two copies $(\mathcal{V}_i\otimes\mathcal{V}_j)$ and $(\mathcal{V}_k\otimes\mathcal{V}_\ell)$ of $\mathcal{W}$. We will use this notation throughout this paper. By construction this $R$-matrix satisfies the generalized Yang-Baxter equation 
	\begin{equation}
		\label{Generalized_YBE}
		\begin{aligned}
			\mathbb{R}_{i,j\tn k,\ell}(u-v,\Delta_{ij},\Delta_{k\ell}) &   \mathbb{R}_{i,j\tn m,n}(u,\Delta_{ij},\Delta_{mn})    \mathbb{R}_{k,\ell\tn m,n}(v,\Delta_{k\ell},\Delta_{mn})=\\    &\mathbb{R}_{k,\ell\tn m,n}(v,\Delta_{k\ell},\Delta_{mn})   \mathbb{R}_{i,j\tn m, n}(u,\Delta_{ij},\Delta_{mn})     \mathbb{R}_{i,j\tn k,\ell}(u-v,\Delta_{ij},\Delta_{k\ell}) \,,
		\end{aligned}
	\end{equation}
	and therefore allows to introduce commuting transfer matrices as in Section~\ref{sec:BasicModel}. Since this construction relies on the properties (\ref{R_props}) of $R(u)$ it is natural to ask which of these are inherited to the $\mathbb{R}$-matrix. It turns out that the $\mathbb{R}$-matrix obeys the following properties
	\begin{subequations}
		\label{Big_props}
		\begin{align}
			\mathbb{R}_{i,j\tn k,\ell}(0,\Delta,\Delta)&= \xi(\Delta)\, \mathbb{P}_{i,j\tn k,\ell},\label{Big_Regu}\\
			\mathbb{R}_{i,j\tn k,\ell}(u,\Delta_{ij},\Delta_{k\ell})\mathbb{R}_{k,\ell\tn i,j}(-u,\Delta_{k\ell},\Delta_{ij})&=\Xi(u,\Delta_{ij},\Delta_{k\ell})\,\mathbf{1}\,,\label{Big_R_Uni}\\
			\mathbb{R}^{t_it_jt_kt_\ell}_{i,j\tn k,\ell}(u,\Delta_{ij},\Delta_{k\ell})&=\mathbb{R}_{k,\ell\tn i,j}(u,-\Delta_{k\ell},-\Delta_{ij})\,,\label{Big_PT}\\
			\mathbb{R}^{t_it_j}_{i,j\tn k,\ell}(u,\Delta_{ij},\Delta_{k\ell})\mathbb{M}_{i,j}\mathbb{R}_{i,j\tn k,\ell}^{t_kt_\ell}(-u-2\eta,-\Delta_{ij},-\Delta_{k\ell})\mathbb{M}^{-1}_{i,j}&=\Xi(u+\eta,\Delta_{ij},\Delta_{k\ell})\,\mathbf{1}\,,\label{Big_Crossing_uni}\\
			\mathbb{M}^{-1}_{i,j}\mathbb{R}_{i,j\tn k,\ell}(u,\Delta_{ij},\Delta_{k\ell})\mathbb{M}_{i,j}&=    \mathbb{M}_{k,\ell}\mathbb{R}_{i,j\tn k,\ell}(u,\Delta_{ij},\Delta_{k\ell})\mathbb{M}^{-1}_{k,\ell}\,,\label{Big_Symmetry_Property} \\
			\mathbb{G}^{-1}_{i,j}\mathbb{R}_{i,j\tn k,\ell}(u,\Delta_{ij},\Delta_{k\ell})\mathbb{G}_{i,j}&=    \mathbb{G}_{k,\ell}\mathbb{R}_{i,j\tn k,\ell}(u,\Delta_{ij},\Delta_{k\ell})\mathbb{G}^{-1}_{k,\ell}\,,\label{Big_G_Symmetry_Property}\\
			\mathbb{R}_{i,j\tn k,\ell}(u+p,\Delta_{ij},\Delta_{k\ell})&=G_{i}G_{j}\mathbb{R}_{i,j\tn k\ell}(u,\Delta_{ij},\Delta_{k\ell})G^{-1}_{i}G^{-1}_{j}\,,\label{Big_Qper}
		\end{align}
	\end{subequations}
	which we discuss in the following order by order. Eq.~(\ref{Big_Regu}) is a  regularity property (with a different normalization) where $\mathbb{P}_{i,j\tn k,\ell}$ is the enlarged permutation operator acting on the state $(a\otimes b)\otimes (c \otimes d) \in \mathcal{W}\otimes\mathcal{W}=(\mathcal{V}_i\otimes\mathcal{V}_j)\otimes(\mathcal{V}_k\otimes\mathcal{V}_\ell)$ as
	\begin{align*}
		\mathbb{P}_{i,j\tn k,\ell}(a\otimes b)\otimes (c \otimes d)=(c \otimes d)\otimes(a\otimes b)\,.
	\end{align*}
	Eq.~(\ref{Big_R_Uni}) is the unitarity condition where the proportionality constant is given by 
	\begin{align*}
		\Xi(u,\Delta_{ij},\Delta_{k\ell})=\xi(u+\Delta_{ij})\xi(u+\Delta_{ij}-\Delta_{k\ell})\xi(u)\xi(u-\Delta_{k\ell})\,.
	\end{align*}
	The third property, Eq.~(\ref{Big_PT}), is a \emph{generalized} $PT$-symmetry of $\mathbb{R}$, where the parameters in $\mathbb{R}$ change sign due to the reordering caused by the transposition: 
	\begin{align*}
		\mathbb{R}^{t_it_jt_kt_\ell}_{i,j\tn k,\ell}(u,\Delta_{ij},\Delta_{k\ell})&=\left(R_{i,\ell}(u+\Delta_{ij})R_{i,k}(u+\Delta_{ij}-\Delta_{k\ell})R_{j,\ell}(u)R_{j,k}(u-\Delta_{k\ell})\right)^{t_it_jt_kt_\ell}\\
		&=R^{t_jt_k}_{j,k}(u-\Delta_{k\ell})R^{t_it_k}_{i,k}(u+\Delta_{ij}-\Delta_{k\ell})R^{t_jt_\ell}_{j,\ell}(u)R^{t_it_\ell}_{i,\ell}(u+\Delta_{ij})\\
		&=R_{k,j}(u-\Delta_{k\ell})R_{k,i}(u+\Delta_{ij}-\Delta_{k\ell})R_{\ell, j}(u)R_{\ell, i}(u+\Delta_{ij})\\
		&=\mathbb{R}_{k,\ell\tn i,j}(u,-\Delta_{k\ell},-\Delta_{ij})\,.
	\end{align*}
	As the $PT$-symmetry is related to the crossing unitary, the  $\mathbb{R}$-matrix satisfies a \emph{generalized} crossing unitarity relation given in the fourth equation where $\mathbb{M}_{i,j}=M_{i}M_{j}$.  \\
	Finally, also the symmetry relation (\ref{Important_Iden_Small_R}) and the quasi periodicity (\ref{R_Qper}) and its implication (\ref{R_G_in}) can be directly transferred to the $\mathbb{R}$-matrix, yielding the last three equations of (\ref{Big_props}) where $\mathbb{G}_{i,j}=G_iG_j$.\footnote{Note that the $\mathbb{R}$-matrix, depending on the choice of the parameters in (\ref{Def_Big_R}), may have an extended symmetry.  This has been discussed recently in the context of the antiferromagnetic Potts model where this construction leads to an integrable model based on the affine $D_2^{(2)}$ Lie algebra starting from the $U_q[\mathfrak{sl}(2)]$ (or $A_1^{(1)}$) invariant $R$-matrix of the six-vertex model \cite{RPJS20}.}

	\subsection{Staggered vertex models with periodic boundary conditions}
	These properties suffice to construct $\mathbb{Z}_2$-staggered models with periodic boundary conditions. Here we show how their formulations in terms of the elementary and the composite $R$-matrices are related. 
	Consider the product of two transfer matrices (\ref{tm1_pbc}) with different spectral parameters corresponding to a staggering in the auxiliary direction ($\bNull$ and $\bbNull$ label different auxiliary spaces) 
	\begin{align}
		\label{tm1_prod}\mathcal{T}^{\mathrm{pbc}}(u,\{\dO,\dOBar,\delta_1,\delta_2\})&=
		\tau^{\mathrm{pbc}}(u+\dO,\{\delta_1,\delta_2\})\,
		\tau^{\mathrm{pbc}}(u+\dOBar,\{\delta_1,\delta_2\})\\
		&=\tr_{\bNull}\bigg(R_{\bNull,2L}(u+\dO+\delta_1)R_{\bNull,2L-1}(u+\dO+\delta_2)\cdots R_{\bNull,1}(u+\dO+\delta_2)\bigg)\nonumber\\
		&\quad\times \tr_{\bbNull}\bigg(R_{\bbNull,2L}(u+\dOBar+\delta_1)R_{\bbNull,2L-1}(u+\dOBar+\delta_2)\cdots R_{\bbNull,1}(u+\dOBar+\delta_2)\bigg)\,.\nonumber
	\end{align}
	By reordering the $R$-matrices we obtain
	\begin{align*}
		\mathcal{T}^{\mathrm{pbc}}(u,&\{\dO,\dOBar,\delta_1,\delta_2\})=\\ \tr_{\bNull\bbNull}\bigg(&R_{\bNull,2L}(u+\dO+\delta_1)R_{\bbNull,2L}(u+\dOBar+\delta_1)R_{\bNull,2L-1}(u+\dO+\delta_2)R_{\bbNull,2L-1}(u+\dOBar+\delta_2)\cdots\\
		&\cdots R_{\bNull,2}(u+\dO+\delta_1)R_{\bbNull,1}(u+\dOBar+\delta_2)R_{\bNull,1}(u+\dO+\delta_2) R_{\bbNull,1}(u+\dOBar+\delta_2)\bigg)\,.
	\end{align*}
	The products of four $R$-matrices appearing in each row can be expressed in terms of the composite $\mathbb{R}$-matrix (\ref{Def_Big_R}). Shifting the spectral parameter as $u\to u-\dOBar-\delta_1$ we obtain a \emph{homogeneous} transfer matrix 
	\begin{align}
		\mathcal{T}^{\mathrm{pbc}}(u,\{\DA,\DQ\})=\tr_{\bNull\bbNull}\bigg(& \mathbb{R}_{\bNull,\bbNull\tn 2L-1,2L}(u,\Delta_{\bNull \bbNull},\Delta_{12})\cdots\mathbb{R}_{\bNull,\bbNull|1,2}(u,\DA,\DQ)\bigg)\,.
	\end{align}
	For the physical interpretation as a lattice model with local (i.e.\ finite range) interactions additional conditions have to be satisfied.  Typically, locality can be derived from the regularity property of the $R$-matrix, i.e.\ that it becomes a permutation operator at a shift point $u=u_0$.  In the present case, (\ref{Big_Regu}), we have $u_0=0$ and need to tune the staggering parameters such that 
	\begin{align}
		\DA=\DQ\equiv\Delta\,.\label{Loc_Con_PBC}
	\end{align}
	With this constraint a Hamiltonian coupling the degrees of freedom from nearest neighbour quantum spaces $\mathcal{W}$ is obtained from
	\begin{align}
		H^{\mathrm{pbc}}\propto\left.\frac{\partial}{\partial u}\log (\mathcal{T}^{\mathrm{pbc}}(u,\Delta,\Delta))\right|_{u=0}\,,
	\end{align}
	where we have assumed, in addition, that $R$ is differentiable at $\pm \Delta$.
	The staggering in the auxiliary direction allows to construct another operator, generating a family of commuting integrals deriving from (\ref{tm1_pbc}): instead of (\ref{tm1_prod}) we can consider the so-called quasi shift operator, given by the quotient of single row transfer matrices
	\begin{align*}
		\frac{\tau^{\mathrm{pbc}}(u+\dOBar,\{\delta_1,\delta_2\})}{\tau^{\mathrm{pbc}}(u+\dO,\{\delta_1,\delta_2\})}\,.
	\end{align*}
	As done for the product of transfer matrices, we shift the spectral parameter $u\to u-\dOBar-\delta_1$, leading to 
	\begin{align}
		\widetilde{\mathcal{Q}}^{\mathrm{pbc}}(u,\{\dO,\dOBar,\delta_1,\delta_2\})
		= \frac{\tau^{\mathrm{pbc}}(u-\delta_1,\{\delta_1,\delta_2\})}{\tau^{\mathrm{pbc}}(u+\DA-\delta_1,\{\delta_1,\delta_2\})}\,.
	\end{align} 
	Restricting the staggering parameters $\delta_1,\delta_2,\delta_0$ and $\delta_{\overline{0}}$ to be compatible with (\ref{Loc_Con_PBC}) and taking the logarithm of this operator at the shift point, $u_0=0$, we obtain the 'quasi momentum'
	\begin{align}
		\mathbb{Q}^{\mathrm{pbc}}=   \log\left[\widetilde{\mathcal{Q}}^{\mathrm{pbc}}(0,\{\delta_1,\delta_2\})\right]=\log \left[\frac{\tau^{\mathrm{pbc}}(-\delta_1,\{\delta_1,\delta_2\})}{\tau^{\mathrm{pbc}}(-\delta_2,\{\delta_1,\delta_2\})}\right]\,.
		\label{QI_PBC}
	\end{align}
	This operator has proven to be particularly useful for the characterization of low energy effective behavior of several staggered vertex models, see e.g.\ \cite{IkJS12,CaIk13,FrSe14,FrHo17,FrSg2022} and has found recently application as a Floquet Hamiltonian \cite{YGK22}.
	
	We now want to generate the quasi momentum from an operator constructed from the composite $\mathbb{R}$-matrix (\ref{Def_Big_R}).  It is straightforward to invert the single row transfer matrix (\ref{tm1_pbc}) in the denominator of (\ref{QI_PBC}) by using regularity and unitarity of the $R$-matrix:
	\begin{align}
		\left(\tau^{\mathrm{pbc}}(-\delta_2,\{\delta_1,\delta_2\})\right)^{-1}&\propto\tr_0\left(R_{1,0}(0)R_{2,0}(-\Dsame)\cdots R_{2L-1,0}(0)R_{2L,0}(-\Dsame) \right)\,.
	\end{align}
	With the definition of
	\begin{align*}
		Q^{\mathrm{pbc}}(u)&=\tr_\bNull\left(R_{1,\bNull }(-u)R_{2,\bNull }(-u-\Delta)\cdots R_{ 2L-1,\bNull}(-u)R_{2L,\bNull}(-u-\Delta) \right)\\&\qquad \times\tr_{\bbNull}\left(R_{\bbNull, 2L}(u)R_{\bbNull ,2L-1}(u-\Dsame)\cdots R_{\bbNull, 2}(u)R_{\bbNull, 1}(u-\Dsame) \right)\,,
	\end{align*}
	we obtain a \emph{product} of $R$-matrices depending on a spectral parameter which after taking the logarithm becomes proportional to (\ref{QI_PBC}) at the shift point.
	Using crossing symmetry (\ref{R_Crossing_Sym}) and expressing the result in terms of the composite $\mathbb{R}$-matrices we find: 
	\begin{equation}
		\begin{aligned}
			Q^{\mathrm{pbc}}(u)&=\tr_{\bNull\bbNull}\big( R_{\bNull, 2L}(u+\Delta-\eta)R_{\bNull, 2L-1}(u-\eta)\cdots R_{\bNull, 2}(u+\Delta-\eta)R_{\bNull, 1}(u-\eta)\\&\qquad\qquad \times R_{\bbNull, 2L}(u)R_{\bbNull, 2L-1}(u-\Dsame)\cdots R_{\bbNull, 2}(u)R_{\bbNull, 1}(u-\Dsame) \big)\\
			&=\tr_{\bNull \bbNull}\Big(\mathbb{R}_{\bNull,\bbNull\tn2L-1,2L}(u,\Dsame-\eta,\Dsame) \cdots \mathbb{R}_{\bNull,\bbNull\tn 1,2}(u,\Dsame-\eta,\Dsame)\Big)\\
			&=\mathcal{T}^{\mathrm{pbc}}\left(u,\{\Delta-\eta,\Delta\}\right)\,.
		\end{aligned}
	\end{equation}
	Note that this becomes the product of single row transfer matrices with arguments differing by the crossing parameter $\eta$ in the homogeneous limit, $\Delta\to0$,
	\begin{align}
		\lim_{\Delta\to 0} Q^{\mathrm{pbc}}(u)=\tr_{\bNull\bbNull}\big( R_{\bNull, 2L}(u-\eta)R_{\bbNull, 2L}(u)\cdots R_{\bNull, 1}(u-\eta)R_{\bbNull, 1}(u)\big)
		= \tau^{\mathrm{pbc}}(u-\eta)\,\tau^{\mathrm{pbc}}(u)\,.
	\end{align} 
	This product can be related to the higher-spin transfer matrices through the $T$-system bilinear functional relations \cite{KuNS11}.
	
	In summary the transfer matrices
	\begin{align}
		\label{TMpbc_uni}
		\mathcal{T}^{\mathrm{pbc}}(u,\{\theta,\Dsame\})&=\tr_{\bNull \bbNull}\bigg(\mathbb{R}_{\bNull,\bbNull|2L-1,2L}(u,\theta,\Dsame) \cdots \mathbb{R}_{\bNull,\bbNull\tn 1,2}(u,\theta,\Dsame)\bigg)
	\end{align}
	provide a unified framework generating both local integrals of motion such as the Hamiltonian under the locality condition (\ref{Loc_Con_PBC}), i.e.\ $\theta=\Delta$, and the quasi momentum (\ref{QI_PBC}) for $\theta=\Dsame-\eta$.
	Note that the third arguments of all $\mathbb{R}$-matrices in (\ref{TMpbc_uni}) coincide. Therefore, commutativity of $\mathcal{T}^{\mathrm{pbc}}\left(u,\{\theta,\Dsame\}\right)$ for different $u$ and $\theta$ (which includes $Q^{\mathrm{pbc}}(u)$)  follows directly from the generalized YBE (\ref{Generalized_YBE}). Moreover, let us note that $\mathbb{Q}^{\mathrm{pbc}}$, unlike the Hamiltonian, is a \emph{non-local} operator.

	\section{Integrable open boundary conditions for staggered models \label{Reversed_Factorization}}
	\subsection{Composite picture for open models \label{Compostive_Picture_for_open_models}}
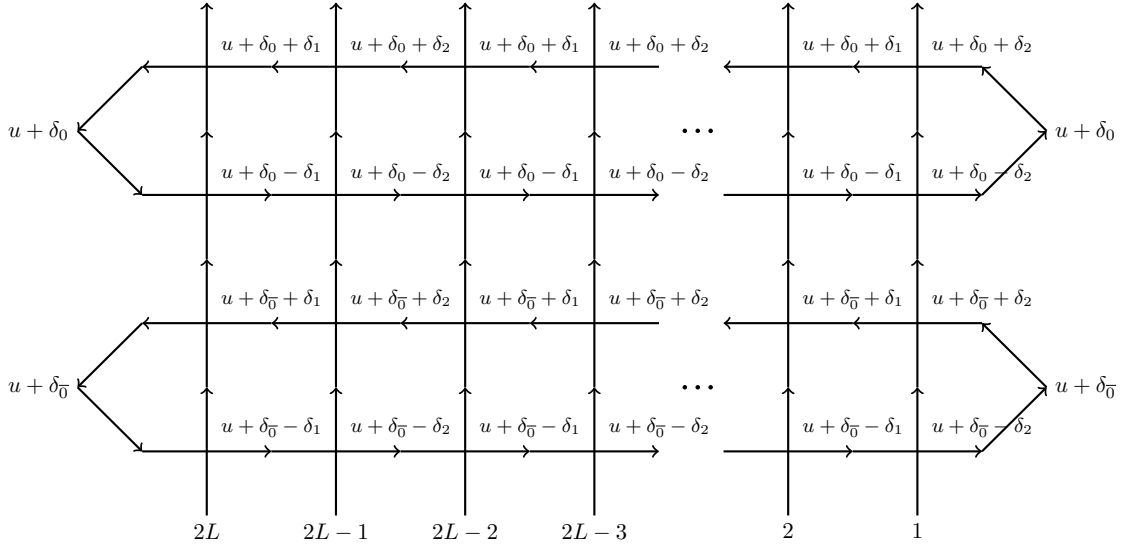
\begin{figure}
		\centering
		\begin{tikzpicture}[scale=0.85, transform shape]
			\ecellr{0}{0}{$u+\dOBar-\delta_1$}
			\ecellr{2}{0}{$u+\dOBar-\delta_2$}
			\ecellr{4}{0}{$u+\dOBar-\delta_1$}
			\ecellr{6}{0}{$u+\dOBar-\delta_2$}
			\draw[->,thick] (-1,2)--(-2,1) node[left] {$u+\dOBar$};
			\draw[->,thick] (-2,1)--(-1,0);
			\ecelll{0}{2}{$u+\dOBar+\delta_1$}
			\ecelll{2}{2}{$u+\dOBar+\delta_2$}
			\ecelll{4}{2}{$u+\dOBar+\delta_1$}
			\ecelll{6}{2}{$u+\dOBar+\delta_2$}
			\fill (7.4,1) circle (1pt);
			\fill (7.6,1) circle (1pt);
			\fill (7.8,1) circle (1pt);
			\ecellr{9}{0}{$u+\dOBar-\delta_1$}
			\ecellr{11}{0}{$u+\dOBar-\delta_2$}
			\ecelll{9}{2}{$u+\dOBar+\delta_1$}
			\ecelll{11}{2}{$u+\dOBar+\delta_2$}
			\draw[<-,thick] (12,2)--(13,1) node[right] {$u+\dOBar$};
			\draw[<-,thick] (13,1)--(12,0);
			
			\ecellr{0}{4}{$u+\dO-\delta_1$}
			\ecellr{2}{4}{$u+\dO-\delta_2$}
			\ecellr{4}{4}{$u+\dO-\delta_1$}
			\ecellr{6}{4}{$u+\dO-\delta_2$}
			\draw[->,thick] (-1,6)--(-2,5) node[left] {$u+\dO$};
			\draw[->,thick] (-2,5)--(-1,4);
			\ecelll{0}{6}{$u+\dO+\delta_1$}
			\ecelll{2}{6}{$u+\dO+\delta_2$}
			\ecelll{4}{6}{$u+\dO+\delta_1$}
			\ecelll{6}{6}{$u+\dO+\delta_2$}
			\fill (7.4,5) circle (1pt);
			\fill (7.6,5) circle (1pt);
			\fill (7.8,5) circle (1pt);
			\ecellr{9}{4}{$u+\dO-\delta_1$}
			\ecellr{11}{4}{$u+\dO-\delta_2$}
			\ecelll{9}{6}{$u+\dO+\delta_1$}
			\ecelll{11}{6}{$u+\dO+\delta_2$}
			\draw[<-,thick] (12,6)--(13,5) node[right] {$u+\dO$};
			\draw[<-,thick] (13,5)--(12,4);

			\node[below] at (0,-1) {$2L$};
			\node[below] at (2,-1) {$2L-1$};
			\node[below] at (4,-1) {$2L-2$};
			\node[below] at (6,-1) {$2L-3$};
			\node[below] at (9,-1) {$2$};
			\node[below] at (11,-1) {$1$};
		\end{tikzpicture}
		\caption{Graphical representation of the product (\ref{tmo_prod}) of two transfer matrices with arbitrary $\mathbb{Z}_2$ staggering by using the conventions defined in Fig.~\ref{Graphical Notation}. }
		\label{Two_Double_Row_TMS}
	\end{figure}

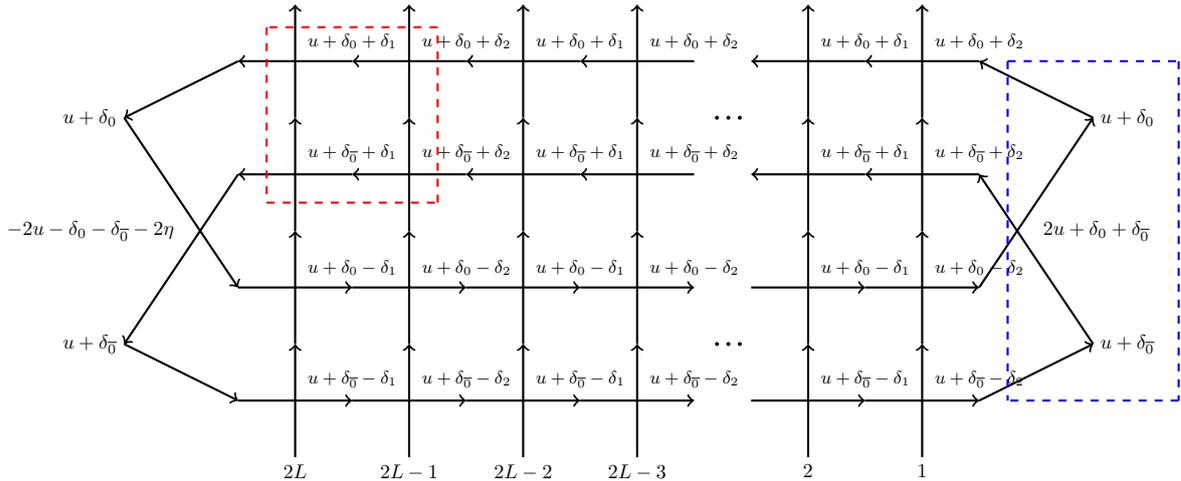
\begin{figure}
		\centering
		\begin{tikzpicture}[scale=0.75, transform shape]
			
			\ecellr{0}{0}{$u+\dOBar-\delta_1$}
			\ecellr{2}{0}{$u+\dOBar-\delta_2$}
			\ecellr{4}{0}{$u+\dOBar-\delta_1$}
			\ecellr{6}{0}{$u+\dOBar-\delta_2$}
			\draw[->,thick] (-1,4)--(-3,1) node[left] {$u+\dOBar$};
			\draw[->,thick] (-3,1)--(-1,0);
			\ecelll{0}{4}{$u+\dOBar+\delta_1$}
			\ecelll{2}{4}{$u+\dOBar+\delta_2$}
			\ecelll{4}{4}{$u+\dOBar+\delta_1$}
			\ecelll{6}{4}{$u+\dOBar+\delta_2$}
			\fill (7.4,1) circle (1pt);
			\fill (7.6,1) circle (1pt);
			\fill (7.8,1) circle (1pt);
			\ecellr{9}{0}{$u+\dOBar-\delta_1$}
			\ecellr{11}{0}{$u+\dOBar-\delta_2$}
			\ecelll{9}{4}{$u+\dOBar+\delta_1$}
			\ecelll{11}{4}{$u+\dOBar+\delta_2$}
			\draw[<-,thick] (12,4)--(14,1) node[right] {$u+\dOBar$};
			\draw[<-,thick] (14,1)--(12,0);
			
			\ecellr{0}{2}{$u+\dO-\delta_1$}
			\ecellr{2}{2}{$u+\dO-\delta_2$}
			\ecellr{4}{2}{$u+\dO-\delta_1$}
			\ecellr{6}{2}{$u+\dO-\delta_2$}
			\draw[->,thick] (-1,6)--(-3,5) node[left] {$u+\dO$};
			\draw[->,thick] (-3,5)--(-1,2);
			\ecelll{0}{6}{$u+\dO+\delta_1$}
			\ecelll{2}{6}{$u+\dO+\delta_2$}
			\ecelll{4}{6}{$u+\dO+\delta_1$}
			\ecelll{6}{6}{$u+\dO+\delta_2$}
			\fill (7.4,5) circle (1pt);
			\fill (7.6,5) circle (1pt);
			\fill (7.8,5) circle (1pt);
			\ecellr{9}{2}{$u+\dO-\delta_1$}
			\ecellr{11}{2}{$u+\dO-\delta_2$}
			\ecelll{9}{6}{$u+\dO+\delta_1$}
			\ecelll{11}{6}{$u+\dO+\delta_2$}
			\draw[<-,thick] (12,6)--(14,5) node[right] {$u+\dO$};
			\draw[<-,thick] (14,5)--(12,2);
			
			\draw[-,dashed,thick,red] (-0.5,6.6)--(-0.5,3.5);
			\draw[-,dashed,thick,red] (-0.5,3.5)--(2.5,3.5);
			\draw[-,dashed,thick,red] (2.5,3.5)--(2.5,6.6);
			\draw[-,dashed,thick,red] (2.5,6.6)--(-0.5,6.6);
			
			\draw[-,dashed,thick,blue] (12.5,6)--(12.5,0);
			\draw[-,dashed,thick,blue] (12.5,6)--(15.5,6);
			
			\draw[-,dashed,thick,blue] (15.5,0)--(12.5,0);
			\draw[-,dashed,thick,blue] (15.5,6)--(15.5,0);
			
			\node[right] at (13,3) {$2u+\dO+\dOBar$};
			\node[left] at (-2,3) {$-2u-\dO-\dOBar-2\eta$};
			
			\node[below] at (0,-1) {$2L$};
			\node[below] at (2,-1) {$2L-1$};
			\node[below] at (4,-1) {$2L-2$};
			\node[below] at (6,-1) {$2L-3$};
			\node[below] at (9,-1) {$2$};
			\node[below] at (11,-1) {$1$};
		\end{tikzpicture}
		\caption{Graphical representation of the product of transfer matrix (\ref{Intermediate_Form_TM}) after the merging procedure. The merged $\mathbb{R}$-matrix here is given by four vertices as in Fig.~\ref{Graphical_Rep_Big_R} as indicated as an example by the red box. Further we see we obtain some enlarged boundary matrices as emphasized by the blue box.  }
		\label{Merged_TM}
	\end{figure}

	We now want to address the question to which extent this procedure can be applied to construct staggered models with open boundary conditions. The strategy is the same: we begin by considering the product of two transfer matrices (\ref{Standard Form}) built out of generic $R$ and $K$-matrices satisfying the Yang-Baxter and reflection equations, respectively, i.e.\
	\begin{equation}
		\label{tmo_prod}
		\begin{aligned}
			\mathcal{T}(u,\{\dO,\dOBar,\delta_1,\delta_2\})&=\tr_{\bNull}\bigg(X_{\bNull}(u+\dO)\bigg)\tr_{\bbNull}\bigg(Y_{\bbNull}(u+\dOBar)\bigg)=\tr_{\bNull\bbNull}\bigg(X_{\bNull}(u+\dO)Y^{t_{\bbNull}}_{\bbNull}(u+\dOBar)\bigg)\,,\\
		\end{aligned}
	\end{equation}
	where we have defined
	\begin{align*}
		X_{\bNull}(u)\equiv&K^{+}_{\bNull}(u)T_{\bNull}(u,\{\delta_1,\delta_2\})K^{-}_{\bNull}(u)T^{-1}_{\bNull}(-u,\{\delta_1,\delta_2\})\,,\\
		Y_{\bbNull}(u)\equiv&T_{\bbNull}(u,\{\delta_1,\delta_2\})K^{-}_{\bbNull}(u)T^{-1}_{\bbNull}(-u,\{\delta_1,\delta_2\})K^{+}_{\bbNull}(u)\,.
	\end{align*}
	(see Figure~\ref{Two_Double_Row_TMS} for a graphical representation of $\mathcal{T}(u)$). 
	Inserting a crossing unitarity (\ref{R_Crossing_Uni})
	and using cyclicity of the trace and $PT$-symmetry we obtain
	\begin{align*}
		&\mathcal{T}(u,\{\dO,\dOBar,\delta_1,\delta_2\})\xi(2u+\dO+\dOBar+\eta)\\
		&\qquad=\tr_{\bNull\bbNull}\bigg(M_{\bNull} R_{\bNull, \bbNull}(-2u-\dO-\dOBar-2\eta)M^{-1}_{\bNull}X_{\bNull}(u+\dO))R_{\bbNull, \bNull}(2u+\dO+\dOBar)Y_{\bbNull}(u+\dOBar)\bigg)\\
		&\qquad=\tr_{\bNull\bbNull}\bigg(M_{\bNull} R_{\bNull, \bbNull}(-2u-\dO-\dOBar-2\eta)M^{-1}_{\bNull}K^{+}_{\bNull}(u+\dO)\\
		&\qquad\qquad\qquad\times
		T_{\bNull}(u+\dO,\{\delta_1,\delta_2\})K^{-}_{\bNull}(u+\dO)T^{-1}_{\bNull}(-u-\dO,\{\delta_1,\delta_2\}) R_{\bbNull, \bNull}(2u+\dO+\dOBar)\\
		&\qquad\qquad\qquad\times
		T_{\bbNull}(u+\dOBar,\{\delta_1,\delta_2\})K^{-}_{\bbNull}(u+\dOBar)T^{-1}_{\bbNull}(-u-\dOBar,\{\delta_1,\delta_2\})K^{+}_{\bbNull}(u+\dOBar)
		\bigg)\,.
	\end{align*}
	Now we use Eq.~(\ref{TRT}) and rearrange the $K$-matrices to get:
	\begin{align*}
		\mathcal{T}(u,\{\dO,\dOBar,\delta_1,\delta_2\})=&\tr_{\bNull\bbNull}\bigg(K^{+}_{\bbNull}(u+\dOBar)M_{\bNull} R_{\bNull, \bbNull}(-2u-\dO-\dOBar-2\eta)M^{-1}_{\bNull}K^{+}_{\bNull}(u+\dO)\\
		&\qquad\times T_{\bNull}(u+\dO,\{\delta_1,\delta_2\})T_{\bbNull}(u+\dOBar,\{\delta_1,\delta_2\})\\
		&\qquad\times K^{-}_{\bNull}(u+\dO) R_{\bbNull, \bNull}(2u+\dO+\dOBar)K^{-}_{\bbNull}(u+\dOBar)\\
		&\qquad\times T^{-1}_{\bNull}(-u-\dO,\{\delta_1,\delta_2\}) T^{-1}_{\bbNull}(-u-\dOBar,\{\delta_1,\delta_2\})
		\bigg)\\
		&\qquad\times \xi^{-1}(2u+\dO+\dOBar+\eta)\,.
	\end{align*}
	Finally, using the expression for the monodromy matrices in terms of the elementary $R$-matrices this transfer matrix can be represented graphically as shown in Fig.~\ref{Merged_TM} (up to a scalar factor).
	Clearly, this can be expressed in terms of the composite $\mathbb{R}$-matrices (\ref{Def_Big_R}) giving
	\begin{equation}
		\label{Intermediate_Form_TM}
		\begin{aligned} &\mathcal{T}(u,\{\dO,\dOBar,\delta_1,\delta_2\})=c_{\tau}(u+\dO,\{\delta_1,\delta_2\}) c_{\tau}(u+\dOBar,\{\delta_1,\delta_2\}) \xi^{-1}(2u+\dO+\dOBar+\eta)\\
			&\qquad \times \tr_{\bNull\bbNull}\bigg(K^{+}_{\bbNull}(u+\dOBar)M_{\bNull} R_{\bNull, \bbNull}(-2u-\dO-\dOBar-2\eta)M^{-1}_{\bNull}K^{+}_{\bNull}(u+\dO)\\
			&\qquad\qquad\qquad\times \mathbb{R}_{\bNull,\bbNull\tn 2L-12L}(u+\dOBar+\delta_1,\DA,\DQ) \cdots \mathbb{R}_{\bNull,\bbNull\tn 1,2}(u+\dOBar+\delta_1,\DA,\DQ)\\
			&\qquad\qquad\qquad\times K^{-}_{\bNull}(u+\dO) R_{\bbNull, \bNull}(2u+\dO+\dOBar)K^{-}_{\bbNull}(u+\dOBar)\\
			&\qquad\qquad\qquad\times \mathbb{R}_{1,2\tn \bbNull, \bNull}(u+\dO-\delta_1,\DQ,\DA) \cdots \mathbb{R}_{2L-1,2L\tn\bbNull, \bNull}(u+\dO-\delta_1,\DQ,\DA)
			\bigg)\,.
		\end{aligned}
	\end{equation}
	
	\subsection{Local interactions I: alternating staggering  \label{Local:Interatcions I}}
\begin{figure}
		\centering
		\begin{tikzpicture}[scale=0.7, transform shape]
			\permucellr{0-0.5}{0}
			\ecellr{0-0.5}{2}{$2\dO$}
			\ecelll{0-0.5}{4}{$-2\dO$}
			\permucelll{0-0.5}{6}
			
			\ecellr{2-0.5}{0}{$-2\dO$}
			\permucellr{2-0.5}{2}
			\permucelll{2-0.5}{4}
			\ecelll{2-0.5}{6}{$2\dO$}
			
			\node[below] at (0-0.5,-1) {$2j-1$};
			\node[below] at (2-0.5,-1) {$2j$};
			\node[below] at (1-0.5,-2) {\Large (b)};
			
			\draw[->,ultra thick] (-2.25-0.5,3)--(-0.75-0.5,3);
			\permucellr{7.5-1}{0}
			\ecellr{7.5-1}{2}{$2\dO$}
			\ecelll{7.5-1}{4}{$-2\dO$}
			\permucelll{7.5-1}{6}
			
			\ecellr{5.5-1}{0}{$-2\dO$}
			\permucellr{5.5-1}{2}
			\permucelll{5.5-1}{4}
			\ecelll{5.5-1}{6}{$2\dO$}

			\ecellr{13-1.5}{0}{$-\frac{p}{2}$}
			\permucellr{13-1.5}{2}
			\permucelll{11-1.5}{4}
			\ecelll{11-1.5}{6}{$-\frac{p}{2}$}
			
			\permucellr{11-1.5}{0}
			\ecellr{11-1.5}{2}{$\frac{p}{2}$}
			\ecelll{13-1.5}{4}{$\frac{p}{2}$}
			\permucelll{13-1.5}{6}

			\node[below] at (5.5-1.0,-1) {$2j-1$};
			\node[below] at (7.5-1.0,-1) {$2j$};
			\node[below] at (6.5-1.5+0.5,-2) {\Large (c)};
			
			\node[below] at (11-1.5,-1) {$2j-1$};
			\node[below] at (13-1.5,-1) {$2j$};
			\node[below] at (12-1.5,-2) {\Large (d)};

			\fill (10.5-1.5,6) circle (0.1);
			\draw (13.5-1.5,6) circle (0.1);
			
			\ecellr{-5.5}{0}{$u+\dOBar-\delta_1$}
			\ecellr{-5.5}{2}{$u+\dO-\delta_1$}
			\ecelll{-5.5}{4}{$u+\dOBar+\delta_1$}
			\ecelll{-5.5}{6}{$u+\dO+\delta_1$}
			
			\ecellr{-3.5}{0}{$u+\dOBar-\delta_2$}
			\ecellr{-3.5}{2}{$u+\dO-\delta_2$}
			\ecelll{-3.5}{4}{$u+\dOBar+\delta_2$}
			\ecelll{-3.5}{6}{$u+\dO+\delta_2$}
			
			\node[below] at (-3.5,-1) {$2j$};
			\node[below] at (-5.5,-1) {$2j-1$};
			\node[below] at (-4.5,-2) {\Large (a)};
			
			\draw[->,ultra thick] (-1.95+16-2,3)--(-0.75+16.25-2,3);
			\draw[->] (16-2,-1)--(16-2,7);
			\draw[->] (18-2,-1)--(18-2,7);
			
			\node[below] at (16-2,-1) {$2j-1$};
			\node[below] at (18-2,-1) {$2j$};
			\node[below] at (17-2,-2) {\Large (e)};
		\end{tikzpicture}
		
		\caption{(a) To establish bulk locality at $u=0$ the parameters $\dO,\dOBar,\delta_1,\delta_2$ have to be fine tuned such that the conditions (\ref{i}-(\ref{iii} hold. The diagrammatic schemes  of the bulk elementary cells at the  shift point $u=0$ for all possible non-trivial choices of staggering are displayed in (b) \&  (c) (alternating case (\ref{stagg_alt})) and (d) (quasi periodic case (\ref{stagg_qper})). Using regularity (\ref{Reg}), unitarity (\ref{R_Uni}), and in (d) quasi periodicity (\ref{R_Qper}) of the elementary $R$-matrix gives the identity (e) in the bulk.}
		\label{Uni_Cell_Identity_Op}
	\end{figure}
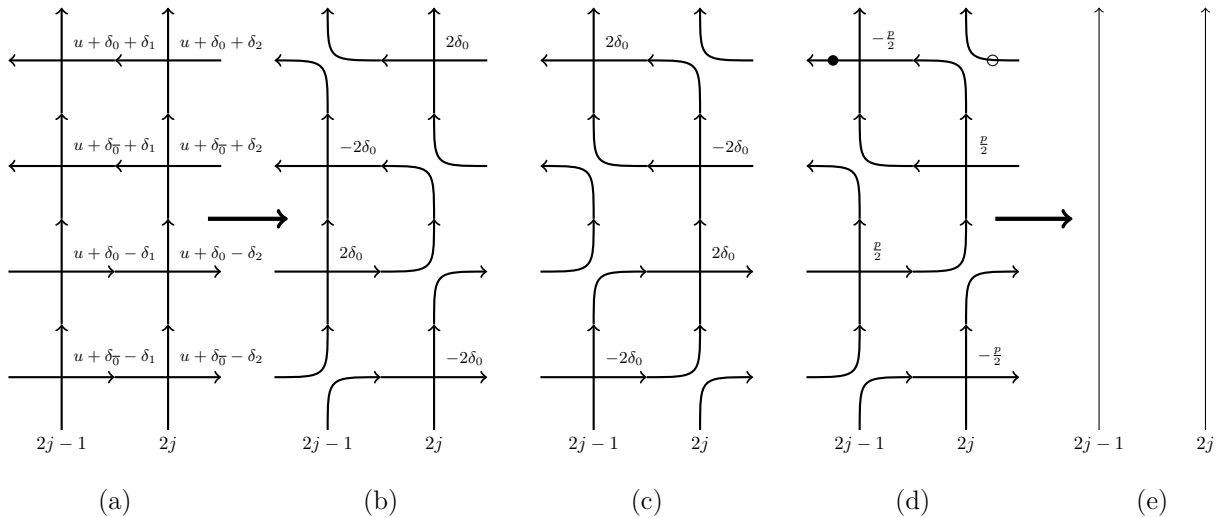

	As for periodic boundary conditions the staggering parameters $\{\dO,\dOBar,\delta_1,\delta_2\}$ have to satisfy constraints to generate local interactions from this open boundary transfer matrix. 
	Nearest neighbour interactions between the composite degrees of freedom of the staggered model are obtained by taking the derivative (assuming all quantities to be differentiable at the corresponding points) of $\mathcal{T}(u)$ with respect to the spectral parameter \cite{Sklyanin:1988yz}
	\begin{align}
		H\propto \left.\frac{\partial}{\partial u} \mathcal{T}\left(u,\{\dO,\dOBar,\delta_1,\delta_2\}\right)\right|_{u=u_0}
		\,.\label{Formula_HOP}
	\end{align}
	Again, locality derives from the regularity of $\mathbb{R}$. To make use of (\ref{Big_Regu}) three conditions are needed to be met: 
	%
	\begin{enumerate}[label=\roman*)]
		\item  \label{i} The $\mathbb{R}$-matrices in (\ref{Intermediate_Form_TM}) need to act on the same auxiliary space e.g.\  $\mathcal{W}_{\bNull\bbNull}$.
		\item  \label{ii}As in the periodic case the staggering parameters have to satisfy the constraint (\ref{Loc_Con_PBC}).  
		\item \label{iii} The staggering parameters have to be chosen such that \emph{all} $\mathbb{R}$ in (\ref{Intermediate_Form_TM}) can simultaneously be evaluated at the shift point $u_0=0$. 
	\end{enumerate}
	For (\ref{i} we use the identity $\mathbb{R}_{i,j\tn\bbNull,\bNull} = P_{\bNull,\bbNull} \mathbb{R}_{i,j\tn \bNull,\bbNull} P_{\bNull,\bbNull}$. Conditions (\ref{ii} and (\ref{iii} are achieved by choosing the staggering parameters to be opposite and equal in both the horizontal and the vertical direction, i.e.\
	\begin{align}
		\dO=-\dOBar=\delta_1=-\delta_2\, \label{stagg_alt}
	\end{align}
	(depicted in Fig. \ref{Uni_Cell_Identity_Op}c) or the equivalent choice of parameters obtained by changing $\dO\to -\dO$ (see Fig. \ref{Uni_Cell_Identity_Op}b).
	These constraints on the staggering parameters imply $\DA=\DQ=2\delta_0$ with $\dO$ remaining as a \emph{free} parameter.  The resulting transfer matrix is
	\begin{align*}
		&\mathcal{T}(u,\{\dO,-\dO,\dO,-\dO\})=c_{\tau}(u+\dO,\{\dO,-\dO\}) c_{\tau}(u-\dO,\{\dO,-\dO\})\\
		&\qquad\tr_{\bNull\bbNull}\bigg(
		\mathbb{K}^+_{\bNull,\bbNull}(u,2\dO)\mathbb{R}_{\bNull, \bbNull\tn 2L-1,2L}(u,2\dO,2\dO) \cdots \mathbb{R}_{\bNull,\bbNull\tn 1,2}(u,2\dO,2\dO)\\
		&\qquad\qquad\times \mathbb{K}^-_{\bNull,\bbNull}(u,2\dO)\mathbb{R}_{1,2\tn \bNull,\bbNull}(u,2\dO,2\dO) \cdots \mathbb{R}_{2L-1,2L\tn\bNull,\bbNull}(u,2\dO,2\dO)
		\bigg),
	\end{align*}
	where we have introduced
	\begin{subequations}
		\label{K_Alternating}
		\begin{align}
			\mathbb{K}^-_{i,j}(u,2\dO)&=P_{i,j}K^-_{j}(u+\dO)R_{i,j}(2u)K^{-}_i(u-\dO)\,,\label{KMinus_Alternating}\\
			\mathbb{K}^+_{i,j}(u,2\dO)&=\frac{1}{\xi(2u+\eta)}P_{i,j}K^+_{j}(u-\dO)M_iR_{i,j}(-2u-2\eta)M^{-1}_iK^{+}_i(u+\dO)\,.\label{KPlus_Alternating} 
		\end{align}
	\end{subequations}
	In terms of the monodromy matrix built from the composite $\mathbb{R}$-matrices,
	\begin{align}
		\mathbb{T}_{\bNull,\bbNull}(u,\DA,\DQ)=\mathbb{R}_{\bNull, \bbNull\tn 2L-1,2L}(u,\DA,\DQ) \cdots \mathbb{R}_{\bNull,\bbNull\tn 1,2}(u,\DA,\DQ)\,,\label{Monodromy_Big_R}
	\end{align}
	the transfer matrix for alternating staggering (\ref{stagg_alt}) is brought into standard form (\ref{Standard Form})\footnote{If one drops the constraint \ref{stagg_alt}, a shift in the of the spectral parameter in (\ref{Intermediate_Form_TM}) leads to a transfer matrix with a moving boundary \cite{NeRe21a}. This transfer matrix does not lead in general, however, to a local Hamiltonian.}  
	\begin{equation}
		\label{TM_Alterning}
		\begin{aligned}
			\mathcal{T}^{\text{alt}}(u,2\dO) &\equiv
			\mathcal{T}(u,\{\dO,-\dO,\dO,-\dO)\\
			&=\tr_{\bNull\bbNull}\bigg(
			\mathbb{K}^+_{\bNull,\bbNull}(u,2\dO)\mathbb{T}_{\bNull ,\bbNull}(u,2\dO,2\dO)\mathbb{K}^-_{\bNull,\bbNull}(u,2\dO)\mathbb{T}^{-1}_{\bNull,\bbNull}(-u,2\dO,2\dO)
			\bigg)\,.
		\end{aligned}
	\end{equation}
	
	\subsection{Local interactions II: quasi periodic staggering}
	\label{ch:local_quasi}
	Interestingly, there exists a second choice of the staggering parameters  leading to a local Hamiltonian (\ref{Formula_HOP}) when the elementary $R$-matrix is quasi periodic (\ref{R_Qper}), namely
	\begin{align}
		\dO=\frac{p}{2}\,,\quad
		\dOBar=0\,,\quad
		\delta_1=0\,,\quad
		\delta_2=\frac{p}{2}\,, \label{stagg_qper}
	\end{align}
	which is displayed in Fig. \ref{Uni_Cell_Identity_Op}d. Again, we have to implement three steps to bring the transfer matrix into a form generating a local Hamiltonian: for step (\ref{i}, i.e.\ switching the auxiliary space $\mathcal{W}_{\bbNull,\bNull}$ to $\mathcal{W}_{\bNull,\bbNull}$, we use the Yang-Baxter equation (\ref{YBE}) for $v=-p/2$ giving 
	\begin{equation}
		\mathbb{R}_{i,j\tn \bbNull,\bNull}\left(u+\frac{p}{2},-\frac{p}{2},\frac{p}{2}\right)\, R_{\bNull,\bbNull}\left(-\frac{p}{2}\right)
		= R_{\bNull,\bbNull}\left(-\frac{p}{2}\right)\, \mathbb{R}_{i,j\tn\bNull,\bbNull}\left(u,-\frac{p}{2},-\frac{p}{2}\right)\,.
	\end{equation}
	For step (\ref{ii}, i.e.\ preparing $\mathbb{R}$ such that the regularity (\ref{Big_Regu}) can be exploited, we use the quasi periodicity of $R$ which implies
	\begin{equation}
		\mathbb{R}_{\bNull,\bbNull\tn i,j}\left(u,\frac{p}{2},-\frac{p}{2}\right)
		=  G_0\, \mathbb{R}_{\bNull,\bbNull\tn i,j}\left(u,-\frac{p}{2},-\frac{p}{2}\right)\,G_0^{-1}\,.
	\end{equation}
	Together with the unitarity condition $R_{\bNull,\bbNull}(-\frac{p}{2}) R_{\bbNull,\bNull}(\frac{p}{2})= \xi(\frac{p}{2})\mathbf{1}$ these identities allow to rewrite the transfer matrix (\ref{Intermediate_Form_TM}) such that also condition (\ref{iii} is satisfied, i.e.\
	
	\begin{equation}
		\begin{aligned}
			&\mathcal{T}(u,\{\frac{p}{2},0,0,\frac{p}{2}\})=
			c_{\tau}(u+\frac{p}{2},\{0,\frac{p}{2}\})\, c_{\tau}(u,\{0,\frac{p}{2}\})\xi^{-1}\left(\frac{p}{2}\right)\\
			&\qquad\tr_{\bNull\bbNull}\bigg(
			\overline{\mathbb{K}}^+_{\bNull,\bbNull}(u,-\frac{p}{2})\mathbb{R}_{\bNull, \bbNull\tn 2L-1,2L}(u,-\frac{p}{2},-\frac{p}{2}) \cdots \mathbb{R}_{\bNull,\bbNull\tn 1,2}(u,-\frac{p}{2},-\frac{p}{2})\\
			&\qquad\qquad\times\overline{\mathbb{K}}^-_{\bNull,\bbNull}(u,-\frac{p}{2})\mathbb{R}_{1,2\tn\bNull,\bbNull}(u,-\frac{p}{2},-\frac{p}{2}) \cdots \mathbb{R}_{2L-1,2L\tn\bNull,\bbNull}(u,-\frac{p}{2},-\frac{p}{2})
			\bigg),
		\end{aligned}
	\end{equation}
	with
	\begin{subequations}
		\label{K_Quasiperiodic}
		\begin{align}
			\overline{\mathbb{K}}^-_{i,j}\left(u,-\frac{p}{2}\right)&=G^{-1}_{i}K^{-}_{i}\left(u+\frac{p}{2}\right) R_{j,i}\left(2u+\frac{p}{2}\right)K^{-}_{j}(u)R_{i,j}\left(-\frac{p}{2}\right)\,,\label{KMinus_Quasiperiodic}\\
			\overline{\mathbb{K}}^+_{i,j}\left(u,-\frac{p}{2}\right)&=\frac{1}{\xi\left(2u+\frac{p}{2}+\eta\right)}R_{j,i}\left(\frac{p}{2}\right)K^{+}_{j}(u)M_{i} R_{i,j}\left(-2u-\frac{p}{2}-2\eta\right)M^{-1}_{i}K^{+}_{i}\left(u+\frac{p}{2}\right)\,G_{i}\,.\label{KPlus_Quasiperiodic} 
		\end{align} 
	\end{subequations}
	
	Using the monodromy matrix (\ref{Monodromy_Big_R}) for the composite $\mathbb{R}$-matrices the transfer matrix for the quasi periodic staggering (\ref{stagg_qper}) can be written as
	\begin{equation}
		\label{TM_QuasiP}
		\begin{aligned}
			\mathcal{T}^{\text{qp}}(u,-\frac{p}{2})  \equiv&\mathcal{T}(u,\{-\frac{p}{2},0,0,-\frac{p}{2}\})\\
			=&\tr_{\bNull\bbNull}\bigg(
			\overline{\mathbb{K}}^+_{\bNull,\bbNull}(u,-\frac{p}{2})\mathbb{T}_{\bNull, \bbNull}(u,-\frac{p}{2},-\frac{p}{2})\overline{\mathbb{K}}^-_{\bNull,\bbNull}(u,-\frac{p}{2})\mathbb{T}^{-1}_{\bNull,\bbNull}(u,-\frac{p}{2},-\frac{p}{2})\bigg)\\
			&\times \xi^{-1}\left(\frac{p}{2}\right)\,.
		\end{aligned}
	\end{equation}
	
\begin{figure}
		\centering
		\begin{tikzpicture}
			
			\ecellr{0}{6}{$u+\delta_1$}
			\ecellr{2}{6}{$u+\delta_2$}
			
			\draw[ultra thick,->] (3.5,6)--(4.5,6) node[midway,sloped,above] {$u\to u-\delta_1$};
			
			\ecellr{0+6}{6}{$u$}
			\ecellr{2+6}{6}{$u+\delta_2-\delta_1$}

			\ecellr{0}{0}{$u-\delta_1$}
			\ecellr{2}{0}{$u-\delta_2$}
			\ecelll{2}{2}{$u+\delta_2$}
			\ecelll{0}{2}{$u+\delta_1$}
			
			\draw[ultra thick,->] (3.5,1)--(4.5,1) node[midway,sloped,above] {$u\to u-\delta_1$};
			
			\ecellr{0+6}{0}{$u-2\delta_1$}
			\ecellr{2+6}{0}{$u-\delta_2-\delta_1$}
			\ecelll{2+6}{2}{$u+\delta_2-\delta_1$}
			\ecelll{0+6}{2}{$u$}
			
			\node[below] at (4,5) {\Large (a)};
			\node[below] at (4,-1) {\Large (b)};
			
		\end{tikzpicture}
		\caption{In the periodic case (a) a shift in the spectral parameter allows to adjust one inhomogeneity to zero  e.g.\  $\delta_1=0$. This is not possible in the open case (b) where each inhomogeneity appears twice with different signs.  }
		\label{Shifting a TM}
	\end{figure}
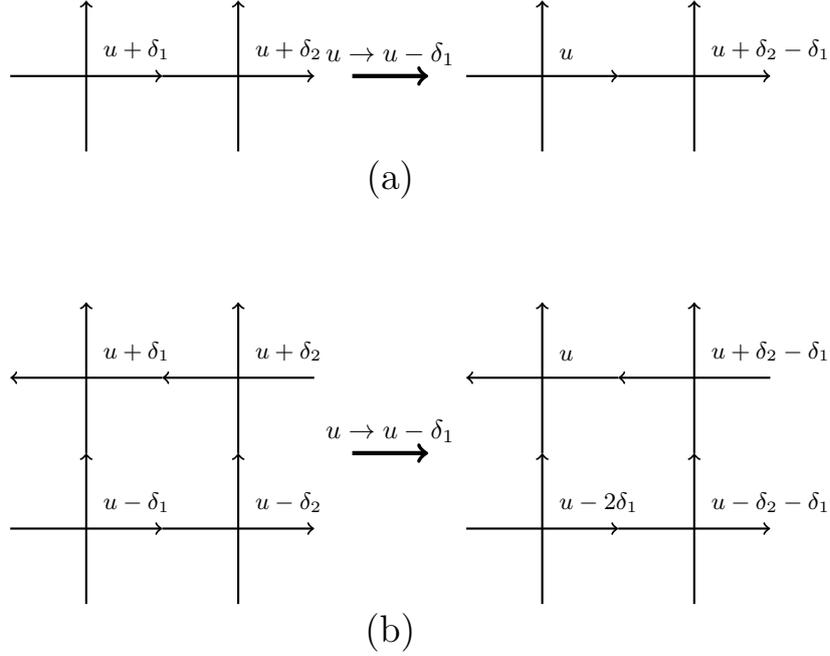

	Note that the composite monodromy matrix $\mathbb{T}$ (\ref{Monodromy_Big_R}) enters in the transfer matrices (\ref{TM_Alterning}) and (\ref{TM_QuasiP}) with identical arguments for the particular choice of $2\dO=-\frac{p}2$.  Hence, the bulk of these models coincides while the reflection matrices $\mathbb{K}$ (\ref{K_Alternating}) and $\overline{\mathbb{K}}$ (\ref{K_Quasiperiodic}) correspond to different boundary conditions.\footnote{This does not lead to different models in the case of periodic boundary conditions where different choices of the horizontal staggering can be related by a shift in the spectral parameter (see Figure \ref{Shifting a TM}).}  This has been discussed recently in the context of a staggered six-vertex (or $A_1^{(1)}$) model \cite{RPJS20,NeRe21a,Li:2022clv}: for $2\dO=-\frac{p}{2}$ the resulting composite model is a vertex model based on the twisted affine Lie algebra $D_2^{(2)}$. The $D_2^{(2)}$ boundary matrices corresponding to $\mathbb{K}$ and $\overline{\mathbb{K}}$ were known previously \cite{NePi18} and can be factorized into objects of the six-vertex model subject to $U_q[\mathfrak{sl}_2]$ boundary conditions.
	
	\subsection{Associated reflection algebras and composite  \texorpdfstring{$\mathbb{K}$-matrices}{K-matrices}}
	We are left to prove that the reflection matrices (\ref{K_Alternating}) and (\ref{K_Quasiperiodic}) are indeed representations of a reflection algebra associated with the $\mathbb{R}$-matrix. 
	We define the following \emph{generalized} reflection algebras:
	\begin{subequations}
		\label{Reflection_algebra_Large_K}
		\begin{align}
			\mathbb{R}_{i,j\tn k,\ell}(u-v,\theta,\theta)& \mathbb{K}^{-}_{i,j} (u,\theta) \mathbb{R}_{k,\ell\tn i,j}(u+v,\theta,\theta)\mathbb{K}^{-}_{k,\ell} (v,\theta)\label{Reflection_algebra_Large_K_Minus}\\&=\mathbb{K}^{-}_{k,\ell} (v,\theta)\mathbb{R}_{i,j\tn k,\ell}(u+v,\theta,\theta)\mathbb{K}^{-}_{i,j} (u,\theta)\mathbb{R}_{k,\ell\tn i,j}(u-v,\theta,\theta)\notag
		\end{align}
		and
		\begin{align}
			&\mathbb{R}_{i,j\tn k,\ell}(-u+v,-\theta,-\theta)\left(\mathbb{K}^{+}_{i,j}(u,\theta)\right)^{t_it_j}\mathbb{M}^{-1}_{i,j}\mathbb{R}_{k,\ell\tn i,j}(-u-v-2\eta,-\theta,-\theta)\mathbb{M}_{i,j}\left(\mathbb{K}^{+}_{k,\ell}(v,\theta)\right)^{t_k t_\ell}=\notag\\&\,\,\left(\mathbb{K}^{+}_{k,\ell}(v,\theta)\right)^{t_kt_\ell}\mathbb{M}_{i,j}\mathbb{R}_{i,j\tn k,\ell}(-u-v-2\eta,-\theta,-\theta)\mathbb{M}^{-1}_{i,j}  \left(\mathbb{K}^{+}_{i,j}(u,\theta)\right)^{t_it_j} \mathbb{R}_{k,\ell\tn i,j}(-u+v,-\theta,-\theta)\label{Reflection_algebra_Large_K_Plus}\,.
		\end{align}
	\end{subequations}
	Note that the sign of the free parameter $\theta$ in the arguments of the composite $\mathbb{R}$-matrices differs between (\ref{Reflection_algebra_Large_K_Minus}) and (\ref{Reflection_algebra_Large_K_Plus}).
	One of the main results is that for given $K^\pm$ satisfying Eqs.~(\ref{Reflection_algebra})
	the matrices $\mathbb{K}$ (\ref{K_Alternating}) and $\overline{\mathbb{K}}$ (\ref{K_Quasiperiodic}) obey the equations (\ref{Reflection_algebra_Large_K}) with the composite $\mathbb{R}$-matrix $\mathbb{R}(u,\theta,\theta)$ for $\theta=2\dO$, $\dO$ arbitrary, and $\theta=-\frac{p}{2}$, respectively.
	The proof for (\ref{KPlus_Alternating}) is given in Appendix~\ref{Proof_KPlus}. The one for (\ref{KMinus_Alternating}) 
	works along the same line, for (\ref{K_Quasiperiodic}) one needs to use multiple times the quasi periodicity in addition. 
	
	Based on the reflection algebra it is straightforward to show the commutativity of the transfer matrices for both alternating and quasi periodic boundary conditions. i.e.
	\begin{align*}
		\kommu{\mathcal{T}(u,\epsilon)}{\mathcal{T}(v,\epsilon)}=0\,, \qquad\epsilon=0,1\,.
	\end{align*}
	
	\subsection{Boundary terms in the Hamiltonian \label{Boundary_terms}}
	Above we have identified two types of staggering, (\ref{stagg_alt}) and (\ref{stagg_qper}), allowing for the construction of a local bulk Hamiltonian from the corresponding transfer matrix of the composite model. For a compact presentation we define
	\begin{align}
		\mathfrak{R}_{i,j\tn k,\ell}(u,\epsilon) =  \begin{cases}
			\mathbb{R}_{i,j\tn k,\ell}(u,-\frac{p}{2},-\frac{p}{2}) & \epsilon=0 \\
			\mathbb{R}_{i,j\tn k,\ell}(u,2\dO,2\dO) & \epsilon=1\end{cases},\qquad  \mathfrak{K}^{\pm}_{i,j}(u,\epsilon)=  \begin{cases}
			\overline{\mathbb{K}}^{\pm}_{i,j}(u,-\frac{p}{2}) & \epsilon=0 \\
			\mathbb{K}^{\pm}_{i,j}(u,2\dO) & \epsilon=1
		\end{cases}
	\end{align}
	where $\epsilon=1$ ($0$) corresponds to the alternating (\ref{TM_Alterning}) and the quasi periodic staggering (\ref{TM_QuasiP}), respectively. Note that $\mathfrak{K}^{-}(0,\epsilon)\propto \mathbf{1}$ by (\ref{K_Uni}) for $\epsilon=1$ and by (\ref{K_G}) for $\epsilon=0$. Hence, we obtain a local \footnote{It is noteworthy, that for $K^-(0)\not\propto \mathbf{1}$ , inducing an alternating staggering is sufficient to define a local Hamiltonian. See also \cite{An00} for a similar approach.} Hamiltonian \cite{Sklyanin:1988yz} via (\ref{Formula_HOP}) whose bulk contribution is found to be
	\begin{align}
		H^{\epsilon}_{bulk}= \frac{2}{\xi(\Delta)}\sum^{L-1}_{j=1}\mathbb{P}_{2j,2j-1|2j+2,2j+1}\mathfrak{R}'_{2j,2j-1|2j+2,2j+1}(0,\epsilon)\,.\label{H_Bulk}
	\end{align}
	Here and in the following the prime indicates the derivative with respect to the first argument where we assume that all quantities are differentiable at the corresponding points. The boundary contributions read
	\begin{equation}
		\begin{aligned}
			H^\epsilon_{left}&=\frac{\tr_{\bNull\bbNull}\bigg(\mathfrak{K}^{'+}_{\bNull,\bbNull}(0,\epsilon)\bigg)}{\tr_{\bNull \bbNull} \bigg(\mathfrak{K}^+_{\bNull, \bbNull}(0,\epsilon)\bigg)}
			+\frac{2\tr_{\bNull\bbNull}\bigg(\mathfrak{K}^{+}_{\bNull,\bbNull}(0,\epsilon)\mathbb{P}_{\bNull,\bbNull\tn 2L-1,2L}\mathfrak{R}'_{\bNull,\bbNull\tn 2L-1,2L}(0,\epsilon)\bigg)}{\tr_{\bNull \bbNull} \bigg(\mathfrak{K}^+_{\bNull, \bbNull}(0,\epsilon)\bigg)\xi(\Delta)}\,,\\
			H^\epsilon_{right}&=\frac{\mathfrak{K}^{-'}_{1,2}(0,\epsilon)}{\mathfrak{K}^{-}_{1,2}(0,\epsilon)}\,.
		\end{aligned}
	\end{equation}
	Note that, to obtain the spectrum of the above Hamiltonians it is sufficient to use the Bethe Ansatz for the \emph{single} double row transfer matrix $\tau(u)$ (\ref{Standard Form}). Knowing the eigenvalue $\Lambda(u)$ of $\tau(u)$ the energies can be calculated via equations (\ref{tmo_prod}) and (\ref{Formula_HOP}) with operators replaced by their eigenvalues. 
	
	\section{Quasi momentum for Open Systems}
	As pointed out for the periodic case above, there exist two families of conserved quantities for the staggered models considered in this paper: in addition to the ones generated to the product of elementary transfer matrices (\ref{tmo_prod}) (or Eqs.\ (\ref{TM_Alterning}) and (\ref{TM_QuasiP}) for the two cases discussed above) one can consider operators such as the quasi momentum generated from the \emph{quotient} of elementary transfer matrices. For the staggered model with open boundary conditions built from arbitrary elementary $R$- and $K$-matrices we replace (\ref{QI_PBC}) by
	\begin{align}
		\mathbb{Q}=\log \left[\frac{\tau(-\delta_1,\{\delta_1,\delta_2\})}{\tau(-\delta_2,\{\delta_1,\delta_2\})}\right]\,.\label{QI_Standard}
	\end{align}
	To express this operator in the composite picture, we adopt the idea from the periodic case: we look for a generating function built out of a \emph{product} of transfer matrices giving (\ref{QI_Standard}) as the leading term. 
	\subsection{Alternating staggering}
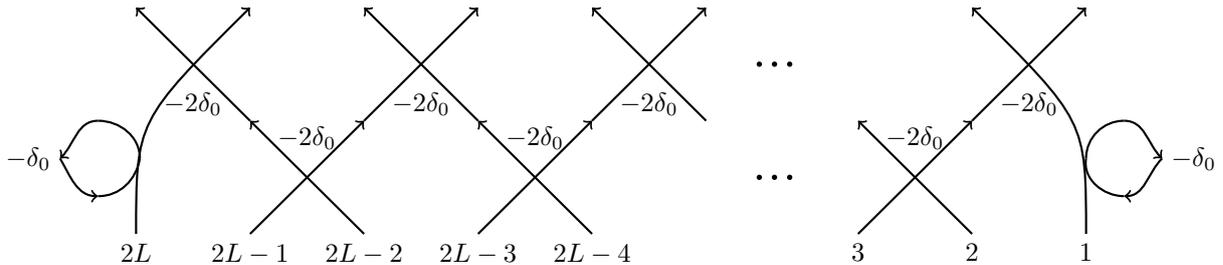
\begin{figure}
		\centering
		\begin{tikzpicture}
			
			\draw[-,thick,out=90,in=0] (0.05,1) to (-0.5,1.5);
			\draw[->,thick,out=180,in=45] (-0.5,1.5) to (-1,1.0);
			\draw[->,thick,out=-60,in=180] (-1,1.0) to (-0.5,0.5);
			\draw[-,thick,out=0,in=-100] (-0.5,0.5) to (0.05,1);
			
			\node[left] at (-1,1) {$-\dO$};

			\draw[->,thick] (0,0) .. controls (0,1.5)  .. (1.5,3);
			\draw[<-,thick] (0,3)--(1.5,1.5);
			\draw[<-,thick] (1.5,1.5)--(3,0);
			
			\draw[->,thick] (1.5,0)--(3,1.5);
			\draw[->,thick] (3,1.5)--(4.5,3);
			
			\draw[<-,thick] (3,3)--(4.5,1.5);
			\draw[<-,thick] (4.5,1.5)--(6,0);
			
			\draw[->,thick] (4.5,0)--(6,1.5);
			\draw[->,thick] (6,1.5)--(7.5,3);
			
			\draw[<-,thick] (6,3)--(7.5,1.5);
			\draw[<-,thick] (7.5+2,1.5)--(9+2,0);
			
			\draw[->,thick] (7.5+2,0)--(9+2,1.5);
			\draw[->,thick] (9+2,1.5)--(10.5+2,3);
			\draw[->,thick] (10.5+2,0) .. controls (10.5+2,1.5)  .. (9+2,3);

			\node[below] at (0.75,2) {$-2\dO$};
			\node[below] at (3.75,2) {$-2\dO$};
			\node[below] at (6.75,2) {$-2\dO$};
			\node[below] at (9.75+2,2) {$-2\dO$};
			
			\node[above] at (2.25,1) {$-2\dO$};
			\node[above] at (5.25,1) {$-2\dO$};
			\node[above] at (8.25+2,1) {$-2\dO$};
			
			\draw[-,thick,out=90,in=180] (12.5,1) to (13,1.5);
			\draw[->,thick,out=0,in=135] (13,1.5) to (13.5,1.0);
			\draw[->,thick,out=-135,in=0] (13.5,1.0) to (13.,0.5);
			\draw[-,thick,out=180,in=-100] (13,0.5) to (12.5,1);
			
			\node[right] at (13.5,1) {$-\dO$};

			\fill (7.7+0.5,2.25) circle (1pt);
			\fill (7.9+0.5,2.25) circle (1pt);
			\fill (8.1+0.5,2.25) circle (1pt);
			\fill (7.7+0.5,0.75) circle (1pt);
			\fill (7.9+0.5,0.75) circle (1pt);
			\fill (8.1+0.5,0.75) circle (1pt);
			

			
			\node[below] at (0,0) {$2L$};
			\node[below] at (1.5,0) {$2L-1$};
			\node[below] at (3,0) {$2L-2$};
			\node[below] at (4.5,0) {$2L-3$};
			\node[below] at (6,0) {$2L-4$};
			
			\node[below] at (7.5+2,0) {$3$};
			\node[below] at (9+2,0) {$2$};
			\node[below] at (10.5+2,0) {$1$};

		\end{tikzpicture}
		\caption{Graphical representation of the quasi momentum operator for alternating staggering in the rotated geometry. One can see that the quasi momentum operator is acting non-locally. The loops at the right and left ending are due to the influences of the boundary matrices.  }
		\label{Figure_QI_Alternating}
	\end{figure}

	For the alternating staggering case the quasi momentum operator can be directly related (up to an additive constant) to a single double row-transfer matrix
	\begin{align}
		\mathbb{Q}^{\mathrm{alt}}=\log\bigg( \tau^2(-\dO,\{\dO,-\dO\}) \bigg).\label{QI_Alt}
	\end{align}
	In this case the quasi momentum can be represented in the rotated geometry as displayed in Figure \ref{Figure_QI_Alternating}. 
	Starting from $\tau(u-\dO,\{\dO,-\dO\})^2$ and repeating the steps in Section~\ref{Compostive_Picture_for_open_models} to reach (\ref{Intermediate_Form_TM}) and then the manipulations (\ref{i}-(\ref{ii} in Section~\ref{Local:Interatcions I}  we obtain another generating functional for the quasi momentum operator: 
	\begin{align}
		Q(u)=\tr_{\bNull\bbNull}\bigg(\mathcal{K}^+_{\bNull,\bbNull}(u,2\dO)\mathbb{T}_{\bNull,\bbNull}(u,0,2\dO)\mathcal{K}^-_{\bNull,\bbNull}(u,2\dO)\mathbb{T}^{-1}_{\bNull,\bbNull}(-(u-2\dO),0,2\dO)\bigg)\,,\label{Q_Gen}
	\end{align}
	where the $\mathcal{K}$-matrices
	\begin{subequations}
		\label{K_QI}
		\begin{align}
			\mathcal{K}^-_{i,j}(u,2\dO)&=P_{i,j}K^-_{j}(u-\dO)R_{i,j}(2u-2\dO)K^-_{i}(u-\dO)\label{KMinus_QI}\,,\\
			\mathcal{K}^+_{i,j}(u,2\dO)&=\frac{1}{\xi(2u-2\dO+\eta)}P_{i,j}K^+_{j}(u-\dO)M_{i}R_{i,j}(-2u+2\dO-2\eta)M^{-1}_{i}K^+_{i}(u-\dO)\label{KPlus_QI}\,,
		\end{align}
	\end{subequations}
	obey the reflection algebras
	\begin{subequations}
		\begin{equation}
			\begin{aligned}
				\mathbb{R}_{i,j\tn k,\ell}&(u-v,0,0)\mathcal{K}^-_{i,j}(u,2\dO)\mathbb{R}_{k,\ell\tn i,j}(u+v-2\dO,0,0)\mathcal{K}^-_{k,\ell}(v,2\dO)=\\
				&\mathcal{K}^-_{k,\ell}(v,2\dO)\mathbb{R}_{i,j\tn k,\ell}(u+v-2\dO,0,0)\mathcal{K}^{-}_{i,j}(u,2\dO)\mathbb{R}_{k,\ell\tn i,j}(u-v,0,0)
			\end{aligned}
		\end{equation}
		and
		\begin{equation}
			\begin{aligned}
				&\mathbb{R}_{i,j\tn k,\ell}(-u+v,0,0)\left(\mathcal{K}^{+}_{i,j}(u,2\dO)\right)^{t_{i}t_j}\mathbb{M}^{-1}_{i,j}\mathbb{R}_{k,\ell\tn i,j}(-u-v+2\dO-2\eta,0,0)\mathbb{M}_{i,j}\left(\mathcal{K}^{+}_{k,\ell}(v,2\dO)\right)^{t_{k}t_{\ell}}=\\
				&\,\,\,\left(\mathcal{K}^{+}_{k,\ell}(v,2\dO)\right)^{t_kt_\ell}\mathbb{M}_{i,j}\mathbb{R}_{i,j\tn k,\ell}(-u-v+2\dO-2\eta,0,0)\mathbb{M}^{-1}_{i,j}\left(\mathcal{K}^{+}_{i,j}(u,2\dO)\right)^{t_{i}t_{j}}\mathbb{R}_{k,\ell\tn i,j}(-u+v,0,0)\,,
			\end{aligned}
		\end{equation}
	\end{subequations}
	respectively.
	Again, the proof is analogous to the one shown in Appendix~\ref{Proof_KPlus} for $\mathbb{K}^+$. 
	Reflection algebras are of this type were introduced by Nepomechie and Retore \cite{NeRe21,NeRe21a}: they describe a moving boundary where reflection of a particle at the boundary not only changes the sign of its rapidity but also leads to the shift by $2\dO$ appearing in the argument of $\mathbb{R}$-matrix containing the sum of $u+v$.
	
	These reflection algebras together with the generalized YBE (\ref{Generalized_YBE}) ensure the commutativity of the $Q$ with itself for different arguments. Finally, we need to prove the commutativity with the transfer matrix (\ref{TM_Alterning}) in the composite picture. 
	For the open chain with alternating staggering this is not obvious  
	because the boundary matrices $\mathbb{K}^\pm$ and $\mathcal{K}^\pm$ are representations of \emph{different} reflection algebras. Remarkably it turns out that they are intertwined by the following relations
	\begin{subequations}
		\label{K_Inter}
		\begin{equation}
			\begin{aligned}
				\mathbb{R}_{i,j\tn k,\ell}&(u-v,0,-2\dO)\mathcal{K}^{-}_{i,j}(u,2\dO)\mathbb{R}_{k,\ell\tn i,j}(u+v-2\dO,2\dO,0)\mathbb{K}^{-}_{k,\ell}(v,2\dO)=\\
				&\mathbb{K}^{-}_{k,\ell}(v,2\dO)\mathbb{R}_{i,j\tn k,\ell}(u+v,0,-2\dO)\mathcal{K}^{-}_{i,j}(u,2\dO)\mathbb{R}_{k,\ell\tn i,j}(u-v-2\dO,2\dO,0)\label{KPlus_Inter}
			\end{aligned}
		\end{equation}
		and
		\begin{equation}
			\begin{aligned}
				&\mathbb{R}_{i,j\tn k,\ell}(-u+v,0,2\dO)\left(\mathcal{K}^{+}_{i,j}(u,2\dO)\right)^{t_it_j}\mathbb{M}^{-1}_{i,j}\mathbb{R}_{k,\ell\tn i,j}(-u-v+2\dO-2\eta,-2\dO,0)\\&\times\mathbb{M}_{i,j}\left(\mathbb{K}_{k,\ell}^{+}(v,2\dO)\right)^{t_kt_\ell}=
				\left(\mathbb{K}^{+}_{k,\ell}(v,2\dO)\right)^{t_kt_\ell}\mathbb{M}_{i,j}\mathbb{R}_{i,j\tn k,\ell}(-u-v-2\eta,0,2\dO)\\ &\times\mathbb{M}^{-1}_{i,j}\left(\mathcal{K}^{+}_{i,j}(u,2\dO)\right)^{t_{i}t_j}\mathbb{R}_{k,\ell\tn i,j}(-u+v+2\dO,-2\dO,0)\,.\label{KMinus_Inter}
			\end{aligned}
		\end{equation}
	\end{subequations}
	Again this can be proven as in Appendix~\ref{Proof_KPlus}.
	Using these algebras one can show on the composite level that
	\begin{align}
		Q(u)\mathcal{T}^{\text{alt}}(v,2\dO)=\mathcal{T}^{\text{alt}}(v,2\dO)Q(u)\,.
	\end{align}
	We want to stress that the intertwining relations (\ref{K_Inter}) ensure the commutativity of transfer matrices with \emph{different} boundary matrices. It would interesting to address whether similar relations between already other known boundary matrices exists.
	
	\subsection{Quasi periodic staggering}
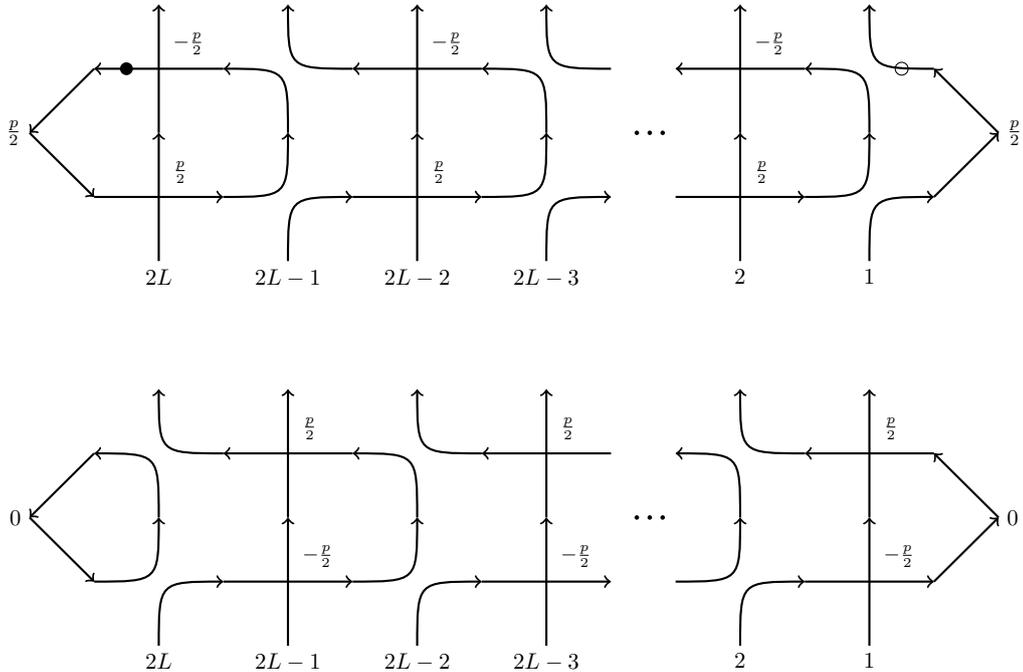
\begin{figure}

		\centering
		\begin{tikzpicture}[scale=0.85, transform shape]
			
			\ecellr{0}{6}{$\frac{p}{2}$}
			\permucellr{2}{6}
			\ecellr{4}{6}{$\frac{p}{2}$}
			\permucellr{6}{6}
			\draw[->,thick] (-1,8)--(-2,7) node[left] {$\frac{p}{2}$};
			\draw[->,thick] (-2,7)--(-1,6);
			\ecelll{0}{8}{$-\frac{p}{2}$}
			\permucelll{2}{8}
			\ecelll{4}{8}{$-\frac{p}{2}$}
			\permucelll{6}{8}
			\fill (7.4,7) circle (1pt);
			\fill (7.6,7) circle (1pt);
			\fill (7.8,7) circle (1pt);
			\ecellr{9}{6}{$\frac{p}{2}$}
			\permucellr{11}{6}
			\ecelll{9}{8}{$-\frac{p}{2}$}
			\permucelll{11}{8}
			\draw[<-,thick] (12,8)--(13,7) node[right] {$\frac{p}{2}$};
			\draw[<-,thick] (13,7)--(12,6);
			\fill (-0.5,8) circle (0.1);
			\draw (11.5,8) circle (0.1);
			
			\node[below] at (0,5) {$2L$};
			\node[below] at (2,5) {$2L-1$};
			\node[below] at (4,5) {$2L-2$};
			\node[below] at (6,5) {$2L-3$};
			\node[below] at (9,5) {$2$};
			\node[below] at (11,5) {$1$};

			\permucellr{0}{0}
			\ecellr{2}{0}{$-\frac{p}{2}$}
			\permucellr{4}{0}
			\ecellr{6}{0}{$-\frac{p}{2}$}
			\draw[->,thick] (-1,2)--(-2,1) node[left] {$0$};
			\draw[->,thick] (-2,1)--(-1,0);
			\permucelll{0}{2}
			\ecelll{2}{2}{$\frac{p}{2}$}
			\permucelll{4}{2}
			\ecelll{6}{2}{$\frac{p}{2}$}
			\fill (7.4,1) circle (1pt);
			\fill (7.6,1) circle (1pt);
			\fill (7.8,1) circle (1pt);
			\permucellr{9}{0}
			\ecellr{11}{0}{$-\frac{p}{2}$}
			\permucelll{9}{2}
			\ecelll{11}{2}{$\frac{p}{2}$}
			\draw[<-,thick] (12,2)--(13,1) node[right] {$0$};
			\draw[<-,thick] (13,1)--(12,0);
			
			\node[below] at (0,-1) {$2L$};
			\node[below] at (2,-1) {$2L-1$};
			\node[below] at (4,-1) {$2L-2$};
			\node[below] at (6,-1) {$2L-3$};
			\node[below] at (9,-1) {$2$};
			\node[below] at (11,-1) {$1$};

		\end{tikzpicture}
		\caption{The two double row transfer matrices defining the quasi momentum as in (\ref{QI_Standard}) for quasi periodic staggering evaluated at the shift point. We see that both at their own become trivial in the bulk and if one assumes that $K^-(0),K^-\left(\frac{p}{2}\right)\propto1$ then the whole transfer matrices become essentially the identity leading to a trivial quasi momentum.}
		\label{Single_TM_at_0}
	\end{figure}

	For the quasi periodic staggering the single ingredients $\tau(-\frac{p}{2},\{0,\frac{p}{2}\})$ and $\tau(0,\{0,\frac{p}{2}\})$  become trivial in the bulk, see Figure \ref{Single_TM_at_0}. 
	Under the assumptions (\ref{K_G})  the quasi momentum operator is trivial 
	\begin{align}
		\mathbb{Q}^{\mathrm{qp}}\propto \mathbf{1}.\label{QI_Triv}
	\end{align}
	Instead of constructing a generating functional in the composite picture for this trivial quasi momentum we consider the next to leading term in the expansion of the corresponding quasi shift operator, i.e.\
	\begin{align}
		\overline{\mathbb{Q}}^{\mathrm{qp}}\propto\left.\frac{\text{d}}{\text{d}u}\frac{\tau(u,\{0,\frac{p}{2}\})}{\tau(u-\frac{p}{2},\{0,\frac{p}{2}\})}\right|_{u=0}\,,\label{QI_Ableitung}
	\end{align}
	where we assume that $\tau$ is differentiable at $u=0,-p/2$.
	Below we study the properties of $\overline{\mathbb{Q}}^{\mathrm{qp}}$ in the special case of the staggered six-vertex model with quasi periodic staggering in the following chapter. 
	
	\section{Example: The staggered \texorpdfstring{$A^{(1)}_1$}{A11} (or \texorpdfstring{$D^{(2)}_2$}{D22}) model}
	In this last section we want to apply our findings to the staggered six-vertex model with elementary  $R$-matrix
	\begin{align}
		R(u)=\begin{pmatrix}
			\sinh{(u+\ri\gamma)}&0&0&0\\
			0&\sinh{(u)}&\sinh{(\ri\gamma)}&0\\
			0&\sinh{(\ri\gamma)}&\sinh{(u)}&0\\
			0&0&0&\sinh{(u+\ri \gamma)}\\
		\end{pmatrix}\,.\label{R-Matrix}
	\end{align}
	Up to different normalizations of the regularity  and crossing symmetry this $R$-matrix obeys the characteristic Eqs.~(\ref{YBE}), (\ref{R_props}) and (\ref{R_Qper})  with
	\begin{equation}
		\begin{aligned}
			M&=\mathbf{1},\qquad \eta=\ri\gamma\,,\qquad f_p=-1\,, \qquad V=\begin{pmatrix}0&1\\-1&0 \end{pmatrix}\,,\\
			p&=\ri\pi\,, \qquad G=\sigma^z\,,\qquad
			\xi(u)=\frac{1}{2}(\cos (2\gamma)-\cosh (2 u))\,.
		\end{aligned}
	\end{equation}
	Moreover, we restrict ourselves to the case of $U_q(\mathfrak{sl}(2))$-invariant boundary conditions
	\begin{align}
		K^-(u)=\begin{pmatrix}
			e^{u}&0\\
			0&e^{-u}
		\end{pmatrix}\label{Kminus}, \qquad K^+(u)=\left(K^{-}(-u-\ri\gamma)\right)^t.
	\end{align}
	The double row transfer matrix (\ref{Standard Form}) 
	can be diagonalized by means of the algebraic Bethe Ansatz \cite{KuSk91}. For the $\mathbb{Z}_2$-staggering (\ref{Z_2_Staggering}) its eigenvalues are given as
	\begin{equation}
		\label{eq:tau_eigenval}
		\begin{aligned}
			\Lambda(u)=&\frac{\sinh(2u+2\ri\gamma)}{\sinh(2u+\ri\gamma)}\bigg(\sinh(u-\delta_1+\ri\gamma)\sinh(u-\delta_2+\ri\gamma)\sinh(u+\delta_1+\ri\gamma)\sinh(u+\delta_2+\ri\gamma)\bigg)^L\\&\quad\times\frac{1}{q\det (T(-u-\frac{\ri\gamma}{2}))}\prod^{M}_{m=1} \frac{\sinh(u-v_m-\frac{\ri\gamma}{2})\sinh(u+v_m-\frac{\ri\gamma}{2})}{\sinh(u-v_m+\frac{\ri\gamma}{2})\sinh(u+v_m+\frac{\ri\gamma}{2})}\\
			&+\frac{\sinh(2u)}{\sinh(2u+\ri\gamma)}\bigg(\sinh(u+\delta_1)\sinh(u+\delta_2)\sinh(u-\delta_1)\sinh(u-\delta_2)\bigg)^L\\&\quad\times\frac{1}{q\det (T(-u-\frac{\ri\gamma}{2}))}\prod^{M}_{m=1} \frac{\sinh(u-v_m+\frac{3\ri\gamma}{2})\sinh(u+v_m+\frac{3\ri\gamma}{2})}{\sinh(u-v_m+\frac{\ri\gamma}{2})\sinh(u+v_m+\frac{\ri\gamma}{2})}\,
		\end{aligned}
	\end{equation}
	in terms of the parameters $v_m$, $m=1\dots M$, solving the Bethe equations
	\begin{align}
		\bigg(\frac{\sinh(v_m-\delta_1+\frac{\ri\gamma}{2})}{\sinh(v_m+\delta_1-\frac{\ri\gamma}{2})}&\frac{\sinh(v_m-\delta_2+\frac{\ri\gamma}{2})}{\sinh(v_m+\delta_2-\frac{\ri\gamma}{2})}\frac{\sinh(v_m+\delta_1+\frac{\ri\gamma}{2})}{\sinh(v_m-\delta_1-\frac{\ri\gamma}{2})}\frac{\sinh(v_m+\delta_2+\frac{\ri\gamma}{2})}{\sinh(v_m-\delta_2-\frac{\ri\gamma}{2})}\bigg)^{L}\label{Flow_BAE}\\=\prod^M_{k=1\neq m}&\frac{\sinh(v_m-v_k+\ri\gamma)\sinh(v_m+v_k+\ri\gamma)}{\sinh(v_m-v_k-\ri\gamma)\sinh(v_m+v_k-\ri\gamma)}\,.\notag
	\end{align}
	In (\ref{eq:tau_eigenval}) the quantum determinant reads
	\begin{equation}
		\label{Quantendeterminate}
		\begin{aligned}
			q\det(T(u))= &\sinh^L \left(u+\delta_1-\frac{\ri\gamma}{2}\right) \sinh^L \left(u+\delta_1+\frac{3\ri\gamma}{2}\right)\\&\times \sinh^L \left(u+\delta_2-\frac{\ri\gamma}{2}\right) \sinh^L \left(u+\delta_2+\frac{3\ri\gamma}{2}\right)\,.
		\end{aligned}
	\end{equation}
	
	From our discussion above we know that only the alternating or quasi periodic staggering leads to a local Hamiltonian. Both cases have been studied extensively in \cite{RoJS21,FrSg2022} and \cite{RoPaJaSa20} respectively. Using the vertex representation of the Temperley-Lieb generators $e_{i,i+1}$:
	\begin{align}
		e_{j,j+1}=\left(\mathds{1}_{\mathbb{C}^2}\right)^{\otimes j-1}\otimes\begin{pmatrix}0&0&0&0\\0&-e^{-\ri\gamma}&1&0\\0&1&-e^{\ri\gamma}&0\\0&0&0&0\end{pmatrix}\otimes \left(\mathds{1}_{\mathbb{C}^2}\right)^{\otimes 2L-j-1}\,,
	\end{align}
	obeying 
	\begin{equation}
		\label{TL Algebra}
		\begin{aligned}
			e_{j,j+1}^2&=-2\cos(\gamma)e_{j,j+1}\,,\\
			e_{j,j+1}e_{j+1,j+2}e_{j,j+1}&=e_{j,j+1}\,,\\
			e_{j+1,j+2}e_{j,j+1}e_{j+1,j+2}&=e_{j+1,j+2}\,,\\
			e_{k,k+1}e_{j,j+1}&=e_{j,j+1}e_{k,k+1}\,, \qquad |k-j|>1\,.
		\end{aligned}    
	\end{equation}
	the Hamiltonian with alternating staggering (\ref{stagg_alt}) can be written as
	\begin{equation*}
		\begin{aligned}
			H^{alt}(2\dO)=&-\frac{1}{\sin(\gamma)\xi(2\dO)}\Bigg(\sum^{2L-1}_{j=1} 2\xi(2\dO) e_{j,j+1}-\sinh(2\dO)\sum^{2L-1}_{j=2} \sinh(2\dO+\ri(-1)^{j+1}\gamma)e_{j,j+1}e_{j-1,j}\\
			&\qquad \qquad \qquad \qquad -\sinh(2\dO)\sum^{2L-1}_{j=2}\sinh(2\dO+\ri(-1)^j\gamma)e_{j-1,j}e_{j,j+1}\Bigg)\,.
		\end{aligned}
	\end{equation*}
	As discussed in Section~\ref{ch:local_quasi} the model where the free staggering parameter is related to the quasi period as $2\dO=\frac{p}2=\frac{\ri\pi}2$ deserves special attention. In this case the Hamiltonian of the staggered model becomes
	\begin{align}
		\label{eq:Heps1_alt}
		H^{alt}\left(\frac{\ri\pi}2\right)=&-\frac{2}{\sin(2\gamma)} \left(2\cos(\gamma) \sum^{2L-1}_{j=1} e_{j,j+1}+\sum^{2L-1}_{j=2} e_{j,j+1}e_{j-1,j}+e_{j-1,j}e_{j,j+1}\right)\,.
	\end{align}
	This expression coincides with the Hamiltonian of the model with quasi periodic staggering (\ref{stagg_qper}) up to boundary terms 
	\begin{align}
		\label{eq:Heps0_qp}
		H^{qper}=& H^{alt}\left(\frac{\ri\pi}2\right)+\frac{2}{\sin(2\gamma)\cos(\gamma)}\left(e_{1,2}+e_{2L-1,2L}\right)\,.
	\end{align}
	As mentioned earlier, the staggered model for $2\dO=\frac{\ri\pi}2$ has an extended underlying $D_2^{(2)}$ symmetry.  In this context the boundary matrices (\ref{K_Alternating}), (\ref{K_Quasiperiodic}) are different representations \cite{MaGu00,NePR17,NePi18} of the corresponding $D^{(2)}_2$ reflection algebra.

	\subsection{Spectral flow between the integrable points}
	Remarkably, the choice of boundary conditions has a profound influence on the low energy properties of the staggered models: 
	the effective theory of (\ref{eq:Heps1_alt}) with anisotropy $\gamma<2\dO<\pi-\gamma$ has been identified to be the $SL(2,\mathbb{R})/U(1)$ sigma model at level $k=\pi/\gamma$ with a non-compact spectrum of conformal weights \cite{RoJS21,FrSg2022}. On the contrary, Robertson \emph{et al.} found that the continuum limit of the model (\ref{eq:Heps0_qp}) is compact \cite{RPJS20}.
	The boundary RG flow between these two critical fixed points has been studied numerically: based on finite size estimates of the gap between the ground state and the lowest excitation when the amplitude of the boundary term in (\ref{eq:Heps0_qp}) is varied between the two integrable points it has been concluded that the fixed points corresponding to (\ref{eq:Heps1_alt}) and (\ref{eq:Heps0_qp}) are unstable and stable, respectively \cite{RoJS21}.
	
	In the setting established in this paper both (\ref{eq:Heps1_alt}) and (\ref{eq:Heps0_qp}) originate from a staggered vertex model with the $A_1^{(1)}$ $R$-matrix (\ref{R-Matrix}). This allows to study the spectral flow between (\ref{eq:Heps1_alt}) and (\ref{eq:Heps0_qp}) in an integrable setting with a fixed choice of the boundary matrices under the variation of the bulk inhomogeneities.
	The price to pay for integrability is giving up locality of the Hamiltonian at the intermediate points.  Using the same staggering in the vertical and horizontal directions of the vertex model, i.e.\ $\{\dO,\dOBar\}=\{\delta_1,\delta_2\}$, we tune $\delta_1$ and $\delta_2$ to interpolate between the integrable models with local interactions. 
	We choose the following  normalization of the non-local 'Hamiltonian' 
	\begin{align}
		H=& \frac{q\det \left(T\left(-\delta_1-\frac{\ri\gamma}{2}\right)\right)q\det \left(T\left(-\delta_2-\frac{\ri\gamma}{2}\right)\right)}{2\ri f(\delta_1)f(\delta_2)}\label{H_Op_Flow}\\
		&\times \left(\left.\frac{\text{d}}{\text{d}u}\right|_{u=0}\mathcal{T}(u,\{\delta_1,\delta_2,\delta_1,\delta_2\})-\left.\frac{\text{d}}{\text{d}u}\right|_{u=0}f(u+\delta_1)f(u+\delta_2)\right)\notag
	\end{align}
	with
	\begin{align}
		f(u)=&\frac{\sinh(2u+2\ri\gamma)}{\sinh(2u+\ri\gamma)}\sinh^L(u-\delta_1+\ri\gamma)\sinh^L(u-\delta_2+\ri\gamma)\label{f_Fkt}\\&\times\sinh^L(u+\delta_1+\ri\gamma)\sinh^L(u+\delta_2+\ri\gamma)\,.\notag
	\end{align}
	Specifically, we choose for the remaining inhomogeneities the parameterization
	\begin{align}
		\delta_1=\frac{\ri\FlowParameter}{2}+\frac{\ri\pi}{4}\,, \qquad     \delta_2=\frac{\ri\FlowParameter}{2}-\frac{\ri\pi}{4}\,, \qquad
		-\frac{\pi}2\leq \FlowParameter \leq 0\,,
		\label{Flow_Staggering}
	\end{align}
	resulting in alternating (quasi periodic) staggering for $\FlowParameter=0$ and $-\pi/2$, respectively.
	
	The eigenvalues of the 'Hamiltonian' (\ref{H_Op_Flow}) in this parameterization are given in terms of the Bethe roots $\{v_m\}$ solving (\ref{Flow_BAE}) as
	\begin{equation}
		\label{eq:EFlow}
		\begin{aligned}
			E=& \left(-4L\cot(2\gamma)+
			L \frac{2\sin(2\FlowParameter)}{\sin(2\gamma)\sin(2(\gamma+\FlowParameter))}
			-\frac{2\sin(\FlowParameter)}{\cos(\gamma)\cos(\gamma+\FlowParameter)}+\frac{2\sin(\FlowParameter)}{\cos(2\gamma)\cos(2\gamma+\FlowParameter)}\right.\\&\qquad \qquad \left.+\frac{2\tan(\gamma)}{\cos(2\gamma)}\right) \times \left(\prod^M_{m=1}\frac{\cos(2(\gamma-\FlowParameter))+\cosh(4 v_m)}{\cos(2(\gamma+\FlowParameter))+\cosh(4 v_m)}-1\right)\\
			&-4\sin(2\gamma)\sum^{M}_{k=1}\left\{\frac{\cos(2\gamma)+\cos(\FlowParameter)\cosh(4v_k)}{(\cos(2(\gamma+\FlowParameter))+\cosh(4v_k))^2}\right\} \times \prod^M_{\substack{m=1{}\\m\neq k}}\frac{\cos(2(\gamma-\FlowParameter))+\cosh(4 v_m)}{\cos(2(\gamma+\FlowParameter))+\cosh(4 v_m)}\,.
		\end{aligned}
	\end{equation}
	As expected, this expression reduces to a sum of bare quasi-particle energies $\epsilon_0(v_m)$ for $\FlowParameter=0,\pm\pi/2$ where the Hamiltonian becomes local.
	Away from these points, the normalization of (\ref{eq:EFlow}) leads to singularities at particular values of the flow parameter: 
	the one at $\FlowParameter=\frac{\pi}{2}-2\gamma$ can be removed by multiplying the Hamiltonian by the $\FlowParameter$-dependent factor $\cos(2\gamma+\FlowParameter)$ while $\FlowParameter=\frac{\pi}2-\gamma>0$ does not lie on the spectral flow (\ref{Flow_Staggering}).  The remaining singularity at $\FlowParameter_{c1}=-\gamma$ depends on the state considered.  In terms of the corresponding Bethe root configuration this can be related to the low energy root configurations of the quasi periodic model (\ref{eq:Heps0_qp}). These consists of pairs of complex conjugate roots with imaginary part $\pm \frac{\pi}{4}$  and an additional root at $\frac{\ri\pi}{4}$ for $M^\mathrm{qper}$ odd \cite{RPJS20}:
	\begin{align}
		v^{\mathrm{qper}}\in \left\{ x_m+\frac{\ri\pi}{4},x_m-\frac{\ri\pi}{4}\bigg| x_m\in \mathbb{R}_{>0}\,, m=1,...,\left\lfloor\frac{M^\mathrm{qper}}{2}\right \rfloor \right\}\cup \left\{\frac{\ri\pi}{4} \right\} \,.\label{BRoots_qper}
	\end{align}
	Here $\lfloor\dots\rfloor$ denotes the  Gaussian bracket.  Exact diagonalization of the Hamiltonian for small systems together with the determination of the corresponding Bethe roots shows that root patterns of this type persist throughout the interval $\FlowParameter=-\pi/2\dots\FlowParameter_{c1}$. At $\FlowParameter_{c1}$, however, several roots become purely imaginary, $v_m=\ri\pi/4$, changing the order of the pole in (\ref{eq:EFlow}). This singularity can be removed by renormalization of the spectrum by a factor $\sin^\nu(\FlowParameter+\gamma)$ with an appropriate choice of an integer $\nu$. 
	
	The spectral flow starting from the alternating model (\ref{eq:Heps1_alt}), $\FlowParameter=0$, can be studied in a similar way: here a class of low energy states (including the ground state) is known \cite{FrSg2022,RoJS21} to be described by configurations consisting of real roots and ones having an imaginary part of $\frac{\pi}{2}$:
	\begin{align}
		v^{\mathrm{alt}}=\left\{
		x_m,y_n+\frac{\ri\pi}{2} \bigg| x_m,y_n\in\mathbb{R}_{>0},\,  m=1,...,M_0,\,
		n=1,...,M_{\frac{\ri\pi}{2}} \right\}
		\,.\label{BRoots_alt}
	\end{align}
	Configurations of this type exist in the interval $\FlowParameter=\FlowParameter_{c2}\dots0$ where roots with vanishing real parts appear for $\FlowParameter_{c2}=\gamma-\pi/2$. These do not, however, lead to singularities in the eigenvalues of (\ref{eq:EFlow}).
	
\begin{figure}[!ht]
		\centering
		\subfigure{\includegraphics[width=0.7\textwidth]{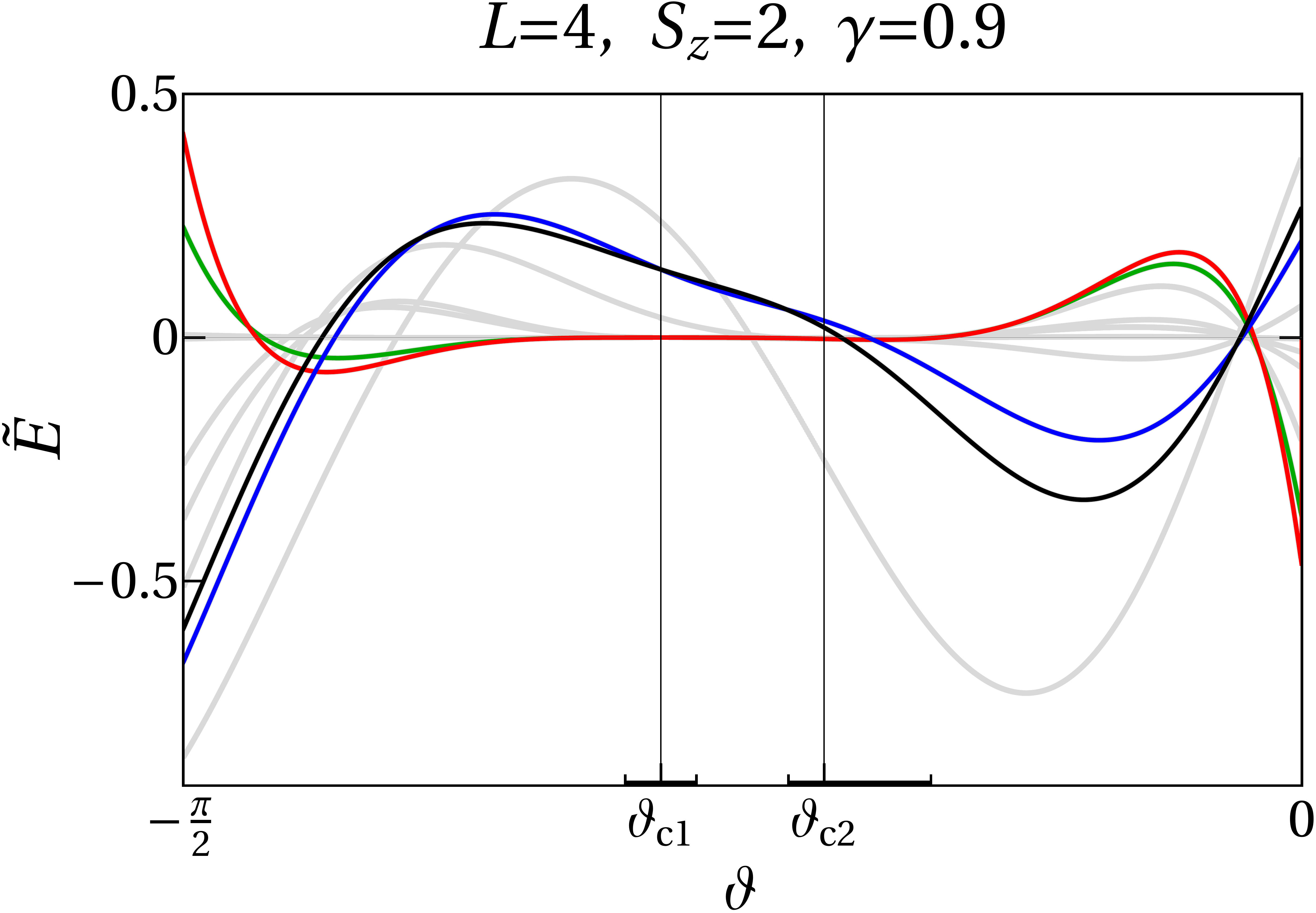}}\\
		\subfigure{\includegraphics[width=0.49\textwidth]{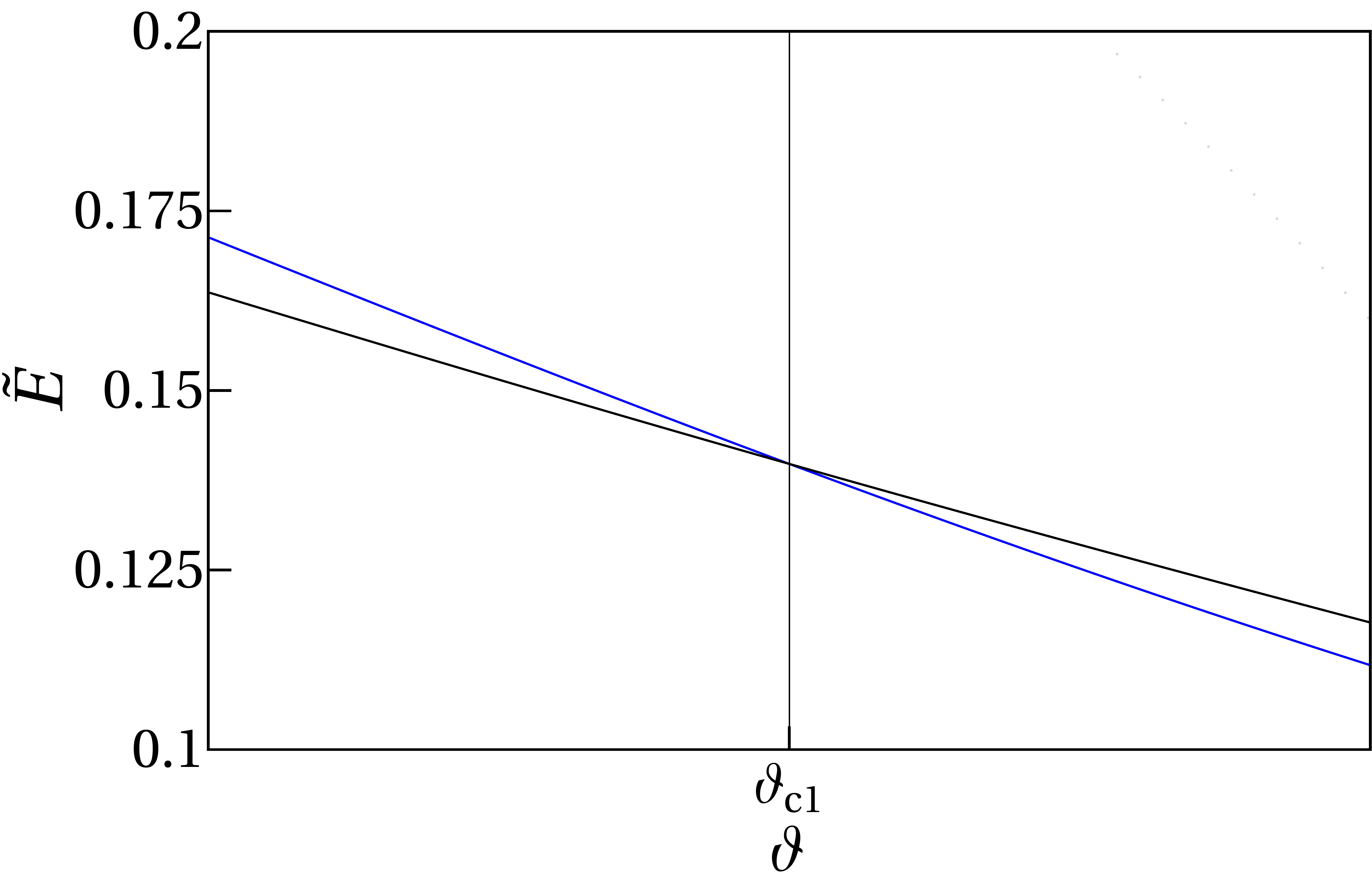}}
		\subfigure{\includegraphics[width=0.49\textwidth]{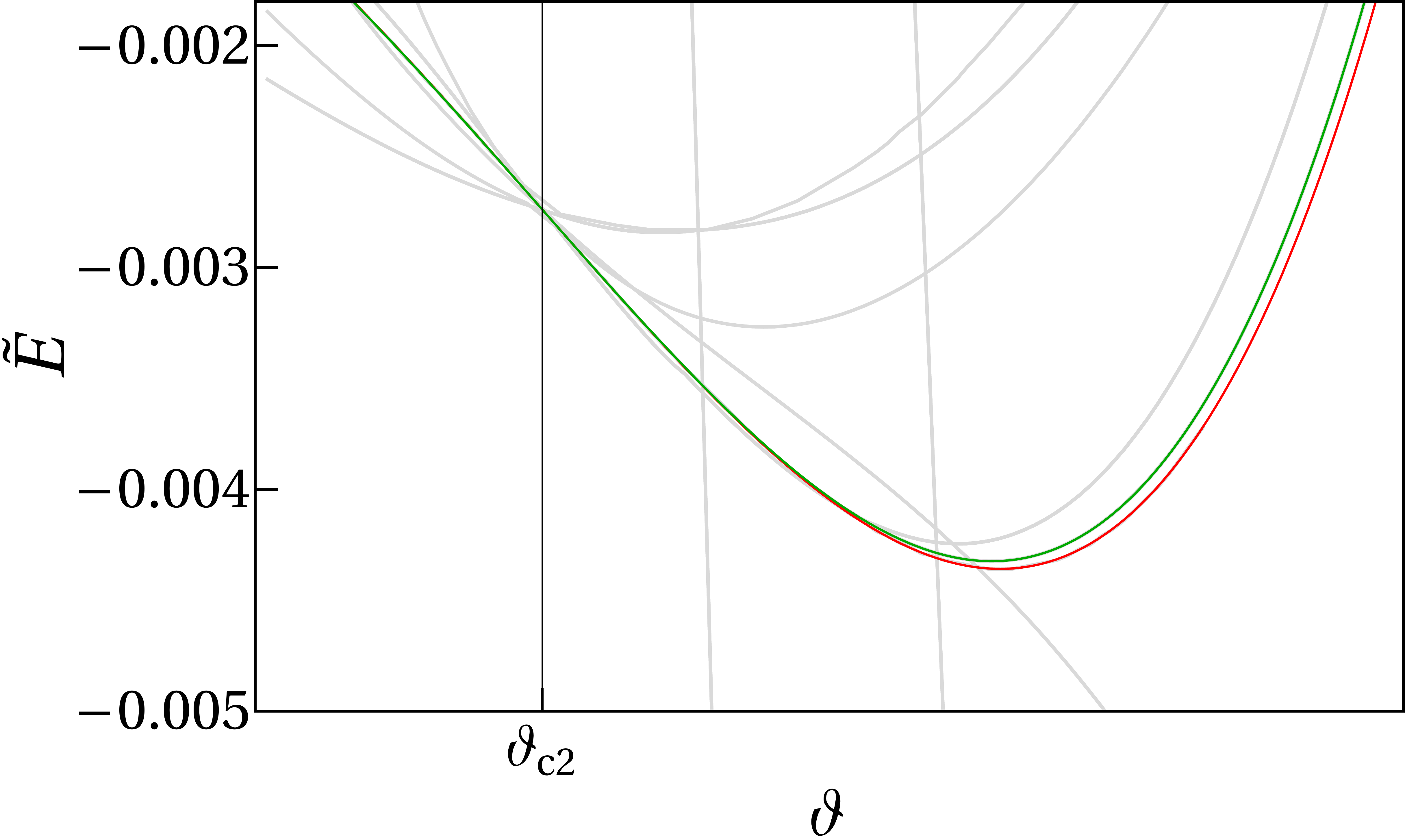}}
		\caption{Rescaled real parts $\tilde{E}$ of the eigenenergies of the staggered six-vertex Hamiltonian with $2L=8$ sites, anisotropy $\gamma=0.9$ for the charge sector $S_z=2$ for the spectral flow (\ref{Flow_Staggering}). Energies have been multiplied with  $-\cos(\FlowParameter+2\gamma)\sin^5(\FlowParameter+\gamma)$ to regularize the singularities as described in the main text. The mapping between low and high energy states in the local models is clearly seen. In the lower plots the level crossings at $\FlowParameter=\FlowParameter_{c1},\FlowParameter_{c2}$ of low lying states evolving from the respective ground states are resolved within the $\FlowParameter$-intervals indicated in the upper image details.
		}
		\label{Spectral_Flow_Exact_Diag}
	\end{figure}

	Both at $\FlowParameter_{c1}$ and $\FlowParameter_{c2}$ the appearance of purely imaginary Bethe roots leads to degeneracies involving many states, as shown in results from exact diagonalization for a spin chain with $2L=8$ sites, anisotropy $\gamma=0.9$ for the charge sector $S_z=2$ shown in Figure~\ref{Spectral_Flow_Exact_Diag}.  Under the spectral flow low energy states of the local Hamiltonian (\ref{eq:Heps1_alt}) are mapped to high energy ones for (\ref{eq:Heps0_qp}) and vice versa. The crossing of a large number of levels indicates the presence of first-order transitions when the flow parameter is $\FlowParameter_{c1}$ or $\FlowParameter_{c2}$.
	
	Support for this interpretation is obtained by studying the spectral flow within the root density formalism \cite{YaYa69}: in the thermodynamic limit the densities of roots in the configuration (\ref{BRoots_alt}) of the alternating model are found to be
	\begin{align}
		\rho^a(v)=\sigma^a(v)+\frac{1}{L}\tau^a(v)\,,\quad a=x,y\,,
	\end{align}
	with bulk and surface contributions
	\begin{align*}
		\sigma^x(v)=\sigma^y(v)=&\frac{\cos\left(\frac{\pi\,\FlowParameter}{\pi-2\gamma}\right)}{\pi-2\gamma}\left[
		\frac{1}
		{\cosh\left(\frac{2\pi v}{\pi-2\gamma}\right)+
			\sin\left(\frac{\pi\,\FlowParameter}{\pi-2\gamma}\right)}
		+\frac{1}
		{\cosh\left(\frac{2\pi v}{\pi-2\gamma}\right)
			-\sin\left(\frac{\pi\,\FlowParameter}{\pi-2\gamma}\right)}
		\right]\,,\\
		\tau^x(v)=
		\tau^y(v)=&\frac{1}{4\pi} \int^{\infty}_{-\infty} \text{d}\omega e^{\ri\omega v}  \frac{\sinh\left(\frac{3\gamma-\pi}{4}\omega\right)}{\sinh\left(\frac{\gamma \omega}{4}\right) \cosh \left(\frac{2\gamma-\pi}{4}\omega\right)}\,.
	\end{align*}
	Similarly, the density $\bar{\rho}(x)$ of root configurations (\ref{BRoots_qper}) of the quasi periodic model is found to be
	\begin{align}
		\bar{\sigma}(x)=&\frac{4}{\pi-2\gamma}\frac{\cos\left(\frac{\pi}{2}\frac{\pi+2\FlowParameter}{\pi-2\gamma}\right)\cosh\left(\frac{2\pi x}{\pi-2\gamma}\right)}{\cosh\left(\frac{4\pi x}{\pi-2\gamma}\right)+\cos\left(\pi \frac{\pi+2\FlowParameter}{\pi-2\gamma}\right)}\,,\quad
		\bar{\tau}(x)=\frac{1}{\pi-2\gamma}\frac{1}{\cosh\left(\frac{2\pi x}{\pi-2\gamma}\right)}\,.\label{Rdensities_qp}
	\end{align}
	That the bulk densities $\sigma^{x,y}(v)$ ($\bar{\sigma}(x)$) vanish at $\FlowParameter_{c2}$ ($\FlowParameter_{c1}$) indicates a transition into a different state in accordance with our results for small system sizes.

	\subsection{Role of the quasi momentum in the quasi periodic model}
	As mentioned above the continuum limit of the alternating model (\ref{eq:Heps1_alt}) is described by a non-compact conformal field theory.  In the lattice model this is manifest in the finite size gaps closing as
	\begin{equation}
		\Delta E^\mathrm{alt}_n = E^\mathrm{alt}_n - L e^\mathrm{alt}_\infty - f^\mathrm{alt}_\infty \sim \frac{\pi v^\mathrm{alt}_F}L \left( \text{const.}+\text{const.}\frac{(dN^\mathrm{alt})^2}{\log(L)^2}\right)\,,
	\end{equation}
	where $Le^\mathrm{alt}_\infty$ and $f^\mathrm{alt}_\infty$ are the bulk and surface contributions to the energy and $v^{\mathrm{alt}}_F$ is the Fermi velocity. On the level of the Bethe configurations (\ref{BRoots_alt}) logarithmic corrections arise when the numbers $M_0$ ($M_{\frac{\ri\pi}{2}}$) of roots with $\text{Im}(v_m^{alt})=0$ ($\frac\pi2$) are different, i.e.\ $dN^\mathrm{alt}=M_0-M_{\frac{\ri\pi}{2}}\neq 0$. For large $L$ the eigenvalues of the quasi momentum operator are proportional to   $dN^\mathrm{alt}/\log (L)$,  which allows a direct identification of the underlying CFT \cite{IkJS12,FrSe14,FrSg2022}.
	
\begin{figure}
		\centering
		\includegraphics[width=0.7\textwidth]{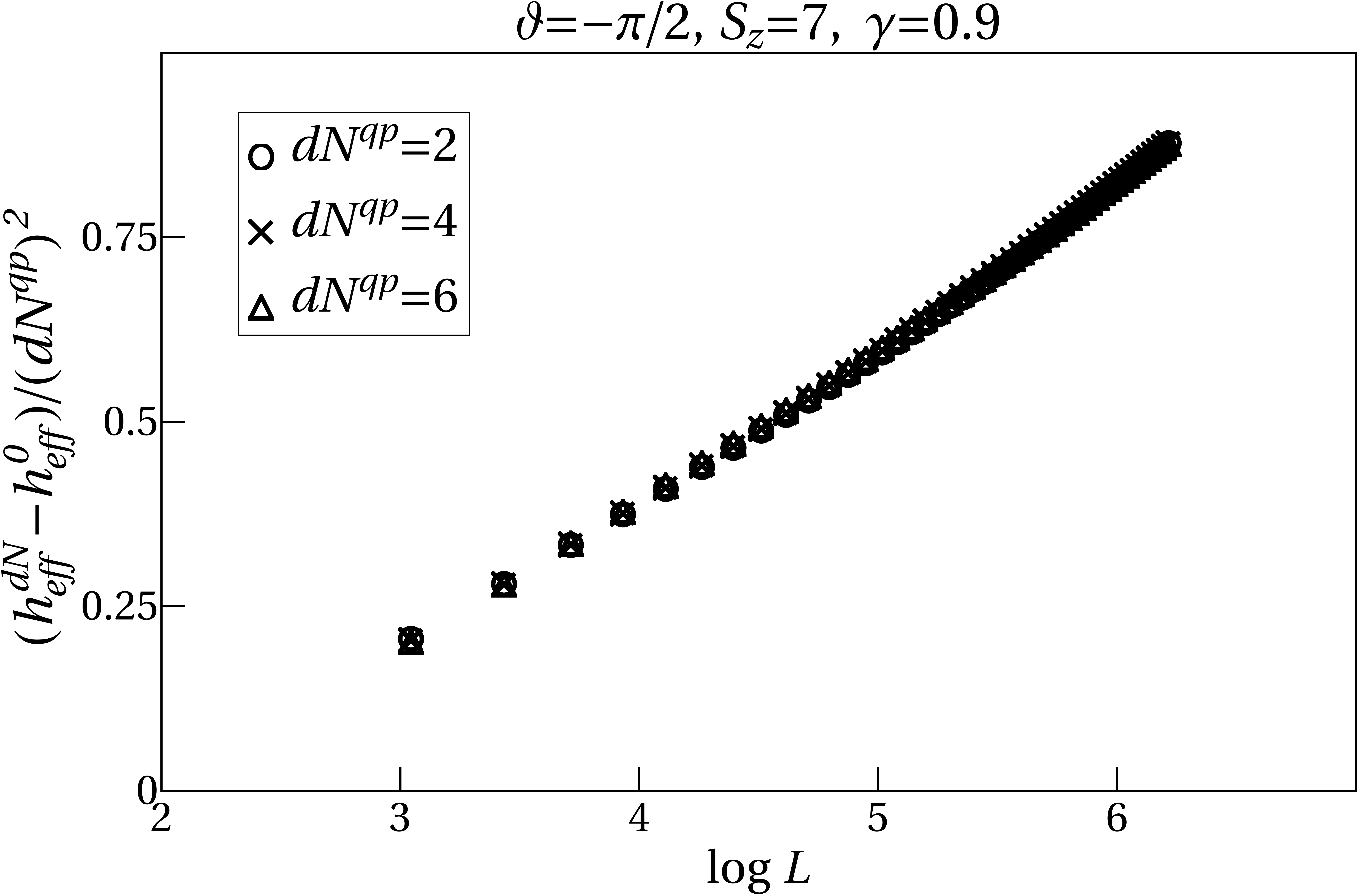}
		\caption{Scaling of energy gaps (\ref{Finite_Size_Formula}) of the quasi periodic chain in the sector $S_z=7$ weighted by $1/\left(dN^\mathrm{qp}\right)^2$ obtained by solving the Bethe Ansatz equations for various $dN^\mathrm{qp}$ and $L$ and fixed $\gamma=0.9$. We see that the scaling dimensions of those state depend on $\left(dN^\mathrm{qp}\right)^2$ and display a clear logarithmic divergence as $L\to \infty$. While solving the Bethe equations numerically, we found that the $dN^{\mathrm{qp}}$ needs to be smaller than $S_z$ for numerical convergence.}
		\label{FS_qper}
	\end{figure}

	In this section, we reconsider the finite-size analysis of the quasi periodic chain using the definition (\ref{QI_Ableitung}).
	Motivated by the insights gained in the alternating model we consider Bethe configurations (\ref{BRoots_qper}) with different numbers $M_{\pm \frac{\ri\pi}{4}}$ of roots on the lines $\text{Im}(v_m^{alt})=\pm \frac{\pi}{4}$.  
	To investigate the scaling behavior of those states, we consider the rescaled energy gaps
	\begin{equation}
		\label{Finite_Size_Formula}
		\begin{aligned}
			h^n_{\text{eff}}&=\frac{L}{\pi v^\mathrm{qp}_F}\left(E^\mathrm{qp}_n - L e^\mathrm{qp}_\infty - f^\mathrm{qp}_\infty\right)\,.
		\end{aligned}    
	\end{equation}
	The energies $e^\mathrm{qp}_\infty$, $f^\mathrm{qp}_\infty$ and Fermi velocity of the quasi periodic model are obtained in the root density formalism using (\ref{Rdensities_qp}) with $\FlowParameter=-\frac{\pi}{2}$ 
	\begin{align}
		e^\mathrm{qp}_{\infty}=2f^\mathrm{qp}_\infty=&-\frac{1}{2}\int^{\infty}_{-\infty}\text{d}\omega \frac{\sinh \left(\frac{\gamma  \omega }{2}\right)}{\sinh \left(\frac{\pi  \omega }{4}\right) \cosh \left(\frac{1}{4} (\pi -2 \gamma ) \omega \right)}\,, \qquad v^\mathrm{qp}_F=\frac{2 \pi }{\pi -2 \gamma }\,.
	\end{align}
	Using the Bethe-Ansatz we have calculated $h^n_{\text{eff}}$ for states with various $dN^\mathrm{qp}=M_{-\frac{\ri\pi}{4}}-M_{\frac{\ri\pi}{4}}$ as displayed in Fig.\  \ref{FS_qper}.
	Differing from the alternating model, we see that the $h^n_{\text{eff}}$ \emph{diverge} logarithmically with an amplitude proportional to  $\left(dN^\mathrm{qp}\right)^2$.
	We interpret these diverging scaling dimensions in the quasi periodic case in the manner that states with $dN^{\mathrm{qp}}\neq 0$ disappear from the low energy sector in the thermodynamic limit. Only states with $dN^\mathrm{qp}=0$ stay in the low energy regime as $L$ tends to infinity which have been extensively studied in \cite{RPJS20}.

\begin{figure}
		\centering
		\includegraphics[width=0.7\textwidth]{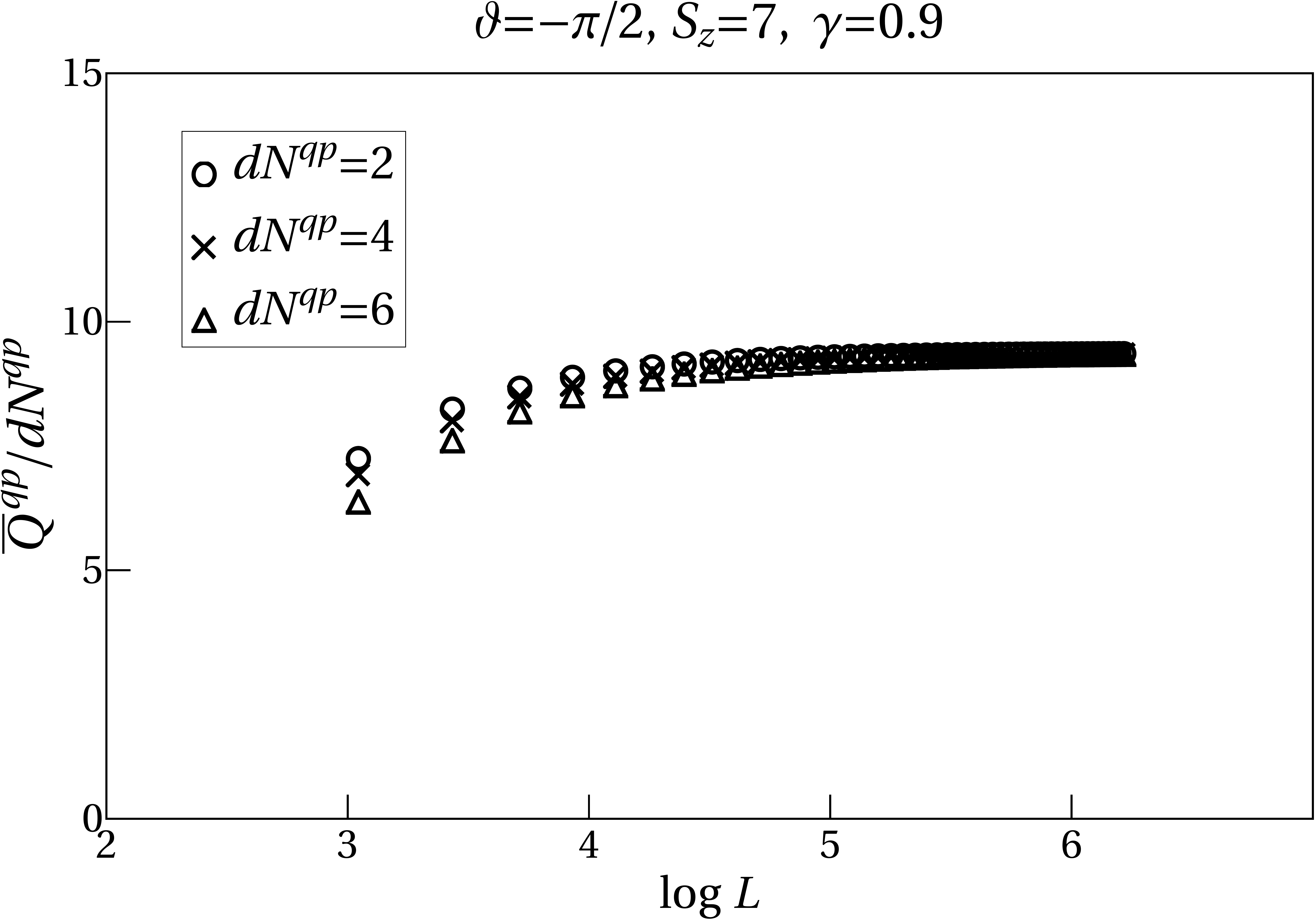}
		\caption{The quasi momentum of Bethe states having a non vanishing quantum number $dN^{\mathrm{qp}}$.}
		\label{QI_QP_Scaling}
	\end{figure}

	To parameterize this behavior in terms of the quasi momentum we consider (\ref{QI_Ableitung}) (recall that the lowest order term in the expansion of the quasi shift operator (\ref{QI_Triv}) is trivial for the quasi periodic model). For the staggered six-vertex model this operator can be expressed in terms of Pauli matrices as 
	\begin{equation}
		\label{QI_QP}
		\begin{aligned}
			\overline{\mathbb{Q}}^{\mathrm{qp}}= \bigg\{&2  \cos (\gamma ) \left(\sum _{j=1}^{2 L-2} (\sigma^{-}_{j}\sigma^{+}_{j+1}\sigma^{z}_{j+2}-\sigma^{+}_{j}\sigma^{-}_{j+1}\sigma^{z}_{j+2}+\sigma^{z}_{j}\sigma^{-}_{j+1}\sigma^{+}_{j+2}-\sigma^{z}_{j}\sigma^{+}_{j+1}\sigma^{-}_{j+2})\right)\\
			&+ \frac{\cos (\gamma )}{\sin (\gamma )} \left(\sum _{j=1}^{2 L-2} (-1)^j (2 (\sigma^{-}_{j+2}\sigma^{+}_{j}+\sigma^{+}_{j+2}\sigma^{-}_{j})+\sigma^{z}_{j}\sigma^{z}_{j+2})\right)\\
			&-2 \ri \sin (\gamma ) (\sigma^{+}_{1}\sigma^{-}_{2}-\sigma^{-}_{1}\sigma^{+}_{2})+2 \ri \sin (\gamma ) (\sigma^{-}_{2L}\sigma^{+}_{2L-1}-\sigma^{+}_{2L}\sigma^{-}_{2L-1})\\
			&-\ri  \sigma^{z}_{1}+\ri  \sigma^{z}_{2}+\ri  \sigma^{z}_{2L-1}-\ri  \sigma^{z}_{2L}\bigg\}\frac{1}{\ri \cos ^2(\gamma )}\,.
		\end{aligned}
	\end{equation}
	Note this this a sum of local operators in contrast to the alternating case, see Fig. \ref{Figure_QI_Alternating}. The normalization of in (\ref{QI_QP}) is chosen such that its eigenvalue $\overline{\mathcal{Q}}^{\mathrm{qp}}$  can be expressed in a simple form in terms of the Bethe roots: 
	\begin{align}
		\overline{\mathcal{Q}}^{\mathrm{qp}}=\sum_{j}q_0(v_j) \quad \text{with} \quad q_0(u)=\frac{16\ri\sin(\gamma)\cosh(2v_j)}{\cosh(4u)-\cos(2\gamma)}\,.\label{EV_QI}
	\end{align}
	Note that $\overline{\mathcal{Q}}^{\mathrm{qp}}$ measures the difference of the number of Bethe roots on the lines $\pm \frac{\ri\pi}{4}$:
	\begin{align}
		q_0(x+\frac{\ri\pi}{4})=-q_0(x-\frac{\ri\pi}{4})\,.
	\end{align}
	This is similar to the role of (\ref{QI_Alt}) in the alternating case.
	In the present case, however, the $\overline{\mathcal{Q}}^{\mathrm{qp}}\propto dN^{\mathrm{qp}}$ as $L\to \infty$, see Fig. \ref{QI_QP_Scaling}.
	Hence, $\overline{\mathcal{Q}}^{\mathrm{qp}}$ does not capture the $L$-dependence observed in Fig. \ref{FS_qper}.

	\section{Conclusion}
	Starting from an elementary solution $R$ of the YBE (\ref{YBE}) we have constructed the composite $\mathbb{R}$-matrix (\ref{Def_Big_R}). In addition to the spectral parameter the composite matrix depends on two free parameters related to the staggering of the elementary vertices (see Figure \ref{Graphical_Rep_Big_R}).  $PT$-symmetry, unitarity, regularity and  crossing unitarity are inherited from the elementary $R$-matrix. Most importantly, the composite $\mathbb{R}$-matrix obeys a generalized YBE (\ref{Generalized_YBE}).  In this picture the commuting transfer matrices of arbitrary $\mathbb{Z}_2$-staggered models can be rewritten as homogeneous ones where the staggering parameters enter through the additional arguments of the composite $\mathbb{R}$-matrices (and reflection matrices in the case of open boundary conditions). Integrals of motion (including the quasi momentum operator) whose natural definition relies on the staggering of the model have been described in the \emph{homogeneous} picture based on the composite $\mathbb{R}$-matrix.  In the case of open boundary conditions the Hamiltonian and the quasi momentum are defined in terms of \emph{different} representations of the reflection algebra intertwined by (\ref{K_Inter}).
	This construction may provide insights into a definition of this operator in homogeneous models (lacking a known factorization of the transfer matrix). This is of particular interest for such models featuring a continuous component of the conformal spectrum at criticality \cite{VeJS14,VeJS16a,FrMa18,FrHM19}. Knowing the quasi momentum operator in these models is expected to foster the identification of the CFT describing the low energy regime.  
	
	Demanding locality in the Hamiltonian limit leads to constraints on the staggering parameters: in the case of periodic boundary conditions they have to be tuned to satisfy (\ref{Loc_Con_PBC}).  For open boundary conditions the staggering has to satisfy Eq.~(\ref{stagg_alt}) for 'alternating staggering'.  Moreover, for quasi periodic $R$- (and $\mathbb{R}$-) matrices the inequivalent choice of 'quasi periodic staggering' (\ref{stagg_qper}) leads to a different Hamiltonian with local interactions. For both cases we have identified the corresponding composite boundary matrices which generalizes the findings of the factorization \cite{Li:2022clv} for the $D^{(2)}_2$ boundary matrices \cite{MaGu00,NePR17,NePi18} to arbitrary algebras. 
	
	Applying our construction to this model we have studied the spectral flow between the alternating and the quasi periodic model. Following this flow along a line of integrable models we find that the endpoints are separated by two first-order transitions which is consistent with the different properties of the corresponding spectra observed previously. The different role of the quasi momentum operator in the alternating and quasi periodic model are briefly discussed.

	\begin{acknowledgments}
		Funding for this work has been provided by the Deutsche Forschungsgemeinschaft under grant No.\ Fr~737/9-2 as part of the research unit   \emph{Correlations in Integrable Quantum Many-Body Systems} (FOR2316).
	\end{acknowledgments}
	\appendix

	\section{Proof of the reflection algebra (\ref{Reflection_algebra_Large_K_Plus}) for \texorpdfstring{$\mathbb{K}^+$}{K+}\label{Proof_KPlus}}
	Using the definitions of the composite quantities in the LHS of the reflection algebra (\ref{Reflection_algebra_Large_K_Plus}) we obtain  (we omit the prefactor of $\mathbb{K}^+$ and set $ijk\ell=1234$ for notational clarity):  
	\begin{align*}
		\mathbb{R}_{1,2\tn 3,4}&(-u+v,-\theta,-\theta)\left(\mathbb{K}^{+}_{1,2}(u,\theta)\right)^{t_1t_2}\mathbb{M}^{-1}_{1,2}\mathbb{R}_{3,4\tn 1,2}(-u-v-2\eta,-\theta,-\theta)\mathbb{M}_{1,2}\left(\mathbb{K}^{+}_{3,4}(v,\theta)\right)^{t_3t_4}\\&=R_{1,4}(-u+v-\theta)R_{1,3}(-u+v)R_{2,4}(-u+v)R_{2,3}(-u+v+\theta)\\
		&\times\left(P_ {1,2}K^+_2(u-\frac{\theta}{2})M_1R_{1,2}(-2u-2\eta)M^{-1}_1K^{+}_1(u+\frac{\theta}{2})\right)^{t_1t_2}M^{-1}_1M^{-1}_2\\
		&\times R_{3,2}(-u-v-2\eta-\theta)R_{3,1}(-u-v-2\eta)R_{4,2}(-u-v-2\eta)R_{4,1}(-u-v-2\eta+\theta)\\
		&\times M_1M_2 \left(P_ {3,4}K^+_4(v-\frac{\theta}{2})M_3R_{3,4}(-2v-2\eta)M^{-1}_3K^{+}_3(v+\frac{\theta}{2})\right)^{t_3t_4}\,.
	\end{align*}
	Resolving the transpositions and reordering the permutation operators gives:
	\begin{align*}
		&\mathbb{R}_{1,2\tn 3,4}(-u+v,-\theta,-\theta)\left(\mathbb{K}^{+}_{1,2}(u,\theta)\right)^{t_1t_2}\mathbb{M}^{-1}_{1,2}\mathbb{R}_{3,4\tn 1,2}(-u-v-2\eta,-\theta,-\theta)\mathbb{M}_{1,2}\left(\mathbb{K}^{+}_{3,4}(v,\theta)\right)^{t_3t_4}\\
		&=P_{3,4}R_{1,3}(-u+v-\theta)R_{1,4}(-u+v)R_{2,3}(-u+v)R_{2,4}(-u+v+\theta)\\
		&\times \left(K^{+}_1\left(u+\frac{\theta}{2}\right)\right)^{t_1}M^{-1}_1R_{2,1}(-2u-2\eta)M_1\left(K^{+}_2\left(u-\frac{\theta}{2}\right)\right)^{t_2}M^{-1}_1M^{-1}_2\\
		&\times R_{4,1}(-u-v-2\eta-\theta)R_{4,2}(-u-v-2\eta)R_{3,1}(-u-v-2\eta)R_{3,2}(-u-v-2\eta+\theta)\\
		&\times M_1M_2\left(K^{+}_4\left(v+\frac{\theta}{2}\right)\right)^{t_4}M^{-1}_4R_{3,4}(-2v-2\eta)M_4\left(K^{+}_3\left(v-\frac{\theta}{2}\right)\right)^{t_3} P_ {1,2}\,.
	\end{align*}
	We present from now on also the graphical proof for maximal clarity: \\[8pt]
	\begin{minipage}[c]{0.45\textwidth}
		\begin{tikzpicture}
			\footnotesize
			\node[right] at (3.95,5) {$2$};
			\draw[>->,thick] (4,5) to(-1,6);
			\draw[->,thick] (-1,6) to (4,7);
			\cross{3.}{5.2}
			\Incross{1.37}{5.55}
			
			\node[right] at (3.95,5.5) {$1$};
			\draw[>->,thick] (4,5.5) to(-1,6.5);
			\draw[->,thick] (-1,6.5) to (4,7.5);
			\cross{3.}{5.7}
			\Incross{-0.2}{6.35}
			\Incross{1.37}{6.05}
			\cross{0.9}{6.135}
			
			\node[right] at (3.95,4) {$4$};
			\draw[>->,thick] (4,4) to(1.5,4.5);
			\draw[->,thick] (1.5,4.5) to (4,9.5);
			\cross{3.}{4.15}
			\Incross{1.75}{4.45}
			
			\node[right] at (3.95,3) {$3$};
			\draw[>->,thick] (4,3) to(1.5,3.5);
			\draw[->,thick] (1.5,3.5) to (4,8.5);
			
			\node[right,rotate=11] at (3.3,7.1) {$-u+v-\theta$};
			\node[right,rotate=11] at (3.0,6.6) {$-u+v$};
			\node[right,rotate=11] at (1.4,7.2) {$-u+v$};
			\node[right,rotate=11] at (0.8,6.6) {$-u+v+\theta$};

			\node[right,rotate=-9] at (2.8,6.) {$-u-v-2\eta$};
			\node[right,rotate=-9] at (2.5,5.5) {$-u-v-2\eta+\theta$};
			\node[right,rotate=-11] at (-0.3,5.5) {$-u-v-2\eta$};
			\node[right,rotate=-8] at (-0.48,6.05) {$-u-v-2\eta-\theta$};
			
			\node[right] at (-1.9,6.25) {$-2u-2\eta$};

			\node[right] at (-1.4,6.8) {$u+\frac{\theta}{2}$};
			\node[right] at (-1.4,5.6) {$u-\frac{\theta}{2}$};
			
			\node[right] at (0.2,4.5) {$v+\frac{\theta}{2}$};
			\node[right] at (0.2,3.5) {$v-\frac{\theta}{2}$};
			\node[right] at (0.,4.) {$-2v-2\eta$};
		\end{tikzpicture}
	\end{minipage}
	\hspace{0.1mm}%
	\begin{minipage}[c]{0.45\textwidth}
		\begin{align*}
			&=P_{3,4}R_{1,3}(-u+v-\theta)R_{1,4}(-u+v)R_{2,3}(-u+v)\\
			&\times R_{2,4}(-u+v+\theta)\left(K^{+}_1\left(u+\frac{\theta}{2}\right)\right)^{t_1}M^{-1}_1\\
			&\times  R_{2,1}(-2u-2\eta)M_1\left(K^{+}_2\left(u-\frac{\theta}{2}\right)\right)^{t_2}\\
			&\times M^{-1}_1M^{-1}_2 R_{4,1}(-u-v-2\eta-\theta)\\
			&\times R_{4,2}(-u-v-2\eta)R_{3,1}(-u-v-2\eta)\\
			&\times R_{3,2}(-u-v-2\eta+\theta) M_1M_2\left(K^{+}_4\left(v+\frac{\theta}{2}\right)\right)^{t_4}\\
			&\times M^{-1}_4R_{3,4}(-2v-2\eta)M_4\left(K^{+}_3\left(v-\frac{\theta}{2}\right)\right)^{t_3} P_ {1,2}\,.
		\end{align*}
	\end{minipage}\\[8pt]
	Canceling the operator insertions gives: \\[10pt]
	\begin{minipage}[c]{0.45\textwidth}
		\begin{tikzpicture}
			\footnotesize
			\node[right] at (3.95,5) {$2$};
			\draw[>->,thick] (4,5) to(-1,6);
			\draw[->,thick] (-1,6) to (4,7);
			\cross{3.}{5.2}
			\Incross{-0.2}{5.85}
			
			\node[right] at (3.95,5.5) {$1$};
			\draw[>->,thick] (4,5.5) to(-1,6.5);
			\draw[->,thick] (-1,6.5) to (4,7.5);
			\cross{3.}{5.7}
			\Incross{-0.2}{6.35}

			\node[right] at (3.95,4) {$4$};
			\draw[>->,thick] (4,4) to(1.5,4.5);
			\draw[->,thick] (1.5,4.5) to (4,9.5);
			\cross{3.}{4.15}
			\Incross{1.75}{4.45}
			
			\node[right] at (3.95,3) {$3$};
			\draw[>->,thick] (4,3) to(1.5,3.5);
			\draw[->,thick] (1.5,3.5) to (4,8.5);
			
			\node[right,rotate=11] at (3.3,7.1) {$-u+v-\theta$};
			\node[right,rotate=11] at (3.0,6.6) {$-u+v$};
			\node[right,rotate=11] at (1.4,7.2) {$-u+v$};
			\node[right,rotate=11] at (0.8,6.6) {$-u+v+\theta$};

			\node[right,rotate=-9] at (2.7,6.) {$-u-v-2\eta$};
			\node[right,rotate=-9] at (2.5,5.5) {$-u-v-2\eta+\theta$};
			\node[right,rotate=-11] at (-0.3,5.5) {$-u-v-2\eta$};
			\node[right,rotate=-8] at (-0.48,6.05) {$-u-v-2\eta-\theta$};
			
			\node[right] at (-1.9,6.25) {$-2u-2\eta$};

			\node[right] at (-1.4,6.8) {$u+\frac{\theta}{2}$};
			\node[right] at (-1.4,5.6) {$u-\frac{\theta}{2}$};
			
			\node[right] at (0.2,4.5) {$v+\frac{\theta}{2}$};
			\node[right] at (0.2,3.5) {$v-\frac{\theta}{2}$};
			\node[right] at (0.,4.) {$-2v-2\eta$};
		\end{tikzpicture}
	\end{minipage}
	\hspace{0.1mm}%
	\begin{minipage}[c]{0.45\textwidth}
		\begin{align*}
			&=P_{3,4}R_{1,3}(-u+v-\theta)R_{1,4}(-u+v)R_{2,3}(-u+v)\\
			&\times R_{2,4}(-u+v+\theta)\left(K^{+}_1\left(u+\frac{\theta}{2}\right)\right)^{t_1}M^{-1}_1\\
			&\times R_{2,1}(-2u-2\eta)\left(K^{+}_2\left(u-\frac{\theta}{2}\right)\right)^{t_2}\\
			&\times M^{-1}_2 R_{4,1}(-u-v-2\eta-\theta)\\
			&\times R_{4,2}(-u-v-2\eta)R_{3,1}(-u-v-2\eta)\\
			&\times R_{3,2}(-u-v-2\eta+\theta) M_1M_2\left(K^{+}_4\left(v+\frac{\theta}{2}\right)\right)^{t_4}\\
			&\times M^{-1}_4R_{3,4}(-2v-2\eta) M_4\left(K^{+}_3\left(v-\frac{\theta}{2}\right)\right)^{t_3} P_ {1,2}\,.
		\end{align*}
	\end{minipage}\\[4pt]
	Using the YBE to pass the weight $-2u-2\eta$ to the right side gives:\\[4pt]
	\begin{minipage}[c]{0.43\textwidth}
		\begin{tikzpicture}
			\footnotesize
			\node[right] at (4,5) {$2$};
			\draw[>->,thick] (4,5) to(-1,6);
			\draw[-,thick,out=170,in=10] (3,5.7) to(-1,6);
			\draw[->,thick,out=0,in=-160] (3,5.7) to (4,7);
			
			\cross{3.}{5.2}
			\Incross{-0.2}{5.85}
			
			\node[right] at (4,5.5) {$1$};
			\draw[<-,thick,out=-10,in=-170] (-1,6.5) to (3,6.8);
			\draw[->,thick] (-1,6.5) to (4,7.5);
			\draw[>-,thick,out=170,in=10] (4,5.5) to (3,6.8);
			
			\cross{3.65}{5.7}
			\Incross{-0.2}{6.35}

			\node[right] at (4,4) {$4$};
			\draw[>->,thick] (4,4) to(1.5,4.5);
			\draw[->,thick] (1.5,4.5) to (4,9.5);
			\cross{3.}{4.2}
			\Incross{1.75}{4.45}
			
			\node[right] at (4,3) {$3$};
			\draw[>->,thick] (4,3) to(1.5,3.5);
			\draw[->,thick] (1.5,3.5) to (4,8.5);
			
			\node[right,rotate=11] at (3.6,7.7) {$-u+v-\theta$};
			\node[right,rotate=11] at (3.2,7.1) {$-u-v-2\eta$};
			\node[right,rotate=11] at (1.4,7.2) {$-u+v$};
			\node[right,rotate=11] at (0.1,6.5) {$-u-v-2\eta-\theta$};

			\node[right,rotate=-11] at (2.35,5.55) {$-u+v$};
			\node[right,rotate=-11] at (2.2,5.) {$-u-v-2\eta+\theta$};
			\node[right,rotate=-11] at (-0.3,5.55) {$-u-v-2\eta$};
			\node[right,rotate=-11] at (0.,6.05) {$-u+v+\theta$};
			
			\node[right] at (3.5,6.25) {$-2u-2\eta$};

			\node[right] at (-1.4,6.8) {$u+\frac{\theta}{2}$};
			\node[right] at (-1.4,5.6) {$u-\frac{\theta}{2}$};
			
			\node[right] at (0.2,4.5) {$v+\frac{\theta}{2}$};
			\node[right] at (0.2,3.5) {$v-\frac{\theta}{2}$};
			\node[right] at (0.,4.) {$-2v-2\eta$};
		\end{tikzpicture}
	\end{minipage}
	\hspace{0.1mm}%
	\begin{minipage}[c]{0.45\textwidth}
		\begin{align*}
			&=P_{3,4}R_{1,3}(-u+v-\theta)R_{1,4}(-u+v) \left(K^{+}_1(u+\frac{\theta}{2})\right)^{t_1} \\
			&\times M^{-1}_1R_{4,1}(-u-v-2\eta-\theta)R_{3,1}(-u-v-2\eta)\\
			&\times R_{2,1}(-2u-2\eta) R_{2,3}(-u+v)R_{2,4}(-u+v+\theta)\\
			&\times \left(K^{+}_2\left(u-\frac{\theta}{2}\right)\right)^{t_2}M^{-1}_2 R_{4,2}(-u-v-2\eta)\\
			&\times R_{3,2}(-u-v-2\eta+\theta) M_1M_2\left(K^{+}_4\left(v+\frac{\theta}{2}\right)\right)^{t_4}\\
			&\times M^{-1}_4R_{3,4}(-2v-2\eta)M_4\left(K^{+}_3\left(v-\frac{\theta}{2}\right)\right)^{t_3} P_ {1,2}\,.
		\end{align*}
	\end{minipage}\\[10pt]
	Using the reflection algebra (\ref{Reflection_algebra_Small_K_Plus}) move the $K^+$-matrix with weight $v+\theta/2$ upwards we obtain:\\[10pt]
	\begin{minipage}[c]{0.40\textwidth}
		\begin{tikzpicture}
			\footnotesize
			\node[right] at (4,5) {$2$};
			\draw[>->,thick] (4,5) to(-1,5.5);
			\draw[-,thick,out=170,in=10] (3,5.7) to(-1,5.5);
			\draw[->,thick,out=0,in=-160] (3,5.7) to (4,7);
			
			\cross{3.}{5.1}
			\Incross{0.5}{5.72}
			
			\node[right] at (4,5.5) {$1$};
			\draw[<-,thick,out=-10,in=-170] (-1,6.5+0.5) to (3,6.8);
			\draw[->,thick] (-1,6.5+0.5) to (4,7.5);
			\draw[>-,thick,out=170,in=10] (4,5.5) to(3,6.8);
			
			\cross{3.65}{5.7}
			\Incross{1.85}{5.2}
			\cross{2.15}{5.8}
			\Incross{-0.2}{6.85}

			\node[right] at (4,4) {$4$};
			\draw[>-,thick,in=-20,out=160] (4,4) to(1.65,4.7);
			\draw[-,thick,out=140,in=-110] (1.65,4.7) to (1.9,5.7);
			\draw[->,thick,out=65,in=-10] (1.9,5.7) to(0.25,6.25);
			\draw[-,thick,out=20,in=-110] (0.25,6.25) to (2.65,6.9);
			\draw[->,thick] (2.65,6.9) to (4,9.5);
			
			\cross{3.4}{4.2}
			\Incross{1.75}{4.685}
			
			\node[right] at (4,3.5) {$3$};
			\draw[>->,thick] (4,3.5) to(1.75,3.75+0.375);
			\draw[->,thick] (1.75,3.75+0.375) to (4,8.5);
			
			\node[right,rotate=5] at (3.6,7.7) {$-u+v-\theta$};
			\node[right,rotate=5] at (3.2,7.1) {$-u-v-2\eta$};
			\node[right,rotate=0] at (1.6,7.6) {$-u+v$};
			\node[right,rotate=0] at (0.1,6.92) {$-u-v-2\eta-\theta$};

			\node[right,rotate=-9] at (2.4,5.5) {$-u+v$};
			\node[right,rotate=-9] at (2.2,5.) {$-u-v-2\eta+\theta$};
			\node[right,rotate=-0] at (-0.3,5.55) {$-u-v-2\eta$};
			\node[right,rotate=-0] at (-0.3,5.05) {$-u+v+\theta$};
			
			\node[right] at (3.5,6.25) {$-2u-2\eta$};
			\node[right] at (0.3,4.45) {$-2v-2\eta$};
			
			\node[right] at (-1.4,7.3) {$u+\frac{\theta}{2}$};
			\node[right] at (-1.4,5.2) {$u-\frac{\theta}{2}$};
			
			\node[right] at (-0.75,6.25) {$v+\frac{\theta}{2}$};
			\node[right] at (0.75,4.1) {$v-\frac{\theta}{2}$};
		\end{tikzpicture}
	\end{minipage}
	\hspace{0.01mm}
	\begin{minipage}[c]{0.45\textwidth}
		\begin{align*}
			&=P_{3,4}R_{1,3}(-u+v-\theta)R_{1,4}(-u+v)\left(K^{+}_1(u+\frac{\theta}{2})\right)^{t_1} \\
			&\times M^{-1}_1R_{4,1}(-u-v-2\eta-\theta)R_{3,1}(-u-v-2\eta)\\
			&\times R_{2,1}(-2u-2\eta) R_{2,3}(-u+v)\left(K^{+}_4\left(v+\frac{\theta}{2}\right)\right)^{t_4}\\
			&\times  M_2  R_{2,4}(-u-v-2\eta)  M^{-1}_2  \left(K^{+}_2\left(u-\frac{\theta}{2}\right)\right)^{t_2} \\
			&\times R_{4,2}(-u+v+\theta)M^{-1}_2R_{3,2}(-u-v-2\eta+\theta)M_1\\
			&\times M_2M^{-1}_4R_{3,4}(-2v-2\eta)M_4\left(K^{+}_3\left(v-\frac{\theta}{2}\right)\right)^{t_3} P_ {1,2}\,.
		\end{align*}
	\end{minipage}\\[1pt]
	Using  (\ref{Reflection_algebra_Small_K_Plus}) again this yields:\\[10pt]
	\begin{minipage}[c]{0.45\textwidth}
		\begin{tikzpicture}
			\footnotesize
			\node[right] at (4,5) {$2$};
			\draw[>->,thick] (4,5) to(-1,5.5);
			\draw[-,thick,out=170,in=10] (3,5.7) to(-1,5.5);
			\draw[->,thick,out=0,in=-160] (3,5.7) to (4,7);
			
			\cross{3.}{5.1}
			\Incross{-0.2}{5.65}
			\cross{3.2}{7.4}

			\node[right] at (4,5.5) {$1$};
			\draw[<-,thick,out=-10,in=-170] (-1,6.5+0.5) to (3,6.8);
			\draw[->,thick] (-1,6.5+0.5) to (4,7.5);
			\draw[>-,thick,out=170,in=10] (4,5.5) to (3,6.8);
			
			\cross{3.65}{5.7}
			\Incross{1.85}{5.2}
			\cross{2.15}{5.8}
			\Incross{-0.2}{7.1}

			\node[right] at (4,4) {$4$};
			\draw[>-,thick,in=-20,out=165] (4,4.1) to(1.85,4.7);
			\draw[-,thick,out=160,in=-120] (1.85,4.7) to (2.5,6.7);
			\draw[-,thick] (2.5,6.7) to (3.,7.5);
			\draw[->,thick,out=70,in=-20] (3.,7.5) to (2.0,8);
			\draw[-,thick,out=20,in=-110] (2.0,8) to (3.5,8.5);
			\draw[->,thick] (3.5,8.5)--(4,9.5);

			\cross{3.4}{4.25}
			\Incross{1.78}{4.725}
			
			\node[right] at (4,3.5) {$3$};
			\draw[>->,thick] (4,3.5) to(1.75,3.75+0.375);
			\draw[->,thick] (1.75,3.75+0.375) to (4,8.5);
			
			\node[right,rotate=5] at (3.6,7.7) {$-u+v-\theta$};
			\node[right,rotate=5] at (3.2,7.1) {$-u-v-2\eta$};
			\node[right,rotate=0] at (0.4,7.6) {$-u-v-2\eta-\theta$};
			\node[right,rotate=0] at (1.4,6.92) {$-u+v$};

			\node[right,rotate=-11] at (2.35,5.55) {$-u+v$};
			\node[right,rotate=-11] at (2.2,5.) {$-u-v-2\eta+\theta$};
			\node[right,rotate=-0] at (-0.3,5.55) {$-u-v-2\eta$};
			\node[right,rotate=-0] at (-0.3,5.05) {$-u+v+\theta$};
			
			\node[right] at (3.5,6.25) {$-2u-2\eta$};
			\node[right] at (0.35,4.5) {$-2v-2\eta$};
			
			\node[right] at (-1.4,7.3) {$u+\frac{\theta}{2}$};
			\node[right] at (-1.4,5.2) {$u-\frac{\theta}{2}$};
			
			\node[right] at (1.0,8.25) {$v+\frac{\theta}{2}$};
			\node[right] at (0.72,4.1) {$v-\frac{\theta}{2}$};
		\end{tikzpicture}
	\end{minipage}
	\hfill%
	\begin{minipage}[c]{0.45\textwidth}
		\begin{align*}
			&=P_{3,4}R_{1,3}(-u+v-\theta)\left(K^{+}_4\left(v+\frac{\theta}{2}\right)\right)^{t_4}  M_1\\
			&\times R_{1,4}(-u-v-2\eta-\theta)M^{-1}_1\left(K^{+}_1\left(u+\frac{\theta}{2}\right)\right)^{t_1}\\
			&\times R_{4,1}(-u+v)  M^{-1}_1R_{3,1}(-u-v-2\eta)\\
			&\times R_{2,1}(-2u-2\eta) R_{2,3}(-u+v)\\
			&\times M_2  R_{2,4}(-u-v-2\eta)M^{-1}_2  \left(K^{+}_2\left(u-\frac{\theta}{2}\right)\right)^{t_2} \\
			&\times R_{4,2}(-u+v+\theta)M^{-1}_2 R_{3,2}(-u-v-2\eta+\theta)M_1 \\
			&\times M_2M^{-1}_4 R_{3,4}(-2v-2\eta)M_4\left(K^{+}_3\left(v-\frac{\theta}{2}\right)\right)^{t_3} P_ {1,2}\,.
		\end{align*}
	\end{minipage}\\[1pt]
	Reshuffling the operator insertions via (\ref{Important_Iden_Small_R}) we get\\[10pt]
	\begin{minipage}[c]{0.45\textwidth}
		\begin{tikzpicture}
			\footnotesize
			\node[right] at (4,5) {$2$};
			\draw[>->,thick] (4,5) to(-1,5.5);
			\draw[-,thick,out=170,in=10] (3,5.7) to(-1,5.5);
			\draw[->,thick,out=0,in=-160] (3,5.7) to (4,7);

			\node[right] at (4,5.5) {$1$};
			\draw[<-,thick,out=-10,in=-170] (-1,6.5+0.5) to (3,6.8);
			\draw[->,thick] (-1,6.5+0.5) to (4,7.5);
			\draw[>-,thick,out=170,in=10] (4,5.5) to (3,6.8);

			\node[right] at (4,4) {$4$};
			\draw[>-,thick,in=-20,out=165] (4,4.1) to(1.85,4.7);
			\draw[-,thick,out=160,in=-120] (1.85,4.7) to (2.5,6.7);
			\draw[-,thick] (2.5,6.7) to (3.,7.5);
			\draw[->,thick,out=70,in=-20] (3.,7.5) to (2.0,8);
			\draw[-,thick,out=20,in=-110] (2.0,8) to (3.5,8.5);
			\draw[->,thick] (3.5,8.5)--(4,9.5);
			
			\cross{3.}{5.1}
			\Incross{-0.2}{5.4}
			\cross{3.65}{5.7}
			\Incross{-0.2}{6.85}
			\cross{3.4}{4.25}
			\Incross{3}{7.6}

			\node[right] at (4,3.5) {$3$};
			\draw[>->,thick] (4,3.5) to(1.75,3.75+0.375);
			\draw[->,thick] (1.75,3.75+0.375) to (4,8.5);
			
			\node[right,rotate=5] at (3.6,7.7) {$-u+v-\theta$};
			\node[right,rotate=5] at (3.2,7.1) {$-u-v-2\eta$};
			\node[right,rotate=0] at (0.4,7.6) {$-u-v-2\eta-\theta$};
			\node[right,rotate=0] at (1.4,6.92) {$-u+v$};

			\node[right,rotate=-11] at (2.35,5.55) {$-u+v$};
			\node[right,rotate=-11] at (2.2,5.) {$-u-v-2\eta+\theta$};
			\node[right,rotate=-0] at (-0.3,5.55) {$-u-v-2\eta$};
			\node[right,rotate=-0] at (-0.3,5.05) {$-u+v+\theta$};
			
			\node[right] at (3.5,6.25) {$-2u-2\eta$};
			\node[right] at (0.35,4.5) {$-2v-2\eta$};
			
			\node[right] at (-1.4,7.3) {$u+\frac{\theta}{2}$};
			\node[right] at (-1.4,5.2) {$u-\frac{\theta}{2}$};
			
			\node[right] at (1.0,8.25) {$v+\frac{\theta}{2}$};
			\node[right] at (0.72,4.1) {$v-\frac{\theta}{2}$};
		\end{tikzpicture}
	\end{minipage}
	\hfill%
	\begin{minipage}[c]{0.45\textwidth}
		\begin{align*}
			&=P_{3,4}R_{1,3}(-u+v-\theta)\left(K^{+}_4\left(v+\frac{\theta}{2}\right)\right)^{t_4}  M^{-1}_4\\
			&\times R_{1,4}(-u-v-2\eta-\theta)\left(K^{+}_1\left(u+\frac{\theta}{2}\right)\right)^{t_1}M^{-1}_1\\
			&\times R_{4,1}(-u+v)  R_{3,1}(-u-v-2\eta)R_{2,1}(-2u-2\eta)\\
			&\times   R_{2,3}(-u+v) R_{2,4}(-u-v-2\eta) \left(K^{+}_2\left(u-\frac{\theta}{2}\right)\right)^{t_2}  \\
			&\times M^{-1}_2R_{4,2}(-u+v+\theta) R_{3,2}(-u-v-2\eta+\theta)\\
			&\times M_1M_2 R_{3,4}(-2v-2\eta)M_4\left(K^{+}_3\left(v-\frac{\theta}{2}\right)\right)^{t_3} P_ {1,2}\,.
		\end{align*}
	\end{minipage}\\[10pt]
	By using the YBE to bring the weight $-2v-2\eta$ to the top we obtain:\\[10pt]
	\begin{minipage}[c]{0.42\textwidth}
		\begin{tikzpicture}
			\footnotesize
			\node[right] at (4,5) {$2$};
			\draw[>->,thick] (4,5) to(-1,5.5);
			\draw[-,thick,out=170,in=10] (3,5.7) to(-1,5.5);
			\draw[->,thick,out=0,in=-160] (3,5.7) to (4,7);
			
			\cross{3.}{5.1}
			\Incross{-0.2}{5.4}
			
			\node[right] at (4,5.5) {$1$};
			\draw[>-,thick,out=170,in=10] (4,5.5) to(3,6.8);
			\draw[<-,thick,out=-10,in=-170] (-1,6.5+0.5) to (3,6.8);
			\draw[->,thick] (-1,6.5+0.5) to (4,7.5);

			\cross{3.65}{5.7}
			\Incross{-0.2}{6.85}
			
			\node[right] at (4,3.5) {$3$};
			\draw[>->,thick,in=-20,out=160] (4,3.5) to (1.5,4.55);
			\draw[-,thick,out=70,in=-120] (1.5,4.55) to (2.5,6.7);
			\draw[->,thick,out=60,in=-150] (2.5,6.7) to (4,8.5);

			\cross{3.4}{4.25}
			\Incross{3.25}{8.6}
			
			\node[right] at (4,4) {$4$};
			\draw[>-,thick,in=-20,out=160] (4,4) to (2.75,4.5);
			\draw[-,thick,in=-110,out=160] (2.75,4.5) to (3.48,7.5);
			\draw[-,thick,in=-110,out=60] (3.48,7.5) to (3.25,8.5);
			\draw[->,thick,out=70,in=-20] (3.25,8.5) to (2.5,9);
			\draw[->,thick,out=20,in=-110] (2.5,9) to (4,9.5);
			
			\node[right,rotate=5] at (3.6,7.7) {$-u-v-2\eta-\theta$};
			\node[right,rotate=5] at (3.4,7.1) {$-u+v$};
			\node[right,rotate=5] at (1.2,7.5) {$-u+v-\theta$};
			\node[right,rotate=0] at (0.6,6.92) {$-u-v-2\eta$};

			\node[right,rotate=-11] at (2.35,5.55) {$-u-v-2\eta$};
			\node[right,rotate=-11] at (2.3,5.) {$-u+v+\theta$};
			\node[right,rotate=-0] at (0.7,5.65) {$-u+v$};
			\node[right,rotate=2] at (-0.8,4.9) {$-u-v-2\eta+\theta$};
			
			\node[right] at (3.5,6.25) {$-2u-2\eta$};
			\node[right] at (1.75,8.) {$-2v-2\eta$};
			
			\node[right] at (-1.4,7.3) {$u+\frac{\theta}{2}$};
			\node[right] at (-1.4,5.2) {$u-\frac{\theta}{2}$};
			
			\node[right] at (1.5,9) {$v+\frac{\theta}{2}$};
			\node[right] at (0.5,4.5) {$v-\frac{\theta}{2}$};
		\end{tikzpicture}
	\end{minipage}
	\hspace{0.1mm}
	\begin{minipage}[c]{0.45\textwidth}
		\begin{align*}
			&=P_{3,4}\left(K^{+}_4\left(v+\frac{\theta}{2}\right)\right)^{t_4}  M^{-1}_4R_{3,4}(-2v-2\eta)\\
			&\times  R_{1,4}(-u-v-2\eta-\theta)R_{1,3}(-u+v-\theta) \\
			&\times\left(K^{+}_1(u+\frac{\theta}{2})\right)^{t_1} M^{-1}_1 R_{3,1}(-u-v-2\eta)R_{4,1}(-u+v) \\
			&\times  R_{2,1}(-2u-2\eta) R_{2,4}(-u-v-2\eta) R_{2,3}(-u+v) \\
			&\times \left(K^{+}_2\left(u-\frac{\theta}{2}\right)\right)^{t_2}M^{-1}_2 R_{3,2}(-u-v-2\eta+\theta)  \\
			&\times R_{4,2}(-u+v+\theta) M_1M_2 M_4\left(K^{+}_3\left(v-\frac{\theta}{2}\right)\right)^{t_3} P_ {1,2}\,.
		\end{align*}
	\end{minipage}\\[10pt]
	Similar as above we use the reflection algebra (\ref{Reflection_algebra_Small_K_Plus}) twice to get\\[10pt]
	\begin{minipage}[c]{0.40\textwidth}
		\begin{tikzpicture}
			\footnotesize
			\node[right] at (4,5) {$2$};
			\draw[>->,thick] (4,5) to(-1,5.5);
			\draw[-,thick,out=170,in=10] (3,5.7) to(-1,5.5);
			\draw[->,thick,out=0,in=-160] (3,5.7) to (4,7);
			
			\cross{3.}{5.1}
			\Incross{-0.2}{5.65}
			
			\node[right] at (4,5.5) {$1$};
			\draw[<-,thick,out=-10,in=-170] (-1,6.5+0.5) to (3,6.8);
			\draw[->,thick] (-1,6.5+0.5) to (4,7.5);
			\draw[>-,thick,out=170,in=10] (4,5.5) to(3,6.8);
			
			\cross{3.65}{5.7}
			\Incross{-0.2}{7.1}
			
			\cross{3.5}{4.15}
			\Incross{2}{5.2}
			\cross{2.25}{5.8}
			\Incross{2.8}{6.75}
			\cross{3}{7.4}
			
			\node[right] at (4,3.5) {$3$};
			
			\draw[>-,thick,in=-20,out=160] (4,3.5) to (1.85,4.2);
			\draw[-,thick,out=160,in=-110] (1.85,4.2) to (2.5,6.7);
			\draw[->,thick,out=60,in=-20] (2.5,6.7) to (2,8);
			\draw[->,thick,out=20,in=-120] (2,8) to (4,8.5);

			\Incross{3.25}{8.6}
			
			\node[right] at (4,4) {$4$};
			\draw[>-,thick,in=-20,out=160] (4,4) to (2.75,4.4);
			\draw[-,thick,in=-110,out=160] (2.75,4.4) to (3.48,7.5);
			\draw[-,thick,in=-110,out=60] (3.48,7.5) to (3.25,8.5);
			\draw[->,thick,out=70,in=-20] (3.25,8.5) to (2.5,9);
			\draw[->,thick,out=20,in=-110] (2.5,9) to (4,9.5);
			
			\node[right,rotate=10] at (3.4,7.6) {$-u-v-2\eta-\theta$};
			\node[right,rotate=5] at (3.4,7.1) {$-u+v$};
			\node[right,rotate=5] at (0.45,7.4) {$-u-v-2\eta$};
			\node[right,rotate=0] at (0.6,6.92) {$-u+v-\theta$};

			\node[right,rotate=-11] at (2.35,5.55) {$-u-v-2\eta$};
			\node[right,rotate=-11] at (2.2,5.) {$-u+v+\theta$};
			\node[right,rotate=-0] at (0.5,5.55) {$-u+v$};
			\node[right,rotate=0] at (-0.5,6.) {$-u-v-2\eta+\theta$};

			\node[right] at (3.5,6.25) {$-2u-2\eta$};
			\node[right] at (1.6,8.35) {$-2v-2\eta$};
			
			\node[right] at (-1.4,7.3) {$u+\frac{\theta}{2}$};
			\node[right] at (-1.4,5.2) {$u-\frac{\theta}{2}$};
			
			\node[right] at (1.5,9) {$v+\frac{\theta}{2}$};
			\node[right] at (1.0,8.0) {$v-\frac{\theta}{2}$};
		\end{tikzpicture}
	\end{minipage}
	\hspace{0.1mm}
	\begin{minipage}[c]{0.42\textwidth}
		\begin{align*}
			&=P_{3,4}\left(K^{+}_4\left(v+\frac{\theta}{2}\right)\right)^{t_4} M^{-1}_4 R_{3,4}(-2v-2\eta)\\
			&\times  \left(K^{+}_3\left(v-\frac{\theta}{2}\right)\right)^{t_3}R_{1,4}(-u-v-2\eta-\theta)M_1\\
			&\times R_{1,3}(-u-v-2\eta)M_1^{-1} \Big(K^{+}_1(u+\frac{\theta}{2})\Big)^{t_1} R_{3,1}(-u+v-\theta)\\
			&\times  M^{-1}_1R_{4,1}(-u+v) R_{2,1}(-2u-2\eta)R_{2,4}(-u-v-2\eta)\\
			&\times   M_2 R_{2,3}(-u-v-2\eta+\theta)M^{-1}_2  \left(K^{+}_2\left(u-\frac{\theta}{2}\right)\right)^{t_2}   \\
			&\times  R_{3,2}(-u+v)M_2^{-1}R_{4,2}(-u+v+\theta) M_1M_2 M_4 P_ {1,2}\,.
		\end{align*}
	\end{minipage}\\[7pt]
	We reshuffle the operator insertions again to obtain:\\[7pt]
	\begin{minipage}[c]{0.45\textwidth}
		\begin{tikzpicture}
			\footnotesize
			\node[right] at (4,5) {$2$};
			\draw[>->,thick] (4,5) to(-1,5.5);
			\draw[->,thick,out=170,in=10] (3,5.7) to(-1,5.5);
			\draw[->,thick,out=0,in=-160] (3,5.7) to (4,7);

			\node[right] at (4,5.5) {$1$};
			\draw[<-,thick,out=-10,in=-170] (-1,6.5+0.5) to (3,6.8);
			\draw[->,thick] (-1,6.5+0.5) to (4,7.5);
			\draw[>-,thick,out=170,in=10] (4,5.5) to(3,6.8);

			\cross{3.}{5.1}
			\cross{3.65}{5.7}
			\Incross{-0.2}{7.1}
			\cross{3.5}{4.15}
			\cross{1.85}{5.6}
			\Incross{2}{5.2}
			\Incross{2.6}{7.05}
			\Incross{1.8}{6.69}
			\cross{3}{7.4}
			\Incross{3.25}{8.6}
			\node[right] at (4,3.5) {$3$};
			
			\draw[>-,thick,in=-20,out=160] (4,3.5) to (1.85,4.2);
			\draw[-,thick,out=160,in=-110] (1.85,4.2) to (2.5,6.7);
			\draw[->,thick,out=60,in=-20] (2.5,6.7) to (2,8);
			\draw[->,thick,out=20,in=-120] (2,8) to (4,8.5);

			\node[right] at (4,4) {$4$};
			\draw[>-,thick,in=-20,out=160] (4,4) to (2.75,4.4);
			\draw[-,thick,in=-110,out=160] (2.75,4.4) to (3.48,7.5);
			\draw[-,thick,in=-110,out=60] (3.48,7.5) to (3.25,8.5);
			\draw[->,thick,out=70,in=-20] (3.25,8.5) to (2.5,9);
			\draw[->,thick,out=20,in=-110] (2.5,9) to (4,9.5);
			
			\node[right,rotate=5] at (3.6,7.7) {$-u-v-2\eta-\theta$};
			\node[right,rotate=5] at (3.4,7.1) {$-u+v$};
			\node[right,rotate=5] at (0.45,7.4) {$-u-v-2\eta$};
			\node[right,rotate=0] at (0.6,6.92) {$-u+v-\theta$};

			\node[right,rotate=-11] at (2.35,5.55) {$-u-v-2\eta$};
			\node[right,rotate=-11] at (2.2,5.) {$-u+v+\theta$};
			\node[right,rotate=-0] at (0.5,5.55) {$-u+v$};
			\node[right,rotate=0] at (-0.5,6.) {$-u-v-2\eta+\theta$};

			\node[right] at (3.5,6.25) {$-2u-2\eta$};
			\node[right] at (1.6,8.35) {$-2v-2\eta$};
			
			\node[right] at (-1.4,7.3) {$u+\frac{\theta}{2}$};
			\node[right] at (-1.4,5.2) {$u-\frac{\theta}{2}$};
			
			\node[right] at (1.5,9) {$v+\frac{\theta}{2}$};
			\node[right] at (1.0,8.0) {$v-\frac{\theta}{2}$};
		\end{tikzpicture}
	\end{minipage}
	\hfill%
	\begin{minipage}[c]{0.45\textwidth}
		\begin{align*}
			&=P_{3,4}\left(K^{+}_4\left(v+\frac{\theta}{2}\right)\right)^{t_4} M^{-1}_4 R_{3,4}(-2v-2\eta)\\
			&\times \left(K^{+}_3\left(v-\frac{\theta}{2}\right)\right)^{t_3}R_{1,4}(-u-v-2\eta-\theta)M_1\\
			&\times R_{1,3}(-u-v-2\eta)M_1^{-1} \left(K^{+}_1\left(u+\frac{\theta}{2}\right)\right)^{t_1}M^{-1}_1 \\
			&\times  M^{-1}_3R_{3,1}(-u+v-\theta)R_{4,1}(-u+v) R_{2,1}(-2u-2\eta)\\
			&\times   R_{2,4}(-u-v-2\eta)R_{2,3}(-u-v-2\eta+\theta)   \\
			&\times M_3\left(K^{+}_2\left(u-\frac{\theta}{2}\right)\right)^{t_2} R_{3,2}(-u+v)M_2^{-1}  \\
			&\times R_{4,2}(-u+v+\theta) M_1M_2 M_4 P_{1,2}\,.
		\end{align*}
	\end{minipage}\\[5pt]
	Now we can use the YBE to bring the weight $-2u-2\eta$ back to left.\\
	\begin{minipage}[c]{0.45\textwidth}
		\begin{tikzpicture}
			\footnotesize
			\node[right] at (4,5) {$2$};
			\draw[>->,thick] (4,5) to(-1,5.5);
			\draw[-,thick,out=-170,in=10] (0.2,5.7) to(-1,5.5);
			\draw[-,thick,out=10,in=-170] (0.2,5.7) to (1,6.7);
			\draw[->,thick,out=10,in=-175] (1,6.7) to (4,7);
			\cross{3.}{5.1}
			
			\node[right] at (4,5.5) {$1$};
			\draw[<-,thick,out=-10,in=170] (-1,6.5+0.5) to (0.2,6.8);
			\draw[->,thick] (-1,6.5+0.5) to (4,7.5);
			\draw[-,thick,out=170,in=-10] (1,5.85) to (0.2,6.8);
			\draw[-<,thick,out=-10,in=170] (1,5.85) to (4,5.5);
			
			\cross{3.65}{5.55}
			\Incross{-0.2}{7.1}

			\node[right] at (4,3.5) {$3$};
			\cross{3.5}{4.15}
			\cross{1.9}{5.45}
			\Incross{2}{5.2}
			\Incross{2.6}{7.05}
			\Incross{-0.2}{6.9}
			
			\cross{3}{7.4}
			\draw[>-,thick,in=-20,out=160] (4,3.5) to (2.25,4.05);
			\draw[-,thick,out=160,in=-120] (2.25,4.05) to (2.5,6.7);
			\draw[->,thick,out=60,in=-20] (2.5,6.7) to (2,8);
			\draw[->,thick,out=20,in=-120] (2,8) to (4,8.5);

			\Incross{3.25}{8.6}
			
			\node[right] at (4,4) {$4$};
			\draw[>-,thick,in=-20,out=160] (4,4) to(2.75,4.35);
			\draw[-,thick,in=-110,out=160] (2.75,4.35) to (3.48,7.5);
			\draw[-,thick,in=-110,out=60] (3.48,7.5) to (3.25,8.5);
			\draw[->,thick,out=70,in=-20] (3.25,8.5) to (2.5,9);
			\draw[->,thick,out=20,in=-110] (2.5,9) to (4,9.5);
			
			\node[right,rotate=5] at (3.6,7.7) {$-u-v-2\eta-\theta$};
			\node[right,rotate=5] at (3.4,7.1) {$-u-v-2\eta$};
			\node[right,rotate=5] at (0.45,7.4) {$-u-v-2\eta$};
			\node[right,rotate=0] at (0,6.99) {$-u-v-2\eta+\theta$};

			\node[right,rotate=-11] at (2.35,5.55) {$-u+v$};
			\node[right,rotate=-11] at (2.2,5.) {$-u+v+\theta$};
			\node[right,rotate=-0] at (-0.,5.55) {$-u+v-\theta$};
			\node[right,rotate=0] at (0.2,5.) {$-u+v$};

			\node[right] at (3.5,6.25) {$-2u-2\eta$};
			\node[right] at (1.6,8.35) {$-2v-2\eta$};
			
			\node[right] at (-1.4,7.3) {$u+\frac{\theta}{2}$};
			\node[right] at (-1.4,5.2) {$u-\frac{\theta}{2}$};
			
			\node[right] at (1.5,9) {$v+\frac{\theta}{2}$};
			\node[right] at (1.0,8.0) {$v-\frac{\theta}{2}$};
		\end{tikzpicture}
	\end{minipage}
	\hspace{1mm}
	\begin{minipage}[c]{0.45\textwidth}
		\begin{align*}
			&=P_{3,4}\left(K^{+}_4\left(v+\frac{\theta}{2}\right)\right)^{t_4} M^{-1}_4 R_{3,4}(-2v-2\eta)\\
			&\times \left(K^{+}_3\left(v-\frac{\theta}{2}\right)\right)^{t_3}R_{1,4}(-u-v-2\eta-\theta)M_1\\
			&\times R_{1,3}(-u-v-2\eta)M_1^{-1} \left(K^{+}_1\left(u+\frac{\theta}{2}\right)\right)^{t_1}M^{-1}_1 \\
			&\times  M^{-1}_3 R_{2,4}(-u-v-2\eta)R_{2,3}(-u-v-2\eta+\theta) \\
			&\times   R_{2,1}(-2u-2\eta)R_{3,1}(-u+v-\theta)R_{4,1}(-u+v)  \\
			&\times M_3\left(K^{+}_2\left(u-\frac{\theta}{2}\right)\right)^{t_2} R_{3,2}(-u+v)M_2^{-1}  \\
			&\times R_{4,2}(-u+v+\theta) M_1M_2 M_4 P_ {1,2}\,.
		\end{align*}
	\end{minipage}\\[10pt]
	Now we reshuffle the operator insertion a last time (\ref{Important_Iden_Small_R}) to obtain finally:\\[10pt]
	\begin{minipage}[c]{0.45\textwidth}
		\begin{tikzpicture}
			\footnotesize
			\node[right] at (4,4.7) {$2$};
			\draw[>->,thick] (4,5) to(-1,6);
			\draw[>->,thick] (-1,6) to (4,7);
			
			\node[right] at (4,5.5) {$1$};
			\draw[>->,thick] (4,5.5) to(-1,6.5);
			\draw[>->,thick] (-1,6.5) to (4,7.5);

			\node[right] at (4,4) {$4$};
			\draw[>->,thick] (4,4) to(1,9);
			\draw[->,thick] (1,9) to (4,10);

			\node[right] at (4,3) {$3$};
			\draw[>->,thick] (4,3) to(1,8);
			\draw[->,thick] (1,8) to (4,9);
			
			\node[right,rotate=11] at (2.1,6.9) {$-u-v-2\eta-\theta$};
			\node[right,rotate=11] at (2.5,6.4) {$-u-v-2\eta$};
			\node[right,rotate=11] at (-0.5,6.8) {$-u-v-2\eta$};
			\node[right,rotate=5] at (-0.7,6.5) {$-u-v-2\eta+\theta$};

			\node[right,rotate=-11] at (2.8,6.) {$-u+v$};
			\node[right,rotate=-11] at (3.2,5.4) {$-u+v+\theta$};
			\node[right,rotate=-11] at (1.5,5.3) {$-u+v$};
			\node[right,rotate=-11] at (0.6,5.9) {$-u+v-\theta$};
			
			\node[right] at (-1.9,6.25) {$-2v-2\eta$};
			\node[right] at (-0.2,8.3) {$-2u-2\eta$};
			
			\node[right] at (-1.4,6.8) {$v+\frac{\theta}{2}$};
			\node[right] at (-1.4,5.6) {$v-\frac{\theta}{2}$};
			
			\node[right] at (-0.2,9.0) {$u+\frac{\theta}{2}$};
			\node[right] at (-0.2,8.0) {$u-\frac{\theta}{2}$};

			\Incross{1.3}{8.5}
			\cross{1.6}{8.}
			
			\cross{3.5}{6.9}
			\cross{3.5}{7.4}
			
			\Incross{1.}{6.9}
			\Incross{1.}{6.4}
			
			\cross{1.}{6.1}
			\Incross{-0.3}{6.4}
			
		\end{tikzpicture}
	\end{minipage}
	\hfill%
	\begin{minipage}[c]{0.45\textwidth}
		\begin{align*}
			&=P_{3,4}\left(K^{+}_4\left(v+\frac{\theta}{2}\right)\right)^{t_4} M^{-1}_4R_{3,4}(-2v-2\eta) M_4\\
			&\times\left(K^{+}_3\left(v-\frac{\theta}{2}\right)\right)^{t_3} M_1M_2 R_{1,4}(-u-v-2\eta-\theta) \\
			&\times R_{1,3}(-u-v-2\eta)R_{2,4}(-u-v-2\eta)\\
			&\times  R_{2,3}(-u-v-2\eta+\theta) M^{-1}_1M^{-1}_2\left(K^{+}_1\left(u+\frac{\theta}{2}\right)\right)^{t_1}  \\
			&\times  M_1^{-1} R_{2,1}(-2u-2\eta)M_1\left(K^{+}_2\left(u-\frac{\theta}{2}\right)\right)^{t_2}  \\
			&\times R_{3,1}(-u+v-\theta)R_{4,1}(-u+v)M_3 R_{3,2}(-u+v) \\
			&\times R_{4,2}(-u+v+\theta) M_1M_2 M_4 P_ {1,2}\,.
		\end{align*}
	\end{minipage}\\[10pt]
	Now we proceed algebraically. We reorder the permutation operators
	\begin{align*}
		\mathbb{R}_{1,2\tn 3,4}&(-u+v,-\theta,-\theta)\left(\mathbb{K}^{+}_{1,2}(u,\theta)\right)^{t_1t_2}\mathbb{M}^{-1}_{1,2}\mathbb{R}_{3,4\tn 1,2}(-u-v-2\eta,-\theta,-\theta)\mathbb{M}_{1,2}\left(\mathbb{K}^{+}_{3,4}(v,\theta)\right)^{t_3t_4}\\
		=&\left(K^{+}_3(v+\frac{\theta}{2})\right)^{t_3} M^{-1}_3 R_{4,3}(-2v-2\eta)M_3\left(K^{+}_4(v-\frac{\theta}{2})\right)^{t_4}P_{3,4}M_1M_2\\
		&\times R_{1,4}(-u-v-2\eta-\theta) R_{1,3}(-u-v-2\eta)R_{2,4}(-u-v-2\eta)R_{2,3}(-u-v-2\eta+\theta)\\
		&\times   M^{-1}_1M^{-1}_2\left(K^{+}_1\left(u+\frac{\theta}{2}\right)\right)^{t_1}M_1^{-1}   R_{2,1}(-2u-2\eta)M_1\left(K^{+}_2\left(u-\frac{\theta}{2}\right)\right)^{t_2} P_ {1,2} \\
		&\times R_{3,2}(-u+v-\theta)R_{4,2}(-u+v)M_3 R_{3,1}(-u+v) R_{4,1}(-u+v+\theta) M_1M_2 M_4 \,,
	\end{align*}
	and  reintroduce the transposition, while using (\ref{R_PT_Sym}) to obtain
	\begin{align*}
		\mathbb{R}_{1,2\tn 3,4}&(-u+v,-\theta,-\theta)\left(\mathbb{K}^{+}_{1,2}(u,\theta)\right)^{t_1t_2}\mathbb{M}^{-1}_{1,2}\mathbb{R}_{3,4\tn 1,2}(-u-v-2\eta,-\theta,-\theta)\mathbb{M}_{1,2}\left(\mathbb{K}^{+}_{3,4}(v,\theta)\right)^{t_3t_4}\\
		=&\left(P_{3,4}K^{+}_4(v-\frac{\theta}{2}) M_3 R_{3,4}(-2v-2\eta)M^{-1}_3K^{+}_3(v+\frac{\theta}{2})\right)^{t_3t_4}M_1M_2\\
		&\times R_{1,4}(-u-v-2\eta-\theta) R_{1,3}(-u-v-2\eta)R_{2,4}(-u-v-2\eta)R_{2,3}(-u-v-2\eta+\theta)\\
		&\times   M^{-1}_1M^{-1}_2\left(P_{1,2}K^{+}_2(u-\frac{\theta}{2})M_1   R_{1,2}(-2u-2\eta)M^{-1}_1K^{+}_1(u+\frac{\theta}{2})\right)^{t_1,t_2} \\
		&\times R_{3,2}(-u+v-\theta)R_{4,2}(-u+v)M_3 R_{3,1}(-u+v) R_{4,1}(-u+v+\theta) M_1M_2 M_4 \\=& \left(\mathbb{K}^{+}_{3,4}(v,\theta)\right)^{t_3t_4} \mathbb{M}_{1,2}\mathbb{R}_{3,4\tn 1,2}(-u-v-2\eta,-\theta,-\theta) \mathbb{M}^{-1}_{1}
		\left(\mathbb{K}^{+}_{1,2}(u,\theta)\right)^{t_1t_2}
		\mathbb{R}_{1,2\tn 3,4}(-u+v,-\theta,-\theta)\,,
	\end{align*}
	which completes the proof.

	
	%
	
\end{document}